\documentclass[twocolumn,aps,pra,longbibliography,superscriptaddress,nofootinbib,10pt]{revtex4-1}
\usepackage[latin1]{inputenc}
\usepackage{graphicx,dcolumn,bm}
\usepackage{subfigure}
\usepackage{multirow}
\usepackage{array}
\usepackage{arydshln}
\usepackage{color}
\usepackage[colorlinks,linkcolor=blue,citecolor=blue,hyperindex]{hyperref}
\usepackage{amsfonts}
\usepackage{amssymb}
\usepackage{amsmath}
\usepackage{latexsym}
\usepackage{simplewick}
\usepackage{epstopdf}
\usepackage{multirow}
\usepackage{float}
\usepackage[table]{xcolor}
\usepackage{multibib}
\usepackage{mathrsfs}
\usepackage[sans]{dsfont}
\usepackage[mathscr]{eucal}
\usepackage{epsfig}
\definecolor{Blue}{rgb}{0.0,0.0,1}
\definecolor{Red}{rgb}{1,0.0,0.0}
\definecolor{Green}{rgb}{0,0.5,0.0}
\usepackage{tikz}
\usetikzlibrary{decorations.pathmorphing}
\usetikzlibrary{arrows}
\usetikzlibrary{intersections,shapes.arrows}
\usetikzlibrary{calc}
\usetikzlibrary{quotes,angles}
\usepackage{nicefrac}
\usepackage{pgfplots}
\usepgfplotslibrary{fillbetween}
\pgfplotsset{compat=1.13,colormap={violetnew}{rgb=(0.293416, 0.0574044, 0.529412) rgb=(0.394818,0.233715,0.671945) rgb =(0.49622,0.410025,0.814477) rgb=(0.588672,0.567494,0.910066) rgb=(0.663226,0.687282,0.911765) rgb=(0.73778,0.807069,0.913465) rgb=(0.807267,0.861883,0.894034) rgb=(0.874222,0.884211,0.864039) rgb=(0.941176, 0.906538, 0.834043)}}
\usepgfplotslibrary{groupplots} 
\usepgfplotslibrary[groupplots] 
\usetikzlibrary{pgfplots.groupplots} 
\usetikzlibrary[pgfplots.groupplots] 
\usepgfplotslibrary{statistics}
\usepackage{pgfplotstable}
\tikzset{jumpdot/.style={mark=*,solid},excl/.append style={jumpdot,fill=white},incl/.append style={jumpdot,fill=black}}
\begin{document}

\title{Relating relative R\'enyi entropies and Wigner-Yanase-Dyson skew information to generalized multiple quantum coherences}
\author{Diego Paiva Pires}
\affiliation{Departamento de F\'{i}sica Te\'{o}rica e Experimental, Universidade Federal do Rio Grande do Norte, 59072-970 Natal, Rio Grande do Norte, Brazil}
\author{Augusto Smerzi}
\affiliation{QSTAR, CNR-INO and LENS, Largo Enrico Fermi 2, I-50125 Firenze, Italy}
\author{Tommaso Macr\`{i}}
\affiliation{Departamento de F\'{i}sica Te\'{o}rica e Experimental, Universidade Federal do Rio Grande do Norte, 59072-970 Natal, Rio Grande do Norte, Brazil}
\affiliation{International Institute of Physics, Federal University of Rio Grande do Norte, Campus Universit\'{a}rio, Lagoa Nova, Natal-RN 59078-970, Brazil}

\begin{abstract}
Quantum coherence is a crucial resource for quantum information processing. By employing the language of coherence orders largely applied in NMR systems, quantum coherence has been currently addressed in terms of multiple quantum coherences (MQCs). Here we investigate the $\alpha$-MQCs, a novel class of multiple quantum coherences which is based on $\alpha$-relative purity, an information-theoretic quantifier analogous to quantum fidelity and closely related to R\'{e}nyi relative entropy of order $\alpha$. Our framework enables linking $\alpha$-MQCs to Wigner-Yanase-Dyson skew information (WYDSI), an asymmetry monotone finding applications in quantum thermodynamics and quantum metrology. Furthermore, we derive a family of bounds on $\alpha$-MQCs, particularly showing that $\alpha$-MQC define a lower bound to quantum Fisher information (QFI). We illustrate these ideas for quantum systems described by single-qubit states, two-qubit Bell-diagonal states, and a wide class of multiparticle mixed states. Finally, we investigate the time evolution of the $\alpha$-MQC spectrum and the overall signal of relative purity, by simulating the time reversal dynamics of a many-body all-to-all Ising Hamiltonian and comment on applications to physical platforms such as NMR systems, trapped ions, and ultracold atoms.
\end{abstract}


\maketitle


\section{Introduction}
\label{sec:introd000_xxx_0001}

Quantum coherence is a primary signature of quantum mechanics. It plays, together with entanglement, a central role in quantum technologies~\cite{LARS_2018} as well as in fundamental physics, including   
quantum thermodynamics~\cite{PhysRevLett.113.150402,PhysRevLett.115.210403}, quantum phase transitions~\cite{PhysRevB.90.104431,PhysRevB.93.184428}, and quantum bio\-logy~\cite{2014_Nature_10_621,10.1063_1.4953243}. Modern approaches include the formulation of quantum coherence within an axiomatic resource theory~\cite{RevModPhys.89.041003}. Quantum coherence can also be addressed through the framework of multiple quantum coherences (MQCs), also known as coherence orders, which were introduced in the eighties in the context of nuclear magnetic resonance (NMR)~\cite{Munowitz,Keeler}. MQCs finds applications ranging from solid-state spectroscopy~\cite{10.1063_1.449344,MUNOWITZ525,10.1021_ja00284a001,KHITRIN1997217} to many-body localization effects induced by decoherence~\cite{PhysRevA.84.012320}, entanglement witnessing~\cite{PhysRevA.78.042301}, and quantum me\-tro\-lo\-gy~\cite{PhysRevA.98.032101}.

Noteworthy, it has been recently proved that MQCs provide a useful criterion 
to probe the buildup of entanglement in quantum many-body systems with long-range interactions~\cite{PhysRevLett.120.040402}. Fur\-ther\-more, MQCs have also contributed to elucidate the role played by coherence orders into the delocalization of quantum information signaled by out-of-time-order correlation functions (OTOCs), recently measured with a quantum simulator implementing the time-reversal dynamics of a fully connected Ising model~\cite{101038nphys4119v01}. Linking MQC and OTOC have triggered experimental investigations ranging from many-body localization in solid-state spin systems~\cite{PhysRevLett.120.070501,PhysRevLett.124.030601,2001_arxiv_2001.02176} to pre\-ther\-ma\-li\-za\-tion effects emerging in the non-equilibrium dynamics in a NMR quantum simulator~\cite{PhysRevLett.123.090605,PhysRevX.7.031011}, and also distinguishing effects of scrambling from decoherence~\cite{10.1038nphys4119}.

Despite the growing interest into MQCs, little is known about its connection with higher order R\'{e}nyi entropies, or even the relation of the second R\'{e}nyi entropy and MQCs. The situation is also unclear for $\alpha$-R\'{e}nyi relative entropies ($\alpha$-RRE), which take an important role in quantum thermodynamics~\cite{Brandao3275,Wei_2017,PhysRevE.99.050101}, quantum communication~\cite{10.1063_1.4960099}, co\-he\-rence quantifiers~\cite{PhysRevLett.117.030401,PhysRevA.94.052336,PhysRevA.93.032136,Streltsov_2018}, and Gaussian states~\cite{10.1063_1.5007167}. So far, promising theoretical achievements discussed the feasibility of probing entanglement by measuring R\'{e}nyi entropies which, up to now, remains a challenge~\cite{PhysRevLett.89.127902,PhysRevLett.106.150404,PhysRevLett.109.020504,2020_arxiv_2007.06305}. Typically, ex\-pe\-ri\-men\-tal results mainly focus on second order R\'{e}nyi entropy by exploiting its relationship with quantum purity of the many-body system~\cite{PhysRevLett.120.050406}. Indeed, significant progress have been made in measuring se\-cond order R\'{e}nyi entropy of a four-site Bose-Hubbard system~\cite{20159_arxiv_1509.01160}, the two-site Fermi-Hubbard model on trapped ion simulator~\cite{PhysRevA.98.052334}, and the quantum long-range XY model~\cite{Brydges260}. Further results include measuring R\'{e}nyi entropy of order $\alpha = 2,3,4$ in the context of quench dynamics of bosons in 1D optical lattices~\cite{PhysRevLett.109.020505}.

Here we promote a study of multiple quantum co\-he\-ren\-ces and the buildup of correlations in quantum many-body systems via $\alpha$-R\'{e}nyi relative entropy. We focus on the so-called $\alpha$-relative purity, a distinguishability measure of quantum states intimately linked to the R\'{e}nyi relative entropy of order $\alpha$ (see details in Sec.~\ref{sec:reviewRRE001}). Motived by the language of coherence orders developed in NMR and recently addressed under the viewpoint of resource theories, here we will present a novel class of MQCs, called $\alpha$-MQCs, which is rooted on $\alpha$-RRE. Our framework unveils the link among MQCs, $\alpha$-RRE, and Wigner-Yanase-Dyson skew information (WYDSI), an information-theoretic quantifier introduced half century ago and which plays a role on the theory of asymmetry. Noteworthy, it has been shown that WYDSI also witnesses the role of classical and quantum fluctuations in many-body systems~\cite{PhysRevB.94.075121}. We derive bounds on $\alpha$-MQCs, proving that $\alpha$-MQCs is upper bounded by quantum Fisher information (QFI), a paradigmatic figure of merit widely applied for enhanced phase estimation~\cite{10.1116_1.5119961}, and the detection the metrologically useful entanglement~\cite{PhysRevLett.102.100401,pezze_smerzi_188_691_2014}.

The paper is organized as follows. In Sec.~\ref{sec:reviewRRE001} we review useful basic concepts regarding R\'{e}nyi relative entropies ($\alpha$-RREs) and highlight their main features. In Sec.~\ref{sec:reviewCoherenceOrders00xx0011} we a\-ddress the concept of coherence orders and derive a novel class of Multiple Quantum Coherences ($\alpha$-MQCs) linked to $\alpha$-RREs. In Sec.~\ref{sec:alphaMQC001} we prove that $\alpha$-RRE is perturbatively linked to WYDSI. Furthermore, we show that WYDSI testifies the coherence encapsulated in a quantum state by proving its connection with $\alpha$-MQCs. In Sec.~\ref{sec:boundWYDSIMQCxxxx000xxx001} we derive a family of upper and lower bounds to the second moment of $\alpha$-MQC and WYDSI. In Sec.~\ref{sec:examples000xxx0001} we illustrate our findings. Sections~\ref{sec:examples000xxx0000000} and~\ref{sec:examples000xxx0002a} provide analytical results for single-qubit states and two-qubit Bell-diagonal states, respectively. In Sec.~\ref{sec:examples000xxx0002} we focus on systems of $N$-particle states, and thus present ana\-ly\-ti\-cal calculations and numerical simulations to su\-pport our theoretical predictions. Section~\ref{sec:examples000xxx0003} examines $\alpha$-MQC in the context of time-reversing the many-body dy\-na\-mics of a long-range Ising model. Finally, in Sec.~\ref{sec:conclusions000xxx001} we summarize our conclusions.


\section{R\'{e}nyi relative entropy: a short review}
\label{sec:reviewRRE001}

In this section we will briefly review some basic pro\-per\-ties of quantum R\'{e}nyi relative entropies. Here we will focus on a physical system described by finite-dimensional Hilbert space $\mathcal{H}$, i.e., $\dim\mathcal{H} = d$. For completeness, let $\mathcal{B}(\mathcal{H})$ be the set of linear operators acting over $\mathcal{H}$. The state of the system will be given by the density matrix $\varrho \in \mathcal{S}$, where $\mathcal{S} = \{\rho \in \mathcal{H} \mid {\rho^{\dagger}} = \rho,~\rho\geq 0,~\text{Tr}(\rho) = 1\}$ denotes the convex space of positive semi-definite density operators. In this setting, given two states $\rho, \varrho\in\mathcal{S}$ and $\alpha \in (0,1)\cup(1,+\infty)$, the quantum $\alpha$-R\'{e}nyi re\-la\-ti\-ve entropy ($\alpha$-RRE) is defined by~\cite{PETZ1985_21_PublResInstKyoto,PETZ198657,IEEE.57.2474.2011,10.10631.4838856}
\begin{equation}
\label{eq:renyiMQC0031}
{\text{D}_{\alpha}}(\rho\|\varrho) =
\begin{cases}
 {(\alpha - 1)^{-1}}\ln\left[ { {{f}_{\alpha}}(\rho,\varrho)  }\right] ~,& \mbox{if $\text{supp}\,\rho \subseteq \text{supp}\,\varrho$} \\
+ \infty ~,& \mbox{otherwise}
\end{cases}
\end{equation}
with the relative purity
\begin{equation}
\label{eq:renyiMQC0032}
{{f}_{\alpha}}(\rho,\varrho) := \text{Tr}\left({\rho^{\alpha}}{\varrho^{1 - \alpha}}\right) ~.
\end{equation}
Here $\text{supp}\, X$ stands for the support of $X\in\mathcal{S}$. In par\-ti\-cu\-lar, for $\alpha \in (0,1)$ the restriction $\text{supp}\,\rho \subseteq \text{supp}\,\varrho$ is e\-qui\-va\-lent to $\rho \not \perp \varrho$, i.e., whenever $\text{supp} \, \rho \, \cap\,  \text{supp} \,\varrho$ contains at least one non-zero vector~\cite{Leditzky2017}. The positivity of $\alpha$-RRE follows from H\"{o}lder's inequality for any $\rho, \varrho\in\mathcal{S}$, and its monotonicity yields that, for $\alpha \in (0,1) \cup (1,2)$, one has ${\text{D}_{\alpha}}(\mathcal{E}(\varrho)\|\mathcal{E}(\omega)) \leq {\text{D}_{\alpha}}(\varrho\|\omega)$, where $\mathcal{E}(\bullet)$ denotes a completely positive and trace preserving (CPTP) map~\cite{Mosonyi2015}. Except for the case $\alpha = 1/2$, R\'{e}nyi relative entropy is not a symmetric information measure and does not define a metric over the space of quantum states. Noteworthy, for $\alpha \geq 1$ R\'{e}nyi $\alpha$-re\-la\-ti\-ve entropy fulfills the Csisz\'{a}r-Pinsker inequality, ${\text{D}_{\alpha}}(\varrho\|\rho) \geq (1/2) {\| \rho - \varrho \|_1^2}$, where the notation ${\|{A}\|_1} = \text{Tr}|{A}|$ stands for the trace norm, with $|A| := \sqrt{{A^{\dagger}}A}$~\cite{csiszar1967,hiai1981,5605338}. Moreover, it has been proved that $\alpha$-RRE also satisfies a family of Pinsker-type inequalities for $\alpha \in (0,1)$~\cite{ARastegin_MathPhysAnalGeom_16_213}. Finally, we also notice the similarity between the $\alpha$-RRE and the so-called sandwiched quantum R\'{e}nyi relative entropy proposed in Refs.~\cite{10.10631.4838856,Winter_2014_331}.

The functional ${f_{\alpha}}(\rho,\varrho)$ defines the $\alpha$-re\-la\-ti\-ve purity and it is bounded as $0 \leq {{f}_{\alpha}}(\rho,\varrho) \leq 1$~\cite{DBLPjournals_qic_Audenaert01}. The property ${f_{1 - \alpha}}(\varrho, \rho) = {f_{\alpha}}(\rho,\varrho)$ for all $\rho, \varrho\in\mathcal{S}$ and $0 < \alpha < 1$ implies that $\alpha$-RRE is skew symmetric for
\begin{equation}
\label{eq:asymmetry000xxx000111}
\alpha\, {\text{D}_{1 - \alpha}}(\rho\|\varrho) = {(1 - \alpha)}\, {\text{D}_{\alpha}}(\varrho\|\rho).
\end{equation} 
In particular, Eq.~\eqref{eq:renyiMQC0032} reduces to ${{f}_{\alpha}}(\rho,{\rho}) = 1$ for all $\alpha$ and $\rho\in\mathcal{S}$, and $\alpha$-RRE is identically zero in such case. Remarkably, for $0 \leq \alpha \leq 1$ one may verify that re\-la\-tive purity is also lower bounded by the trace norm (or Schatten 1-norm) as ${{f}_{\alpha}}(\rho,{\varrho})  \geq 1 - ({1}/{2})\,{\|{\rho - \varrho}\|_1}$~\cite{PhysRevLett.98.160501}, which collapses into the Powers-St{\o}rmer's ine\-qua\-li\-ty for $\alpha = 1/2$~\cite{Powers1970,doi:10.1142_S0129055X12300026}.

We summarize some limiting cases of $\alpha$-RRE. For $\alpha = 1$, Eq.~\eqref{eq:renyiMQC0031} recovers the so-called Umegaki's re\-la\-tive entropy, ${\text{D}_{1}}(\rho\|\varrho) = \text{Tr}[\rho\,(\ln\rho - \ln\varrho)]$, also known as quantum relative entropy or Kullback-Leibler divergence~\cite{umegaki1962,6832827ErvenHarremos}. Furthermore, $\alpha = 0$ sets the min-relative entropy ${\text{D}_{\text{min}}}(\rho\|\varrho)  = -\ln \, [ \text{Tr}({{\Xi}_{\rho}}\varrho) ]$, with ${{\Xi}_{\rho}}$ being the projector onto the support of $\rho$, while the max-entropy ${\text{D}_{\text{max}}}(\rho\|\varrho)  = \inf\{ \lambda \in \mathbb{R} \mid \rho \leq \exp(\lambda)\varrho \}$ is obtained for $\alpha \rightarrow \infty$ if the kernel of $\varrho$ is contained in the kernel of state $\rho$~\cite{4957651_Datta}.


\section{$\alpha$-Multiple Quantum Coherences}
\label{sec:reviewCoherenceOrders00xx0011}

In the following we will present the framework to address a novel family of multiple quantum coherences which is related to the relative purity ${f_{\alpha}}(\rho,\varrho)$ defined in Eq.~\eqref{eq:renyiMQC0032}. Unless otherwise stated, from now on we will set $0 < \alpha < 1$. Let us define the density operator 
\begin{equation}
\label{eq:renyiMQC00202}
{\rho^{(\alpha)}} := {c_{\alpha}} \, {\rho^{\alpha}} ~,
\end{equation}
where 
${c_{\alpha}^{-1}} = \text{Tr}({\rho^{\alpha}})$
is a positive real number. Using the spectral decomposition $\rho = {\sum_l}\,{p_l}|{\psi_l}\rangle\langle{\psi_l}|$, with $\langle{\psi_l}|{\psi_r}\rangle = {\delta_{l,r}}$,  $0 < {p_l} < 1$, and ${\sum_l}\, {p_l} = 1$, one may readily conclude that ${c_{\alpha}^{-1}} = {{\sum_l}\,{p_l^{\alpha}}} > 0$. 

To formulate the concept of co\-he\-ren\-ce orders, we need first to fix some preferred basis of states~\cite{Munowitz,Keeler}. Thus, given the observable $\hat{A} \in \mathcal{B}(\mathcal{H})$, let us denote by ${\{|{{\ell}}\rangle\}_{\ell = 1,\ldots,d}}$ its complete set of eigenstates, and ${\{ {\lambda_{\ell}} \}_{\ell = 1,\ldots,d}}$ the corresponding set of discrete eigenvalues. In the remainder of the paper, we will refer to this basis of states as the {\it reference basis}. We furthermore assume that the spacing of the eigenvalues of the spectrum of $\hat{A}$ is an integer $m \in \mathbb{Z}$ 
\begin{equation}
{\lambda_j} - {\lambda_{\ell}} = m,
\end{equation}
for all $j,\ell \in \{1,\ldots,d\}$. The coherence order decomposition of the density operator ${\rho^{(\alpha)}}$ reads
\begin{equation}
\label{eq:renyiMQC004}
{\rho^{(\alpha)}} = {\sum_m}\,{\rho_m^{(\alpha)}} ~,
\end{equation}
where we define
\begin{equation}
\label{eq:renyiMQC005}
{\rho_m^{(\alpha)}} := {\sum_{{\lambda_j} - {\lambda_{\ell}} = m}}\, \langle{j}|{\rho^{(\alpha)}}|{\ell}\rangle |{j}\rangle \langle{\ell}| ~.
\end{equation}
One should note that Eq.~\eqref{eq:renyiMQC005} allows us to write down the density matrix ${\rho^{(\alpha)}}$ as a sum of non-Hermitian blocks ${\rho_m^{(\alpha)}}$ in terms of the reference basis. In other words, ${\rho_m^{(\alpha)}}$ contains all coherences between eigenstates $|{j}\rangle$ and $|{\ell}\rangle$ of $\hat{A}$ such that ${\lambda_j} - {\lambda_{\ell}} = m$, with $m\in\mathbb{Z}$. 

Noteworthy, ${\rho_m^{(\alpha)}}$ satisfies three crucial pro\-per\-ties. 
\begin{enumerate}
\item[(1)] The block ${\rho_m^{(\alpha)}}$ is asymmetric with respect to index $m$ under conjugate transposition, i.e.,
\begin{align}
\label{eq:renyiMQC0020aaaa1aa}
{({\rho_m^{(\alpha)}})^{\dagger}} = {\rho_{-m}^{(\alpha)}}  ~.
\end{align}
\item[(2)] The blocks ${\rho_m^{(\alpha)}}$ and ${\rho_n^{(\beta)}}$ are orthogonal according to the Hilbert-Schmidt inner product as
\begin{equation}
\label{eq:renyiMQC0020}
{\langle {\rho_m^{(\alpha)}} , {\rho_n^{(\beta)}} \rangle_{\text{HS}}} = {\delta_{m,n}} {\langle {\rho_m^{(\alpha)}} , {\rho_m^{(\beta)}} \rangle_{\text{HS}}} ~.
\end{equation}
where we define ${\langle{A,B}\rangle_{\text{HS}}} : = \text{Tr}({A^{\dagger}}B)$, for $A,B\in \mathcal{B}(\mathcal{H})$. 
\item[(3)] By considering the observable $\hat{A}$ which generates the translationally-covariant o\-pe\-ra\-tion ${\mathcal{U}_{\phi}}(\bullet) := {e^{-i\phi\hat{A}}} \bullet {e^{i\phi\hat{A}}}$, with $\phi \in (0,2\pi]$, thus block ${\rho_m^{(\alpha)}}$ acquires a phase shift that reads
\begin{equation}
\label{eq:renyiMQC0027}
{\mathcal{U}_{\phi}} ({\rho_m^{(\alpha)}}) = {e^{- i m \phi}} \, {\rho_m^{(\alpha)}}  ~.
\end{equation}
Note that ${\rho_0^{(\alpha)}}$ is incoherent under such a phase encoding process, i.e., the subspace related to the mode of co\-he\-ren\-ce $m = 0$ is translationally symmetric with respect to $\hat{A}$~\cite{PhysRevA.94.052324}. For details in the proof of Eqs.~\eqref{eq:renyiMQC0020aaaa1aa}--\eqref{eq:renyiMQC0027}, see Appendix~\ref{sec:appendix0001a}.
\end{enumerate}

In the following we will discuss how the relative purity ${{f}_{\alpha}}(\rho,{\rho_{\phi}})$ of states $\rho$ and ${\rho_{\phi}} = {\mathcal{U}_{\phi}}(\rho)$ behaves under the framework of coherence orders. From Eq.~\eqref{eq:renyiMQC0027}, one may verify that
\begin{equation}
\label{eq:renyiMQC0028xx000xx111}
{\rho_{\phi}^{\alpha}} = {c_{\alpha}^{-1}}\, {\sum_m} \, {\mathcal{U}_{\phi}}({\rho_m^{(\alpha)}}) = {c_{\alpha}^{-1}}\, {\sum_m}\,{e^{- i m \phi}} \, {\rho_m^{(\alpha)}}  ~,
\end{equation}
where we used the property ${\rho_{\phi}^{\alpha}} = {[\,{\mathcal{U}_{\phi}}(\rho)]^{\alpha}} = {\mathcal{U}_{\phi}}({\rho^{\alpha}})$, which holds for $0 < \alpha < 1$~\cite{Bathia_Rajendra,PhysRevA.91.042330}. Cru\-cially, Eq.~\eqref{eq:renyiMQC0028xx000xx111} implies that the unitary evolution imprints a phase shift on each block ${\rho_m^{(\alpha)}}$ built from the coherence orders decomposition of the probe state $\rho$. Hence, from Eqs.~\eqref{eq:renyiMQC004},~\eqref{eq:renyiMQC0020}, and~\eqref{eq:renyiMQC0028xx000xx111}, the relative purity becomes
\begin{align}
\label{eq:renyiMQC0029}
{{f}_{\alpha}}(\rho,{\rho_{\phi}}) &=  {( {{c_{\alpha}}\, {c_{1 - \alpha}}} )^{-1}} \, {\sum_{m,n}} \,{e^{- i m \phi}} \, \text{Tr}\left( {\rho_n^{(\alpha)}} {\rho_m^{(1 - \alpha)}} \right) \nonumber\\ 
&= {( {{c_{\alpha}}\, {c_{1 - \alpha}}} )^{-1}} \, {\sum_m} \,{e^{- i m \phi}} \, {I_m^{\alpha}}(\rho) ~,
\end{align}
where ${I_m^{\alpha}}(\rho)$ is the $\alpha$-Multiple-Quantum Intensity ($\alpha$-MQI) defined as
\begin{equation}
\label{eq:renyiMQC0023}
{I_m^{\alpha}}(\rho) = \text{Tr}\left( {({\rho_m^{(\alpha)}})^{\dagger}}{\rho_m^{(1 - \alpha)}} \right) ~.
\end{equation}
The set ${\{ {I_m^{\alpha}}(\rho) \}_{m \in \mathbb{Z}}}$ is called $\alpha$-MQI spectrum. 
Quite remarkably, the asymmetry property presented in Eq.~\eqref{eq:renyiMQC0020aaaa1aa} implies that $\alpha$-MQI satisfies the following algebraic identities
\begin{equation}
\label{eq:renyiMQC0024b}
{[{I_m^{\alpha}}(\rho)]^*} = {I_m^{1 - \alpha}}(\rho) = {I_{-m}^{\alpha}}(\rho) ~.
\end{equation}
Furthermore, setting $\phi = 0$ into Eq.~\eqref{eq:renyiMQC0029}, it is straightforward to verify that the sum of all $\alpha$-MQI relative to state $\rho$, fulfills the normalization constraint 
\begin{equation}
\label{eq:sumconstraint000xxx000xxx0001}
{\sum_m}\, {I_m^{\alpha}}(\rho) = {c_{\alpha}}\, {c_{1 - \alpha}} ~.
\end{equation}
We emphasize that one may access the $\alpha$-MQI ${I_m^{\alpha}}(\rho)$ by Fourier transforming Eq.~\eqref{eq:renyiMQC0029} with respect to $\phi \in (0,2\pi]$, which reads
\begin{equation}
\label{eq:renyiMQC0030xxx0000xxx011111}
{I_m^{\alpha}}(\rho) = \frac{{{c_{\alpha}}\, {c_{1 - \alpha}}}}{2\pi}\, {\int_0^{2\pi}}\, d\phi\, {e^{i m \phi}} \, {{f}_{\alpha}}(\rho,{\rho_{\phi}}) ~.
\end{equation}
It should be noted that $\alpha$-MQI defined in Eq.~\eqref{eq:renyiMQC0023} is analogous to the standard MQI addressed by G\"{a}rttner {\it et al.}~\cite{101038nphys4119v01,PhysRevLett.120.040402}. However, it turns out the framework developed here covers the subtle case of coherence orders involving rational powers $\rho^{\alpha}$ of the density operator, with $0 < \alpha < 1$.

It is worth mentioning that relative purity ${f_{\alpha}}(\rho,{\rho_{\phi}})$ implies a nontrivial constraint involving $\alpha$-MQI and $\alpha$-R\'{e}nyi relative entropy. Indeed, by substituting Eq.~\eqref{eq:renyiMQC0029} into Eq.~\eqref{eq:renyiMQC0031}, one obtains
\begin{align}
\label{eq:renyiMQC0030xxx0000xxx01133333333}
{\text{D}_{\alpha}}(\rho\|{\rho_{\phi}}) &=  \frac{\alpha}{\alpha - 1} \, {S_{1 - \alpha}}(\rho) - {S_{\alpha}}(\rho) \nonumber\\
&+ \frac{1}{\alpha - 1}\ln\left( {\sum_m} \,{e^{- i m \phi}} \, {I_m^{\alpha}}(\rho) \right) ~,
\end{align}
where ${S_{\alpha}}(\rho)$ is the standard R\'{e}nyi entropy 
\begin{equation}
\label{eq:renyiMQC0030xxx0000xxx011333444444444}
{S_{\alpha}}(\rho) := \frac{1}{1 - \alpha}\ln \left[ \text{Tr}({\rho^{\alpha}}) \right] ~.
\end{equation}
In summary, Eq.~\eqref{eq:renyiMQC0030xxx0000xxx01133333333} means that, to distinguish states $\rho$ and $\rho_{\phi}$ through $\alpha$-R\'{e}nyi relative entropy, one requires the knowledge of R\'{e}nyi entropy ${S_{\alpha}}(\rho)$ and the $\alpha$-MQI spectrum of state $\rho$ with respect to the reference basis of generator $\hat{A}$. 


\section{Bridging R\'{e}nyi relative entropy, $\alpha$-MQC, and Wigner-Yanase-Dyson skew information}
\label{sec:alphaMQC001}

In this Section we study the connection between the $\alpha$-RRE, the WDSI, and the $\alpha$-MQC.

\subsection{$\alpha$-RRE and WYDSI}

\begin{figure}[t]
\centering
\includegraphics[scale=0.9]{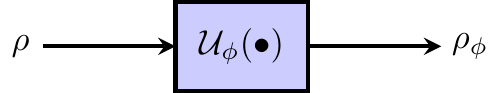}
\caption{(Color online) Schematic depiction of the quantum protocol.}
\label{fig:picturepaperv00222222}
\end{figure}

Let us consider the protocol of  Fig.~\ref{fig:picturepaperv00222222}, where the parameter $\phi$ is imprinted on the probe state $\rho \in \mathcal{S}$ through the unitary evolution ${\mathcal{U}_{\phi}}(\bullet) := {e^{-i\phi\hat{A}}} \bullet {e^{i\phi\hat{A}}}$, where $\hat{A} \in \mathcal{B}(\mathcal{H})$ is a generic observable. 
In general, the problem of estimating the phase shift $\phi$ is addressed via the so-called Cram\'{e}r-Rao bound~\cite{PhysRevLett.96_010401,RevModPhys.90.035005}, which relates the inverse of quantum Fisher information (QFI) to the maximum phase sensitivity achievable for state $\rho$ undergoing the referred physical process. 
Furthermore, estimating such unknown parameter is also a task related to the ability of distinguishing both states $\rho$ and ${\rho_{\phi}} = {\mathcal{U}_{\phi}}(\rho)$~\cite{T_th_2014}. 
In this context, one typically introduces the Bures distance or another suitable {\it bona fide} quantifier also related to the Uhlmann-Jozsa fidelity~\cite{1994_JModOpt_41_2315}. 
Here we will adopt the $\alpha$-RRE introduced in Sec.~\ref{sec:reviewRRE001} as a figure of merit to distinguish quantum states. 
By performing a Taylor expansion of $\alpha$-RRE up to second order in $\phi$ around $\phi = 0$, one obtains
\begin{equation}
\label{eq:renyiMQC0059a}
{\text{D}_{\alpha}}(\rho\|{\rho_{\phi}}) \approx - \frac{\phi^2}{\alpha - 1}\, {\mathcal{I}_{\alpha}}(\rho,\hat{A}) + {O}({\phi^3}) ~,
\end{equation}
where we define 
\begin{equation}
\label{eq:renyiMQC0055}
{\mathcal{I}_{\alpha}}(\rho,\hat{A}) := - \frac{1}{2}\,\text{Tr}\left( [ {\hat{A}}, {\rho^{\alpha}} ] \, [{\hat{A}},{\rho^{1 - \alpha}}] \right) ~.
\end{equation}

Interestingly, Eq.~\eqref{eq:renyiMQC0055} defines the so-called Wigner-Yanase-Dyson skew information (WYDSI)~\cite{Wigner910}. WYDSI is positive, ${\mathcal{I}_{\alpha}}(\rho,\hat{A}) \geq 0$, and a convex quantity~\cite{LIEB1973267,PhysRevLett.30.434}, i.e., ${\mathcal{I}_{\alpha}}(\gamma\rho + (1 - \gamma)\varrho,\hat{A}) \leq \gamma\,{\mathcal{I}_{\alpha}}(\rho,\hat{A}) + (1 - \gamma)\, {\mathcal{I}_{\alpha}}(\varrho,\hat{A})$, for all $0 < \alpha < 1$ and $0 \leq \gamma \leq 1$, with $\rho,\varrho\in\mathcal{S}$ and $\hat{A}\in\mathcal{B}(\mathcal{H})$. Furthermore, WYDSI is additive for pro\-duct states, i.e., ${\mathcal{I}_{\alpha}}({\rho_1}\otimes{\rho_2},{\hat{A}_1}\otimes\mathbb{I} + \mathbb{I}\otimes{\hat{A}_2}) = {\mathcal{I}_{\alpha}}({\rho_1},{\hat{A}_1}) + {\mathcal{I}_{\alpha}}({\rho_2},{\hat{A}_2})$~\cite{Takagi_SciRep14562}. Physically, WYDSI quantifies the non-commutativity of operator $\hat{A}$ regarding to the quantum state $\rho$. Noteworthy, WYDSI has been also recognized as an asymmetry measure~\cite{Marvian_nature,PhysRevA.94.052324}. Moreover, WYDSI also appears in a slightly modified quantum version of the work dissipation fluctuation relation in nonequilibrium quantum thermodynamics~\cite{PhysRevLett.123.230603,PhysRevResearch.2.023377}. In particular, for $\alpha = 1/2$, WYDSI reduces to the so-called Wigner-Yanase skew information (WYSI), which is defined as ${\mathcal{I}_{1/2}}(\rho,\hat{A}) = - ({1}/{2})\,\text{Tr}\,({[\sqrt{\rho},\hat{A}\, ]^2})$. 
In Appendix~\ref{sec:criticalcase000xxx000111} we show that, for $\alpha \rightarrow 1$, Eq.~\eqref{eq:renyiMQC0059a} is well behaved and reduces to ${\text{D}_{1}}(\rho\|{\rho_{\phi}}) = { \lim_{\alpha \rightarrow 1}} \, {\text{D}_{\alpha}}(\rho\|{\rho_{\phi}}) \approx {\phi^2}\left( \text{Tr}({\hat{A}^2}\rho\ln\rho) - \text{Tr}(\hat{A}\rho\hat{A}\ln\rho)  \right) + {O}({\phi^3})$.

The proof of Eq.~\eqref{eq:renyiMQC0059a} goes as follows. Given the states $\rho$ and ${\rho_{\phi}} = {e^{-i\phi\hat{A}}} \rho\, {e^{i\phi\hat{A}}}$, with $\text{supp}\,\rho \subseteq \text{supp}\,{\rho_{\phi}}$, we know from Sec.~\ref{sec:reviewRRE001} that ${\text{D}_{\alpha}}(\rho\|{\rho_{\phi}}) = {(\alpha - 1)^{-1}}\ln\left[ { {{f}_{\alpha}}(\rho,{\rho_{\phi}})}\right]$, with ${{f}_{\alpha}}(\rho,{\rho_{\phi}}) = \text{Tr}({\rho^{\alpha}}{\rho_{\phi}^{1 - \alpha}})$. The Taylor expansion of $\alpha$-RRE up to second order in $\phi$, around $\phi = 0$, is given by
\begin{align}
\label{eq:expansRRE00001}
{\text{D}_{\alpha}}(\rho\|{\rho_{\phi}}) \approx&~{\left[ {\text{D}_{\alpha}}(\rho\|{\rho_{\phi}}) \right]_{\phi = 0}} + \phi {\left[ {\text{D}'_{\alpha}}(\rho\|{\rho_{\phi}}) \right]_{\phi = 0}} \nonumber\\
&+ \frac{\phi^2}{2}{\left[ {\text{D}''_{\alpha}}(\rho\|{\rho_{\phi}}) \right]_{\phi = 0}} + {O}({\phi^3}) ~,
\end{align}
where the notation $\mathcal{A}'$, $\mathcal{A}''$ stand for the derivatives $d\mathcal{A}/d\phi$, and ${d^2}\mathcal{A}/{d\phi^2}$, respectively.
We notice that ${\left[{\text{D}_{\alpha}}(\rho\|{\rho_{\phi}}) \right]_{\phi = 0}} = 0$ since ${\rho_0} = \rho$. 
Moreover, both the first and the second order derivatives of $\alpha$-RRE 
with respect to $\phi$ can be written as
\begin{equation}
\label{eq:renyiMQC003200003aaaa}
{\left[ {\text{D}'_{\alpha}}(\rho\|{\rho_{\phi}}) \right]_{\phi = 0}} = \frac{1}{\alpha - 1}  {\left({{f}'_{\alpha}}(\rho,{\rho_{\phi}})\right)_{\phi = 0}}  ~,
\end{equation}
and
\begin{align}
\label{eq:renyiMQC003200003}
&{\left[ {\text{D}''_{\alpha}}(\rho\|{\rho_{\phi}}) \right]_{\phi = 0}} = \frac{1}{\alpha - 1} {\left[ {f''_{\alpha}}(\rho,{\rho_{\phi}})   - {{\left( {{f}'_{\alpha}}(\rho,{\rho_{\phi}}) \right)^2} } \right]_{\phi = 0}}  ~,
\end{align}
where we have used that ${\lim_{\phi \rightarrow 0}} \, {f_{\alpha}}(\rho,{\rho_{\phi}}) = 1$. 
In order to compute Eqs.~\eqref{eq:renyiMQC003200003aaaa} and~\eqref{eq:renyiMQC003200003}, 
we need to evaluate the derivatives ${{f}'_{\alpha}}(\rho,{\rho_{\phi}})$ and ${{f}''_{\alpha}}(\rho,{\rho_{\phi}})$ 
at $\phi = 0$. To do so, one may prove that the quantum state ${\rho_{\phi}} = {e^{-i\phi\hat{A}}} \rho {e^{i\phi\hat{A}}}$ 
evolving unitarily implies that 
${\rho_{\phi}^{s}} = {[\,{\mathcal{U}_{\phi}}(\rho)]^{s}} = {\mathcal{U}_{\phi}}({\rho^{s}})$, for $ 0 < s < 1$ 
(see Appendix A in Ref.~\cite{PhysRevA.91.042330}). 
Therefore, it follows that the $k$th order derivative of state $\rho_{\phi}^s$ becomes
\begin{equation}
\label{eq:renyiMQC003601a}
\frac{d^k}{d{\phi^k}}{\rho_{\phi}^s} = {(-i)^k} \, \underbrace{[\hat{A},[\hat{A},\ldots,[\hat{A},{\rho_{\phi}^s}]\ldots]]}_{k~\text{times}} ~.
\end{equation}
Hence, starting from Eq.~\eqref{eq:renyiMQC003601a}, both first and second order derivatives of the relative purity at the vicinity of $\phi = 0$ are given by
\begin{align}
\label{eq:renyiMQC0039}
 {\left( {{f}'_{\alpha}}(\rho,{\rho_{\phi}}) \right)_{\phi = 0}} &= i\, {\left[ \text{Tr}\left( {\hat{A}} \, [{\rho^{\alpha}},{\rho_{\phi}^{1 - \alpha}}]  \right) \right]_{\phi = 0}} \nonumber\\
&= 0 ~,
\end{align}
and
\begin{align}
\label{eq:renyiMQC0053}
 {\left( {f''_{\alpha}}(\rho,{\rho_{\phi}}) \right)_{\phi = 0}} &=  {\left[\text{Tr}\left( [ {\hat{A}}, {\rho^{\alpha}} ] \, [{\hat{A}},{\rho_{\phi}^{1 - \alpha}}] \right)\right]_{\phi = 0}} \nonumber\\
 &= -2\, {\mathcal{I}_{\alpha}}(\rho,\hat{A}) ~,
\end{align}
respectively, where ${\mathcal{I}_{\alpha}}(\rho,\hat{A})$ is the WYDSI defined in Eq.~\eqref{eq:renyiMQC0055}. 
Substituting Eqs.~\eqref{eq:renyiMQC0039} and~\eqref{eq:renyiMQC0053} into Eqs.~\eqref{eq:renyiMQC003200003aaaa} and~\eqref{eq:renyiMQC003200003}, it yields ${\left[{\text{D}'_{\alpha}}(\rho\|{\rho_{\phi}}) \right]_{\phi = 0}} = 0$ and also
\begin{equation}
\label{eq:renyiMQC0059aapp}
{\left[ {\text{D}''_{\alpha}}(\rho\|{\rho_{\phi}}) \right]_{\phi = 0}} = - \frac{2}{\alpha - 1}\, {\mathcal{I}_{\alpha}}(\rho,\hat{A}) ~.
\end{equation}
Finally, by plugging these results into Eq.~\eqref{eq:expansRRE00001}, one recovers the Taylor expansion of $\alpha$-RRE aforementioned in Eq.~\eqref{eq:renyiMQC0059a}. It should be noted that a similar conclusion was previously reported in the context of resource theory of asymmetry, but focusing on the Taylor expansion of relative purity~\cite{MarvianThesis}.

\subsection{WYDSI and $\alpha$-MQC}

Remarkably, WYDSI captures information about the coherence order decomposition of state $\rho$ with respect to the reference basis of the observable $\hat{A}$. 
Indeed, one may prove that
\begin{equation}
\label{eq:renyiMQC1000mainresult}
{4\, {c_{\alpha}}\, {c_{1 - \alpha}}}\, {\mathcal{I}_{\alpha}}(\rho,\hat{A}) = {{F_I^{\alpha}}(\rho,\hat{A})} ~,
\end{equation}
where here ${F_I^{\alpha}}(\rho,\hat{A})$ denotes the second moment of the $\alpha$-MQC spectrum defined by
\begin{equation}
\label{eq:renyiMQC0082}
{F_I^{\alpha}}(\rho,\hat{A}) := 2\,{\sum_{m}}\, {m^2} \, {I_m^{\alpha}}(\rho) ~.
\end{equation}
In order to prove such statement, we will take advantage from the framework of coherence orders discussed in Sec.~\ref{sec:reviewCoherenceOrders00xx0011}. Starting from the definition of WYDSI in Eq.~\eqref{eq:renyiMQC0055}, one may write down 
\begin{equation}
\label{eq:renyiMQC0055app03}
{\mathcal{I}_{\alpha}}(\rho,\hat{A}) = - \frac{1}{2\, {c_{\alpha}}{c_{1 - \alpha}}}\, {\sum_{m,n}}\, \text{Tr}\left( [ {\hat{A}}, {\rho_n^{(\alpha)}} ] \, [{\hat{A}},{\rho_m^{(1 - \alpha)}}] \right) ~,
\end{equation}
where we have used that $\rho = {c_s^{-1}}\, {\sum_{m}}\,{\rho_m^{(s)}}$ (see Eqs.~\eqref{eq:renyiMQC00202} and~\eqref{eq:renyiMQC004}). Now, note that each commutator in Eq.~\eqref{eq:renyiMQC0055app03} can be conveniently simplified according to the identity below
\begin{align}
\label{eq:renyiMQC0055app04}
[{\hat{A}}, {\rho_m^{(s)}}] &= {\sum_{{\lambda_j} - {\lambda_{\ell}} = m}}\, \langle{j}|{\rho^{(s)}}|{\ell}\rangle \, [\hat{A}, |{j}\rangle \langle{\ell}| \, ] \nonumber\\
&= {\sum_{{\lambda_j} - {\lambda_{\ell}} = m}}\, ({\underbrace{{\lambda_j} - {\lambda_{\ell}}}_{m}}) \, \langle{j}|{\rho^{(s)}}|{\ell}\rangle |{j}\rangle \langle{\ell}| \nonumber\\
&= m\, {\rho_m^{(s)}} ~,
\end{align}
which descends from $\hat{A}|{j}\rangle = {\lambda_j}|{j}\rangle$. Moreover, from Eqs.~\eqref{eq:renyiMQC0020} and~\eqref{eq:renyiMQC0023} we also know that
\begin{align}
\label{eq:renyiMQC0055app08}
\text{Tr}\left( {\rho_n^{(\alpha)}} {\rho_m^{(1 - \alpha)}}\right) &= {\delta_{n,-m}} \text{Tr}\left( {\rho_{-m}^{(\alpha)}} \, {\rho_m^{(1 - \alpha)}}\right) \nonumber\\
&= {\delta_{n,-m}} \, {I^{\alpha}_m}(\rho) ~.
\end{align}
Finally, by substituting Eqs.~\eqref{eq:renyiMQC0055app04} and~\eqref{eq:renyiMQC0055app08} into Eq.~\eqref{eq:renyiMQC0055app03}, one arrives to the result indicated in Eq.~\eqref{eq:renyiMQC1000mainresult}. 

We point out that one could obtain the same result as in Eq.~\eqref{eq:renyiMQC1000mainresult} by simply taking the second order derivative of $\alpha$-RRE in Eq.~\eqref{eq:renyiMQC0030xxx0000xxx01133333333} at $\phi = 0$. Quite interestingly, it is possible to verify that ${F_I^{\alpha}}(\rho,\hat{A})$ is a real number. Indeed, we know from Eq.~\eqref{eq:renyiMQC0024b} that ${[{I_m^{\alpha}}(\rho)]^*} = {I_{-m}^{\alpha}}(\rho)$, Therefore, by taking the complex conjugate of Eq.~\eqref{eq:renyiMQC0082}, one obtains
\begin{align}
\label{eq:renyiMQC00820a}
{[{F_I^{\alpha}}(\rho,\hat{A})]^*} &= 2\,{\sum_{m}}\, {m^2} \, {[{I_m^{\alpha}}(\rho)]^*} \nonumber\\
&= 2\,{\sum_{m}}\, {m^2} \, {I_{- m}^{\alpha}}(\rho) \nonumber \\
&= {F_I^{\alpha}}(\rho,\hat{A}) ~.
\end{align}
where we applied the substitution $m \rightarrow -m$ over the summation label.

Equation~\eqref{eq:renyiMQC1000mainresult} is one of the main results of the paper. To be more specific, in Refs.~\cite{101038nphys4119v01,PhysRevLett.120.040402,PhysRevA.99.052354} the se\-cond moment of MQC spectrum is obtained from quantum fidelity, also called relative purity, i.e., the overlap between states $\rho_0$ and $\rho_{\phi}$, which in turn define a lower bound on quantum Fisher information (QFI). Notwithstanding, addressing quantum relative R\'{e}nyi entropy as a {\it bona fide} distinguishability measure of mixed states, here we derive the novel class of $\alpha$-Multiple-Quantum Intensity, ${I_m^{\alpha}}(\rho)$ (see Eq.~\eqref{eq:renyiMQC0023}). In turn, $\alpha$-MQI implies the second moment of $\alpha$-MQC spectrum, ${{F_I^{\alpha}}(\rho,\hat{A})}$ (see Eq.~\eqref{eq:renyiMQC0082}), which play the role of $\alpha$-curvature. We also proved that ${{F_I^{\alpha}}(\rho,\hat{A})}$ is related to WYDSI (see Eq.~\eqref{eq:renyiMQC1000mainresult}), a widely established asymmetry measure in the context of resource theories~\cite{Marvian_nature,PhysRevA.94.052324}, which also captures the signature of quantum fluctuations in many-body systems at finite temperature~\cite{PhysRevB.94.075121}. This means that, by bri\-dging R\'{e}nyi relative entropy, $\alpha$-MQC, and Wigner-Yanase-Dyson skew information, one provides an alternative perspective to the understanding of quantum fluctuations and quantum correlations. 


\section{Bounds on $\alpha$-MQC}
\label{sec:boundWYDSIMQCxxxx000xxx001}

In this section we will establish a novel class of bounds on WYDSI that naturally holds for the second moment of $\alpha$-Multiple Quantum Intensity. We introduce the lower bound 
\begin{equation}
\label{eq:renyiMQC0055c2a4bound000000x0n}
{F_I^{\alpha}}(\rho,\hat{A}) \geq 8\, \alpha(1 - \alpha)\, {c_{\alpha}}{c_{1 - \alpha}} \, {\mathcal{I}^L}(\rho,\hat{A}) ~,
\end{equation}
where
\begin{equation}
\label{eq:renyiMQC00550d02d32}
{\mathcal{I}^L}(\rho,\hat{A}) := - \frac{1}{4} \, \text{Tr}({[\rho,\hat{A}]^2}) ~.
\end{equation}
that we prove in Appendix~\ref{sec:boundsWYDSIxxx000xxx111}.
For the case $\alpha = 1/2$, Eq.~\eqref{eq:renyiMQC0055c2a4bound000000x0n} becomes
\begin{equation}
\label{eq:renyiMQC00550d02d33}
{F_I^{1/2}}(\rho,\hat{A}) \geq 2\, {c_{1/2}^2} \, {\mathcal{I}^L}(\rho,\hat{A}) ~.
\end{equation}
Importantly, quantifier ${\mathcal{I}^L}(\rho,\hat{A})$ have been introduced in the context of quantum coherence characterization, thus defining a lower bound on Wigner-Yanase skew information, i.e., ${\mathcal{I}_{1/2}}(\rho,\hat{A}) \geq {\mathcal{I}^L}(\rho,\hat{A})$~\cite{PhysRevLett.113.170401}. 
Recently, a detection scheme to measure ${\mathcal{I}^L}$ was implemented in an all-optical
experiment~\cite{PhysRevA.96.042327,e19030124}. Eq.~\eqref{eq:renyiMQC00550d02d33} 
generalizes this bound by providing a less tight lower bound to the quantity ${F_I^{1/2}}(\rho,\hat{A})$. 
To see this, note that Eq.~\eqref{eq:renyiMQC1000mainresult} 
becomes ${F_I^{1/2}}(\rho,\hat{A}) = 4\, {c_{1/2}^2}\,{\mathcal{I}_{1/2}}(\rho,\hat{A})$ for $\alpha = 1/2$, 
which allow us to recast Eq.~\eqref{eq:renyiMQC00550d02d33} 
into the form ${\mathcal{I}_{1/2}}(\rho,\hat{A}) \geq (1/2)\,{\mathcal{I}^L}(\rho,\hat{A})$. 
Hence, the latter inequality differs from the bound ${\mathcal{I}_{1/2}}(\rho,\hat{A}) \geq {\mathcal{I}^L}(\rho,\hat{A})$ by a factor $1/2$ and does not set the tightest lower bound.

An upper bound on WYDSI and thus on the second moment of $\alpha$-Multiple Quantum Intensity can be derived 
using the inequalities ${\mathcal{I}_{\alpha}}(\rho,\hat{A}) \leq {\mathcal{I}_{1/2}}(\rho,\hat{A}) \leq {\mathcal{V}_{1/2}}(\rho,\hat{A})$, and ${\mathcal{I}_{\alpha}}(\rho,\hat{A}) \leq {\mathcal{V}_{\alpha}}(\rho,\hat{A}) \leq {\mathcal{V}_{1/2}}(\rho,\hat{A})$, respectively~\cite{Yanagi_2010}. 
Therefore ${F_I^{\alpha}}(\rho,\hat{A})$ fulfills the two inequalities
\begin{equation}
\label{eq:renyiMQC1000mainresult01}
\frac{{F_I^{\alpha}}(\rho,\hat{A})}{4\, {c_{\alpha}}\, {c_{1 - \alpha}}}  \leq {\mathcal{I}_{1/2}}(\rho,\hat{A}) \leq {\mathcal{V}_{1/2}}(\rho,\hat{A}) ~,
\end{equation}
and
\begin{equation}
\label{eq:renyiMQC00550a02}
\frac{{F_I^{\alpha}}(\rho,\hat{A})}{4\, {c_{\alpha}}\, {c_{1 - \alpha}}}  \leq {\mathcal{V}_{\alpha}}(\rho,\hat{A}) \leq {\mathcal{V}_{1/2}}(\rho,\hat{A})  ~,
\end{equation}
where ${\mathcal{V}_{\alpha}}(\rho,\hat{A})$ denotes the $\alpha$-variance
\begin{equation}
\label{eq:renyiMQC00550c}
 {\mathcal{V}_{\alpha}}(\rho,\hat{A}) := \sqrt{ {[V(\rho,\hat{A})]^2} - {[V(\rho,\hat{A}) -  {\mathcal{I}_{\alpha}}(\rho,\hat{A})]^2}} ~,
\end{equation}
and $V(\rho,\hat{A})$ stands for the variance
\begin{equation}
\label{eq:renyiMQC00550d}
V(\rho,\hat{A}) = \text{Tr}(\rho{\hat{A}^2}) - {[ \text{Tr}(\rho\hat{A}) ]^2} ~.
\end{equation}
It is worth emphasizing that inequalities in Eqs.~\eqref{eq:renyiMQC00550a02} and~\eqref{eq:renyiMQC00550c} cannot be recasted in a single inequality. In fact, upon varying $\alpha$,
there exists intervals over the range $0 < \alpha < 1$ in which ${\mathcal{I}_{1/2}}(\rho,\hat{A}) \geq {\mathcal{V}_{\alpha}}(\rho,\hat{A})$, and others in which ${\mathcal{I}_{1/2}}(\rho,\hat{A}) \leq {\mathcal{V}_{\alpha}}(\rho,\hat{A})$. For more details, see Sec.~\ref{sec:examples000xxx0002}, particularly panels in Figs.~\ref{fig:examples02},~\ref{fig:examples00003} and~\ref{fig:examples0000320101}.

We are now in position to derive a novel class of hie\-rar\-chi\-cal bounds on the second moment of $\alpha$-Multiple Quantum Intensity. In fact, one may bring together ine\-qua\-lities given in Eqs.~\eqref{eq:renyiMQC0055c2a4bound000000x0n},~\eqref{eq:renyiMQC1000mainresult01} and~\eqref{eq:renyiMQC00550a02} and thus combine them to produce a general family of bounds on ${F_I^{\alpha}}(\rho,\hat{A}) $. Therefore, one straightforwardly gets
\begin{equation}
\label{eq:renyiMQC0055c2a4bound000000x0p}
 2\, \alpha(1 - \alpha) \, {\mathcal{I}^L}(\rho,\hat{A}) \leq  \frac{{F_I^{\alpha}}(\rho,\hat{A})}{4\, {c_{\alpha}}\, {c_{1 - \alpha}}}  \leq {\mathcal{I}_{1/2}}(\rho,\hat{A}) \leq {\mathcal{V}_{1/2}}(\rho,\hat{A}) ~,
\end{equation}
and also
\begin{equation}
\label{eq:renyiMQC0055c2a4bound000000x0q}
2\, \alpha(1 - \alpha)\, {\mathcal{I}^L}(\rho,\hat{A}) \leq  \frac{{F_I^{\alpha}}(\rho,\hat{A})}{4\, {c_{\alpha}}\, {c_{1 - \alpha}}} \leq {\mathcal{V}_{\alpha}}(\rho,\hat{A}) \leq {\mathcal{V}_{1/2}}(\rho,\hat{A}) ~.
\end{equation}

\subsection{Bounds on the Quantum Fisher Information}

From now on we shall prove that ${F_I^{\alpha}}(\rho,\hat{A})$ defines a lower bound on quantum Fisher information (QFI). We begin by recalling the standard setup for phase-estimation based on QFI. Given a finite-dimensional quantum system undergoing a unitary evolution to the output state ${\rho_{\phi}} = {e^{-i\phi\hat{A}}} \rho \, {e^{i\phi\hat{A}}}$, generated by the observable $\hat{A}$, then QFI related to estimating the phase shift $\phi$ encoded into the probe state $\rho = {\sum_j}\,{p_j}|{\psi_j}\rangle\langle{\psi_j}|$ reads~\cite{PhysRevLett.72.3439,e19030124}
\begin{equation}
\label{sec:qfi000xxx001}
{\mathcal{F}_Q}(\rho,\hat{A}) = \frac{1}{2}\, {\sum_{j,l = 1}^d} \frac{{({p_j} - {p_l})^2}}{{p_j} + {p_l}} {|{\langle{\psi_j}|\hat{A}|{\psi_l}\rangle}|^2} ~,
\end{equation}
where $d = \dim\mathcal{H}$ is the Hilbert space dimension, $0 < {p_j} < 1$, and the sum runs over all the indeces $\{j,l\}$ such that ${p_j} + {p_l} \neq 0$. In comparison to standard definitions of QFI (see Refs.~\cite{PhysRevLett.120.040402,PhysRevLett.72.3439,e19030124}), note that Eq.~\eqref{sec:qfi000xxx001} includes an extra normalizing factor $1/4$ and guarantees that QFI recovers the variance of generator $\hat{A}$ for pure states~\cite{Ingemar_Bengtsson_Zyczkowski,Luoproc1322004,Gibilisco2008137}. To proceed deriving the novel upper bound to ${F_I^{\alpha}}(\rho,\hat{A})$, we point out that ${\mathcal{I}_{1/2}}(\rho,\hat{A})$ is also related to QFI according to the inequality ${\mathcal{I}_{1/2}}(\rho,\hat{A}) \leq {\mathcal{F}_Q}(\rho,\hat{A}) \leq 2\, {\mathcal{I}_{1/2}}(\rho,\hat{A})$~\cite{Luoproc1322004,Gibilisco2008137}. 
Therefore, by substi\-tu\-ting the latter into Eq.~\eqref{eq:renyiMQC0055c2a4bound000000x0p}, it is possible to show a strict bound involving the second moment of $\alpha$-MQI and QFI, which reads
\begin{equation}
\label{eq:renyiMQC0055c2a4bound000000x0p000003}
 \frac{{F_I^{\alpha}}(\rho,\hat{A})}{4\, {c_{\alpha}}\, {c_{1 - \alpha}}}  \leq {\mathcal{I}_{1/2}}(\rho,\hat{A}) \leq  {\mathcal{F}_Q}(\rho,\hat{A}) \leq 2 \, {\mathcal{I}_{1/2}}(\rho,\hat{A}) ~.
\end{equation}
Equation~\eqref{eq:renyiMQC0055c2a4bound000000x0p000003} is one of the main results of the paper. It provides a family of lower bounds on QFI, ${\mathcal{F}_Q}(\rho,\hat{A})$, which in turn depend on Wigner-Yanase skew information, ${\mathcal{I}_{1/2}}(\rho,\hat{A})$, and also on the second moment of $\alpha$-MQC spectrum, ${F_I^{\alpha}}(\rho,\hat{A})$. Importantly, this result paves the way for a discussion of entanglement characterization by using $\alpha$-multiple quantum coherences from $\alpha$-R\'{e}nyi relative entropies. More in general, Eq.~\eqref{eq:renyiMQC0055c2a4bound000000x0p000003} defines a novel criterion for detecting entanglement in a mixed many-body state.


\section{Examples}
\label{sec:examples000xxx0001}

In this Section we present some e\-xam\-ples to illustrate our main findings. In Sec.~\ref{sec:examples000xxx0000000}, by considering the paradigmatic case of a single qubit state, we obtain analytical expressions for $\alpha$-MQI spectrum ${I_m^{\alpha}}(\rho)$ and also ${F_I^{\alpha}}(\rho,\hat{A})$. In Sec.~\ref{sec:examples000xxx0002a} we discuss $\alpha$-MQCs for the case of two-qubit Bell-diagonal states. Mo\-ving to the multiparticle scenario, Sec.~\ref{sec:examples000xxx0002} presents numerical analysis for $\alpha$-MQI spectrum related to a class of mixed entangled states, viz., uniform superposition state, Greenberger-Horne-Zeilinger state (GHZ-state), and Werner state (W-state). Finally, Sec.~\ref{sec:examples000xxx0003} discusses $\alpha$-MQI spectrum for a physical scenario in which the referred class of multiparticle states evolves under a time-reversal quantum protocol.


\subsection{Single qubit state}
\label{sec:examples000xxx0000000}

Let us consider the quantum system described by $\rho = (1/2)(\mathbb{I} + \vec{r}\cdot\vec{\sigma})$, i.e., the Bloch sphere representation of the single qubit mixed state, where $\vec{\sigma} = ({\sigma_x},{\sigma_y},{\sigma_z})$ is the vector of Pauli matrices, $\vec{r} = r\,\hat{r}$ is the Bloch vector, with $\hat{r} = \{\sin\theta\cos\varphi,\sin\theta\sin\varphi, \cos\theta \}$, $0 < r < 1$, $\theta \in [0,\pi]$ and $\varphi \in [0,2{\pi}[$, while $\mathbb{I}$ is the $2\times 2$ identity matrix. Here we will choose the o\-pe\-ra\-tor $\hat{A} = (1/2)(\hat{n}\cdot\vec{\sigma})$ as the generator of the phase encoding protocol, where $\hat{n} = \{ {n_x},{n_y},{n_z}\}$ is a unit vector with ${n_x^2} + {n_y^2} + {n_z^2} = 1$. In this case, the reference basis is composed by the eigenstates $\{ |{+}\rangle\rangle, |{-}\rangle\rangle \}$ of $\hat{A}$ defined as
\begin{equation}
\label{eq:examplequbit000xxx000}
|\pm\rangle\rangle = \frac{1}{\sqrt{2}}\left(\pm \sqrt{1 \pm {n_z}} \, |{0}\rangle + \frac{{n_x} + i \, {n_y}}{\sqrt{1 \pm {n_z}}} \, |{1}\rangle\right) ~,
\end{equation}
where $|0\rangle = {[ 1 \quad 0 ]^{\textsf{T}}}$ and $|1\rangle = {[0 \quad 1]^{\textsf{T}}}$ are the vectors defining the computational basis states in the complex two-dimensional vector space ${\mathbb{C}^2}$, where we have that $\hat{A}|{\pm}\rangle\rangle = {\lambda_{\pm}}|{\pm}\rangle\rangle$, with eigenvalues ${\lambda_{\pm}} = \pm 1/2$. 

One may verify that, for $0 < \alpha < 1$, operator ${\rho^{(\alpha)}}$ in Eq.~\eqref{eq:renyiMQC00202} is given by
\begin{equation}
\label{eq:examplequbit000xxx001}
{\rho^{(\alpha)}} = \frac{1}{2}\left[\mathbb{I} + \left(1 - {2^{1 - \alpha}}\, {(1 - r)^{\alpha}}\, {c_{\alpha}}\right)(\hat{r}\cdot\vec{\sigma})\right] ~,
\end{equation}
and 
\begin{equation}
\label{eq:examplequbit000xxx002}
{c_{\alpha}^{-1}} = {2^{-\alpha}}\left[{(1 + r)^{\alpha}} + {(1 - r)^{\alpha}}\right] ~.
\end{equation}
\begin{figure}[t]
\centering
\includegraphics[scale=0.8]{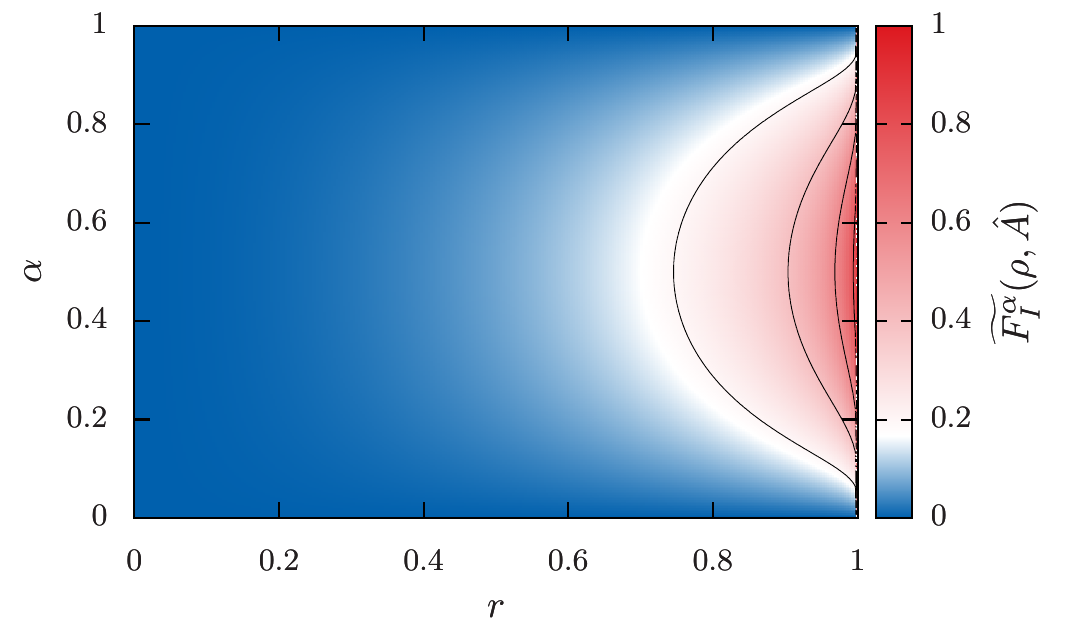}
\caption{(Color online) Density plot of figure of merit $\widetilde{F_I^{\alpha}}(\rho,\hat{A})$ for $\hat{n} = \{0,0,1 \}$ and $\hat{r} = \{\cos\varphi,\sin\varphi,0\}$, with $\widetilde{\textsf{G}} := ( \textsf{G} - \min\{\textsf{G}\} )/(\max\{ \textsf{G} \} - \min\{ \textsf{G} \} )$, and ${F_I^{\alpha}}(\rho,\hat{A})$ is given in Eq.~\eqref{eq:examplequbit000xxx007}. In this case, since vectors $\hat{n}$ and $\vec{r}$ are orthogonal, ${F_I^{\alpha}}(\rho,\hat{A})$ does not depend on the azimuthal angle $\varphi$, and thus it is solely function of $r$ and $\alpha$.}
\label{fig:examples02a}
\end{figure}
Starting from Eq.~\eqref{eq:examplequbit000xxx001}, the coherence orders decomposition 
reads ${\rho^{(\alpha)}} = {\sum_m}\,{\rho_m^{(\alpha)}}$ with $m = \{ -1,0,+1 \}$.
The non-Hermitian matrix blocks ${\rho_{m}^{(\alpha)}}$ are given by
\begin{align}
\label{eq:examplequbit000xxx003}
{\rho_{\pm 1}^{(\alpha)}} = \frac{1}{4}\, (1 - {2^{1 - \alpha}} {(1 - r)^{\alpha}} \, {c_{\alpha}})&\left[  (\hat{n}\times\vec{\sigma})\cdot(\hat{n}\times\hat{r}) \right. \nonumber\\ &\left. \pm \, i (\hat{n}\times\hat{r})\cdot\vec{\sigma} \, \right] ~,
\end{align}
and
\begin{equation}
\label{eq:examplequbit000xxx004}
{\rho_{0}^{(\alpha)}} = \frac{1}{2}\left[\mathbb{I} + (1 - {2^{1 - \alpha}} {(1 - r)^{\alpha}} \, {c_{\alpha}})(\hat{n}\cdot\hat{r})(\hat{n}\cdot\vec{\sigma}) \right] ~.
\end{equation}
Based on Eqs.~\eqref{eq:examplequbit000xxx003} and~\eqref{eq:examplequbit000xxx004}, one readily conclude that $\text{Tr}({\rho_m^{(\alpha)}}) = {\delta_{m,0}}$ and ${\rho_{- 1}^{(\alpha)}} = {({\rho_{+ 1}^{(\alpha)}})^{\dagger}}$. Therefore, $\alpha$-Multiple Quantum Intensity defined in Eq.~\eqref{eq:renyiMQC0023} becomes
 \begin{equation}
 \label{eq:examplequbit000xxx005}
 {I^{\alpha}_{\pm 1}}(\rho) = \frac{1}{4}\, (2\, {c_{\alpha}}{c_{1 - \alpha}} - 1) \left[1 - ({\hat{n}\cdot\hat{r})^2} \right] ~,
 \end{equation}
 and 
  \begin{equation}
  \label{eq:examplequbit000xxx006}
 {I^{\alpha}_{0}}(\rho) = \frac{1}{2}\left[ 1 + \left(2\, {c_{\alpha}}{c_{1 - \alpha}} - 1\right) ({\hat{n}\cdot\hat{r})^2}\right] ~.
 \end{equation}
Furthermore, note that if vectors $\hat{n}$ and $\hat{r}$ are parallel, then we have ${I^{\alpha}_{\pm 1}}(\rho)=0$ and ${I^{\alpha}_{0}}(\rho) = {c_{\alpha}}{c_{1 - \alpha}}$. Conversely, if vectors $\hat{n}$ and $\hat{r}$ are orthogonal, it follows that ${I^{\alpha}_{\pm 1}}(\rho) = (1/4)(2\,{c_{\alpha}}{c_{1 - \alpha}} - 1)$ and ${I^{\alpha}_{0}}(\rho) = 1/2$. Finally, from Eqs.~\eqref{eq:examplequbit000xxx005} and~\eqref{eq:examplequbit000xxx006} the second moment of $\alpha$-MQI (see Eq.~\eqref{eq:renyiMQC0082}) is written as
\begin{equation}
\label{eq:examplequbit000xxx007}
{F_I^{\alpha}}(\rho,\hat{A}) = (2\, {c_{\alpha}}{c_{1 - \alpha}} - 1) \left[ 1 - ({\hat{n}\cdot\hat{r})^2} \right] ~.
\end{equation}

Let us now analyze the behaviour of the second moment of $\alpha$-MQI in Eq.~\eqref{eq:examplequbit000xxx007}. Naturally, ${F_I^{\alpha}}(\rho,\hat{A})$ inherits some properties from $\alpha$-MQI. On the one hand, when vectors $\hat{n}$ and $\hat{r}$ are orthogonal, i.e., $\hat{n}\cdot\hat{r} = 0$, thus ${F_I^{\alpha}}(\rho,\hat{A})$ depends uniquely on Bloch sphere radius $r$ and the parameter $\alpha$. For instance, this case is illustrated in Fig.~\ref{fig:examples02a} choosing vector $\hat{n} = \{0,0,1\}$ related to the generator $\hat{A} = (1/2){\sigma_z}$, and $\hat{r} = \{\cos\varphi,\sin\varphi,0\}$ denoting the single qubit mixed state lying in the equatorial $xy$-plane of the Bloch sphere. On the other hand, when vectors $\hat{n}$ and $\hat{r}$ parallel, we have that ${F_I^{\alpha}}(\rho,\hat{A})$ vanishes.
\begin{table}[t]
\caption{Analytical expressions for the family of theoretical-information quantifiers related the single qubit mixed state.}
\begin{center}
\begin{tabular}{cc}
\hline\hline
Quantifier & Analytical value  \\
\hline
${\mathcal{I}^L}(\rho,\hat{A})$ & $({r^2}/8)\left[1 - {(\hat{n}\cdot\hat{r} )^2}\right]$ \\
${\mathcal{F}_Q}(\rho,\hat{A})$ & $({r^2}/4)\left[1 - {(\hat{n}\cdot\hat{r} )^2} \right]$  \\
$V(\rho,\hat{A})$ & $(1/4)\left[1 - {(\hat{n}\cdot\vec{r}\, )^2} \right]$  \\
${\mathcal{I}_{1/2}}(\rho,\hat{A})$ & $(1/4)\left(1 - \sqrt{1 - {r^2}}\, \right)\left[1 - {(\hat{n}\cdot\hat{r})^2}\right]$  \\
\hline\hline
\end{tabular}
\label{tab:quantity000xxx000xxx001}
\end{center}
\end{table}
For completeness, in Table~\ref{tab:quantity000xxx000xxx001} 
we summarize analytical expressions, obtained by using the single qubit state $\rho = (1/2)(\mathbb{I} + \vec{r}
\cdot\vec{\sigma})$ and generator $\hat{A} = (1/2)(\hat{n}\cdot\vec{\sigma})$, 
for the functional ${\mathcal{I}^L}(\rho,\hat{A})$, quantum Fisher information ${\mathcal{F}_Q}(\rho,\hat{A})$, 
standard variance $V(\rho,\hat{A})$, and also Wigner-Yanase skew information ${\mathcal{I}_{1/2}}(\rho,\hat{A})$.


\subsection{Bell-diagonal states}
\label{sec:examples000xxx0002a}

Let us now consider the class of two-qubit states with maximally mixed marginals represented by the Bell-diagonal states~\cite{PhysRevA.54.1838}
\begin{equation}
\label{eq:belldiag000xxx001}
{\rho_{\text{BD}}} = \frac{1}{4}\left(\mathbb{I}\otimes\mathbb{I} + {\sum_{j = x,y,z}}\, {a_j}\, {\sigma_j}\otimes{\sigma_j}\right) ~,
\end{equation}
where $\mathbb{I}$ is the $2\times 2$ identity matrix, $\sigma_j$ is the $j$th Pauli matrix, and the coefficients ${a_j} = \text{Tr}[\rho({\sigma_j}\otimes{\sigma_j})] \in [-1,1]$ denote the triple $\vec{a} = \{{a_x},{a_y},{a_z}\}$, which uniquely 
identifies the Bell-diagonal state. 
In particular, for $|{a_x}| + |{a_y}| + |{a_z}| \leq 1$ we thus have $\rho$ as a separable state~\cite{RevModPhys.81.865}. 
Here we will choose the generator $\hat{A} = \hat{n}\cdot\vec{S}$, where $\hat{n} = \{{n_x},{n_y},{n_z}\}$ 
is a unit vector with ${n_x^2} + {n_y^2} + {n_z^2} = 1$, and $\vec{S} = \{{\hat{S}_x},{\hat{S}_y},{\hat{S}_z}\}$ 
is the angular momentum vector, with ${\hat{S}_j} = (1/2)({\sigma_j}\otimes\mathbb{I} + \mathbb{I}\otimes{\sigma_j})$ 
for $j \in \{x,y,z\}$. The re\-fe\-ren\-ce basis $\{ |\ell\rangle\rangle\}_{\ell = 1,\ldots,4}$ contains the 
eigenstates of $\hat{A}$ given by
\begin{align}
|1\rangle\rangle &= \frac{1}{\sqrt{2}}(|0,1\rangle - |1,0\rangle) \nonumber\\
|2\rangle\rangle &= -\frac{1}{\sqrt{2}}\left[\frac{{n_-}\left({n_-} |0,0\rangle - \sqrt{2}\,{n_z}|\texttt{L}\rangle\right)}{\sqrt{1 - {n_z^2}}} - \sqrt{1 - {n_z^2}} \, |1,1\rangle \right] \nonumber\\
|3\rangle\rangle &= \frac{1}{2}\left(\frac{{n_-^2}}{1 + {n_z}} \, |0,0\rangle - \sqrt{2}\, {n_-}|\texttt{L}\rangle + (1 + {n_z}) \, |1,1\rangle\right) \nonumber\\
|4\rangle\rangle &= \frac{1}{2}\left(\frac{n_-^2}{1 - {n_z}} \, |0,0\rangle + \sqrt{2}\, {n_-}|\texttt{L}\rangle + (1 - {n_z}) \, |1,1\rangle\right)
\end{align}
with ${n_{\pm}} := {n_x} \pm i {n_y}$, and $|\texttt{L}\rangle := ({1}/{\sqrt{2}})(|0,1\rangle + |1,0\rangle)$. Note that $\hat{A}|\ell\rangle\rangle = {\lambda_{\ell}}|\ell\rangle\rangle$, where ${\lambda_1} = {\lambda_2} = 0$, ${\lambda_3} = -1$, ${\lambda_4} = 1$, and thus one obtains $m \in \{\pm 2, \pm 1, 0\}$.

Given the Bell-diagonal state, one may verify that, for $0 < \alpha < 1$, the operator ${\rho_{\text{BD}}^{(\alpha)}} = {c_{\alpha}}{({\rho_{\text{BD}}})^{\alpha}}$ (cf. Eq.~\eqref{eq:renyiMQC00202}) becomes
\begin{equation}
\label{eq:bellsquareroot000xxx000xxx001}
{\rho_{\text{BD}}^{(\alpha)}} = \frac{1}{4}\left(\mathbb{I}\otimes\mathbb{I} + {\sum_{j = x,y,z}}\, {\eta_{\alpha,j}} {\sigma_j}\otimes{\sigma_j} \right) ~,
\end{equation}
with
\begin{align}
\label{eq:bellsquareroot000xxx000xxx002}
{\eta_{\alpha,j}} &:= {c_{\alpha}} \left[ -{\upsilon_1^{\alpha}} + (1 - 2\, {\delta_{j,z}}){\upsilon_2^{\alpha}} \right. \nonumber\\ 
&\left. + (1 - 2\, {\delta_{j,y}}){\upsilon_3^{\alpha}} + (1 - 2\, {\delta_{j,x}}){\upsilon_4^{\alpha}} \right] ~,
\end{align}
for $j \in \{x,y,z\}$, and also
\begin{equation}
\label{eq:bellsquareroot000xxx000xxx003}
{c_{\alpha}^{-1}} = {\upsilon_1^{\alpha}} + {\upsilon_2^{\alpha}} + {\upsilon_3^{\alpha}} + {\upsilon_4^{\alpha}} ~.
\end{equation}
Here $\{ {\upsilon_r} \}_{r = 1,\ldots, 4}$ denote the set of eigenvalues of the two-qubit Bell-diagonal state, where
\begin{align}
&{\upsilon_r} = \frac{1}{4} \left[1 - (1 - 2{\delta_{r,2}} - 2{\delta_{r,3}}){a_x} \right.\\ \nonumber 
&\left. + (1 - 2{\delta_{r,1}} - 2{\delta_{r,3}}){a_y} + (1 - 2{\delta_{r,1}} - 2{\delta_{r,2}}){a_z} \right] ~.
\end{align}
Based on Eq.~\eqref{eq:bellsquareroot000xxx000xxx001}, one may evaluate the non-Hermitian blocks ${({\rho_{\text{BD}}^{(\alpha)}})_m}$ appearing into the coherence orders decomposition ${\rho_{\text{BD}}^{(\alpha)}} = {\sum_m}\, {({\rho_{\text{BD}}^{(\alpha)}})_m}$, and thus determine the $\alpha$-Multiple Quantum Intensity spectrum $\{ {I_m^{\alpha}}({\rho_{\text{BD}}}) \}$, with $m \in \{ 0, \pm 1, \pm 2 \}$. We will not show them here as the expressions are cumbersome. 
After a lengthy calculation the expression for $\alpha$-MQI yields
\begin{equation}
\label{eq:generalBDxxxx000xxxx1111}
{F_I^{\alpha}}({\rho_{\text{BD}}},\hat{A}) = {\sum_{j\neq k \neq l}}\, {n_j^2} \, ({\eta_{\alpha,k}} - {\eta_{\alpha,l}})({\eta_{1 - \alpha,k}} - {\eta_{1 - \alpha,l}}) ~,
\end{equation}
where the sum run over index $j,k,l \in \{ x,y,z\}$. Noteworthy, Eq.~\eqref{eq:generalBDxxxx000xxxx1111} collapses into the particular cases (i) ${F_I^{\alpha}}({\rho_{\text{BD}}},{\hat{S}_x}) = ({\eta_{\alpha,y}} - {\eta_{\alpha,z}})({\eta_{1 - \alpha,y}} - {\eta_{1 - \alpha,z}})$ for $\hat{n} = \{1,0,0\}$; (ii) ${F_I^{\alpha}}({\rho_{\text{BD}}},{\hat{S}_y}) = ({\eta_{\alpha,x}} - {\eta_{\alpha,z}})({\eta_{1 - \alpha,x}} - {\eta_{1 - \alpha,z}})$ for $\hat{n} = \{0,1,0\}$; and (iii) ${F_I^{\alpha}}({\rho_{\text{BD}}},{\hat{S}_z}) = ({\eta_{\alpha,x}} - {\eta_{\alpha,y}})({\eta_{1 - \alpha,x}} - {\eta_{1 - \alpha,y}})$ for $\hat{n} = \{0,0,1\}$.

In Table~\ref{tab:quantity000xxx000xxx002} we list the analytical expressions obtained for the functional ${\mathcal{I}^L}({\rho_{\text{BD}}},\hat{A})$, quantum Fisher information ${\mathcal{F}_Q}({\rho_{\text{BD}}},\hat{A})$, standard variance $V({\rho_{\text{BD}}},\hat{A})$, and also Wigner-Yanase skew information ${\mathcal{I}_{1/2}}({\rho_{\text{BD}}},\hat{A})$.
\begin{table}[t]
\caption{Analytical expressions for the family of theoretical-information quantifiers related to the Bell-diagonal state and generator $\hat{A} = \hat{n}\cdot\vec{S}$, where $\hat{n} = \{{n_x},{n_y},{n_z}\}$ is a unit vector with ${n_x^2} + {n_y^2} + {n_z^2} = 1$, and $\vec{S} = \{{\hat{S}_x},{\hat{S}_y},{\hat{S}_z}\}$ is the angular momentum vector, with ${\hat{S}_j} = (1/2)({\sigma_j}\otimes\mathbb{I} + \mathbb{I}\otimes{\sigma_j})$ for $j \in \{x,y,z\}$. Note that sum runs over index $j,k,l \in \{ x,y,z\}$, and ${|\vec{a}|^2} = {a_x^2} + {a_y^2} + {a_z^2}$.}
\begin{center}
\begin{tabular}{cc}
\hline\hline
 Quantifier & Analytical value  \\
\hline
${\mathcal{I}^L}({\rho_{\text{BD}}},\hat{A})$ & $\frac{1}{16}\left(2\,{|\vec{a}|^2} - {\sum_{j \neq k \neq l}}\, ({a_j^2} + 2{a_k}{a_l}){n_j^2} \right)$ \\
${\mathcal{F}_Q}({\rho_{\text{BD}}},\hat{A})$ & $\frac{1}{4}\, {\sum_{j\neq k \neq l} }\, \frac{\left( ({a_k} - {a_l}){n_j}\right)^2}{(1 + {a_j})}$  \\
$V({\rho_{\text{BD}}},\hat{A})$ & $\frac{1}{2}\left(1 + {\sum_j}\, {a_j}{n_j^2} \right)$  \\
${\mathcal{I}_{1/2}}({\rho_{\text{BD}}},\hat{A})$ & $\frac{1}{8}\, {\sum_{j\neq k \neq l}}\, {n_j^2}{({\eta_{1/2,k}} - {\eta_{1/2,l}})^2}$  \\
\hline\hline
\end{tabular}
\label{tab:quantity000xxx000xxx002}
\end{center}
\end{table}


\subsection{Multiparticle states}
\label{sec:examples000xxx0002}
\begin{table*}[thb]
\caption{Family of theoretical-information quantifiers ${\mathcal{I}^L}$, quantum Fisher information ${\mathcal{F}_Q}$, variance $V$, and Wigner-Yanase skew information ${\mathcal{I}_{1/2}}$. Here we have evaluated these quantities by considering the following operator pairs (i) $({\rho_{\text{eqn}}}, {\hat{S}_z})$, (ii) $({\rho_{\text{GHZ}}}, {\hat{S}_z})$, and (iii) $({\rho_{\text{W}}}, {\hat{S}_x})$, where $\rho_{\text{eqn}}$, $\rho_{\text{GHZ}}$ and $\rho_{\text{W}}$ denote the $N$-particle states in Eqs.~\eqref{eq:renyiMQCexample000001ex000001},~\eqref{eq:renyiMQCexample000003ex0000001}, and~\eqref{eq:renyiMQCexample000005x000001000xxx}, respectively, with ${d = 2^N}$. The collective spin operators ${\hat{S}_x}$, ${\hat{S}_z}$ are defined in Eq.~\eqref{eq:renyiMQCexample000007ex000001}.}
\begin{center}
\begin{tabular}{cccc}
\hline\hline
 Quantifier & $({\rho_{\text{eqn}}},{\hat{S}_z})$ & $({\rho_{\text{GHZ}}},{\hat{S}_z})$ & $({\rho_{\text{W}}},{\hat{S}_x})$  \\
\hline
${\mathcal{I}^L}$ & $\frac{1}{8}\, N\, {p^2}$ & $\frac{1}{8}\, {N^2}\, {p^2}$ & $\frac{1}{8}\left(4 + 3\left({N - 2}\right)\right){p^2}$ \\
${\mathcal{F}_Q}$ & $N\frac{d\,{p^2}}{4\left(2 + (d - 2)p \right)}$ & ${N^2}\frac{d\,{p^2}}{4\left(2 + (d - 2)p \right)}$ & $(3N - 2) \frac{{d}\,{p^2}}{4\left(2 + \left({d - 2}\right)p \right) }$  \\
$V$ & $\frac{1}{4}\,{N}$ & $\frac{1}{4}\,{N^2}$ & $\frac{1}{4}\left(N + 2\left(N - 1\right){p} \right)$  \\
${\mathcal{I}_{1/2}}$ & $\frac{N}{4d}{\left(\sqrt{1 + (d - 1)p} - \sqrt{1 - p} \, \right)^2}$ & $\frac{{N^2}}{4d}{\left(\sqrt{1 + (d - 1)p} - \sqrt{1 - p}\, \right)^2}$ & $\frac{\left(3N - 2\right)}{4\, d}{\left(\sqrt{1 + (d - 1)p} - \sqrt{1 - p}\, \right)^2}$  
\\
\hline\hline
\end{tabular}
\label{tab:unifiedtablequantities000xxx000xxx001}
\end{center}
\end{table*}

\begin{figure}[!htb]
\begin{center}
\includegraphics[scale=0.9]{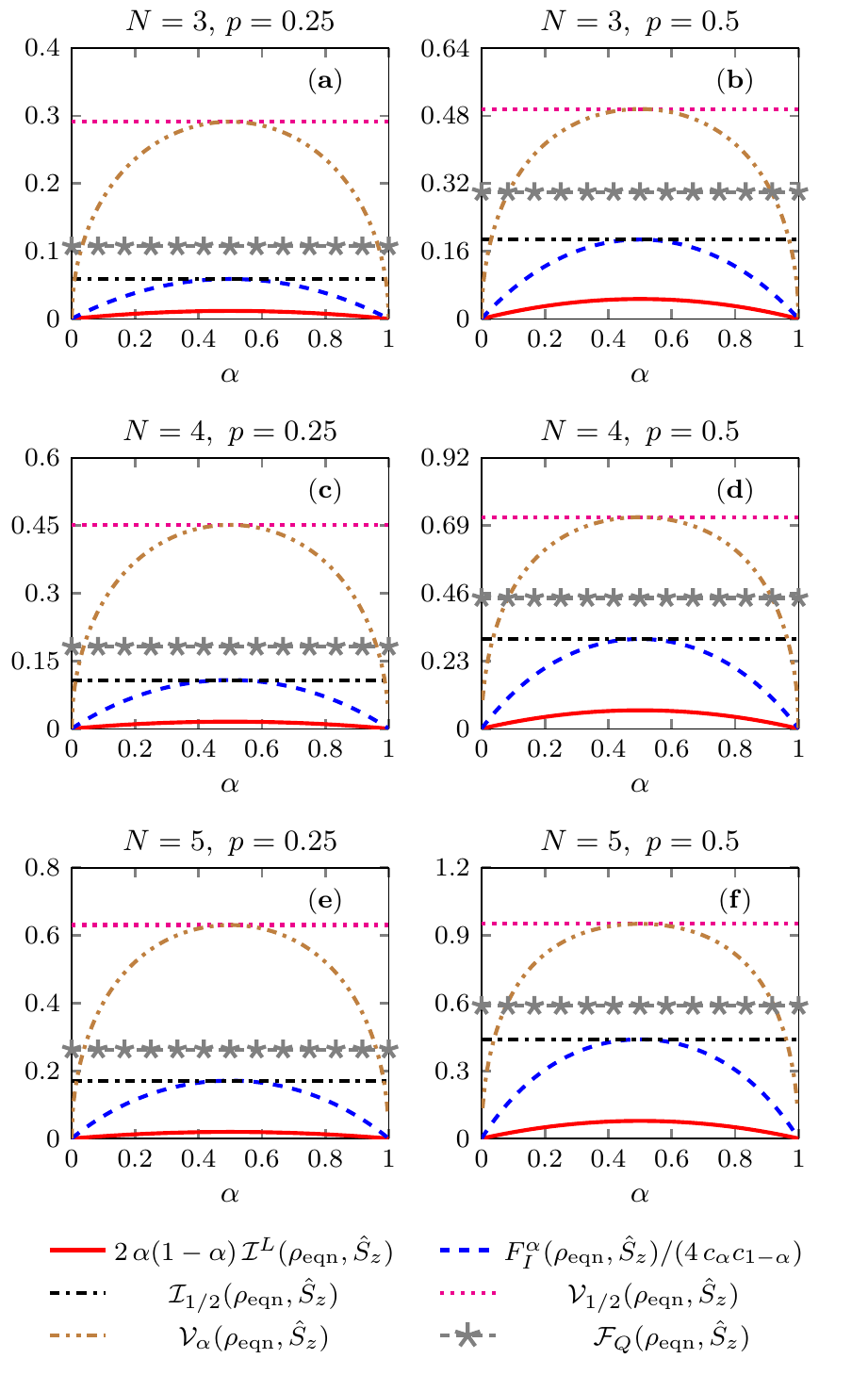}
\caption{(Color online) Plot of quantity $2\, \alpha(1 - \alpha)\, {\mathcal{I}^L}({\rho_{\text{eqn}}},{\hat{S}_z})$ (red solid line), $\alpha$-MQI ${F_I^{\alpha}}({\rho_{\text{eqn}}},{\hat{S}_z})/(4{c_{\alpha}}{c_{1 - \alpha}})$ (blue dashed line), Wigner-Yanase skew information ${\mathcal{I}_{1/2}}({\rho_{\text{eqn}}},{\hat{S}_z})$ (black dot dashed line), $1/2$-variance ${\mathcal{V}_{1/2}}({\rho_{\text{eqn}}},{\hat{S}_z})$ (magenta dotted line), $\alpha$-variance ${\mathcal{V}_{\alpha}}({\rho_{\text{eqn}}},{\hat{S}_z})$ (brown dashed and double-dotted line), and quantum Fisher information ${\mathcal{F}_Q}({\rho_{\text{eqn}}},{\hat{S}_z})$ (gray star dashed line). Here we choose the mixed state ${\rho_{\text{eqn}}} = \left({(1 - p)}/{2^N}\right)\mathbb{I} + p\,{\left(|+\rangle\langle{+}|\right)^{\otimes N}}$, with $|{+}\rangle = ({1}/{\sqrt{2}} \, )\left(|0\rangle + |1\rangle \right)$, and the generator ${\hat{S}_z} = ({1}/{2}) \,{\sum_{l=1}^{N}} ~{\mathbb{I}^{\, \otimes{l - 1}}}\otimes{\sigma_l^z}\otimes{\mathbb{I}^{\, \otimes{N - l}}}$, for values (a) $N = 3$ and $p = 0.25$; (b) $N = 3$ and $p = 0.5$; (c) $N = 4$ and $p = 0.25$; (d) $N = 4$ and $p = 0.5$; (e) $N = 5$ and $p = 0.25$; and (f) $N = 5$ and $p = 0.5$. In each panel, the plots successfully fulfill the constraints imposed by the chain of bounds given in Eqs.~\eqref{eq:renyiMQC0055c2a4bound000000x0p},~\eqref{eq:renyiMQC0055c2a4bound000000x0q}, and~\eqref{eq:renyiMQC0055c2a4bound000000x0p000003}.}
    \label{fig:examples02}
    \end{center}
    \end{figure}

In this Section we study multiparticle systems of $N$-qubit states belonging to the $d$-dimensional Hilbert space ${\mathcal{H}_d}$, with $d = {2^N}$. We consider three prototypical examples of states which are well known in 
quantum information.
From now on we will choose the collective spin operator $\hat{A} = \hat{n}\cdot\vec{S}$, where $\hat{n} = \{{n_x},{n_y},{n_z}\}$ is a unit vector with ${n_x^2} + {n_y^2} + {n_z^2} = 1$, and $\vec{S} = \{{\hat{S}_x},{\hat{S}_y},{\hat{S}_z}\}$ is the angular momentum vector, with
\begin{equation}
\label{eq:renyiMQCexample000007ex000001}
{\hat{S}_{x,y,z}} = \frac{1}{2}\,{\sum_{l=1}^{N}} ~{\mathbb{I}^{\, \otimes{l - 1}}}\otimes{\sigma_l^{x,y,z}}\otimes{\mathbb{I}^{\, \otimes{N - l}}} ~.
\end{equation}
Let us first set $\hat{n} = \{0,0,1\}$, i.e., $\hat{A} = {\hat{S}_z}$, and consider the probe state 
\begin{equation}
\label{eq:renyiMQCexample000001ex000001}
{\rho_{\text{eqn}}} = \left(\frac{1 - p}{d}\right)\mathbb{I} + p\,{\left(|+\rangle\langle{+}|\right)^{\otimes N}} ~,
\end{equation}
with $d = 2^N$, $0 < p < 1$, and $|{+}\rangle = ({1}/{\sqrt{2}} \, )\left(|0\rangle + |1\rangle \right)$ is the equal superposition state. For $0 < \alpha < 1$, we obtain
\begin{equation}
\label{eq:alphaMQCUnifSup000xxx000xxx003}
{\rho_{\text{eqn}}^{(\alpha)}} = {c_{\alpha}} {\left(\frac{1 - p}{d}\right)^{\alpha}}\mathbb{I} + {\xi_{\alpha}}(p,d){\left(|+\rangle\langle{+}|\right)^{\otimes N}} ~,
\end{equation}
where we define 
\begin{equation}
\label{eq:alphaMQCUnifSup000xxx000xxx004}
{c_{\alpha}^{-1}} = (d - 1){\left(\frac{1 - p}{d}\right)^{\alpha}} + {\left(\frac{1 + (d - 1)\, p}{d}\right)^{\alpha}}
\end{equation}
and 
\begin{equation}
\label{eq:alphaMQCUnifSup000xxx000xxx000xxx1111}
{\xi_{\alpha}}(d,p) := 1 - {c_{\alpha}}\, {d^{1 - \alpha}}{{(1 - p)^{\alpha}}} ~.
\end{equation}
The coherence orders decomposition ${\rho_{\text{eqn}}^{(\alpha)}} = {\sum_m}\, {({\rho_{\text{eqn}}^{(\alpha)}})_m}$ into non-Hermitian blocks originates cumbersome expressions that we do not report here. 
It turns out that the corresponding expressions for the $\alpha$-MQI take simple forms. 
For $m = 0$ one obtains
\begin{equation}
\label{eq:state1main0000_xxxx_000001}
{I_0^{\alpha}}({\rho_{\text{eqn}}}) = \frac{1}{d}\left[1 + \left(\frac{(2N)!}{d\,{(N!)^2}} - 1\right){\xi_{\alpha}}(d,p)\, {\xi_{1 - \alpha}}(d,p)\right] ~,
\end{equation}
while, for $m \neq 0$, we have
\begin{equation}
\label{eq:state1main0000_xxxx_000002}
{I_m^{\alpha}}({\rho_{\text{eqn}}}) = \frac{{g_{N,m}}}{d^2}\, {\xi_{\alpha}}(d,p)\, {\xi_{1 - \alpha}}(d,p) ~,
\end{equation}
where 
\begin{equation}
\label{eq:state1main0000_xxxx_000003}
{g_{N,m}} = \frac{(2N)!}{{(N - m)!}\,{(N + m)!}}
\end{equation}
is the degeneracy of each block.
Therefore, from Eqs.~\eqref{eq:state1main0000_xxxx_000001} and~\eqref{eq:state1main0000_xxxx_000002} one may write down 
\begin{equation}
\label{eq:alphaMQCUnifSup000xxx000xxx001}
{F_I^{\alpha}}({\rho_{\text{eqn}}},{S_z}) = N\, {\xi_{\alpha}}(d,p)\, {\xi_{1 - \alpha}}(d,p) ~.
\end{equation}
In Table~\ref{tab:unifiedtablequantities000xxx000xxx001} we list the expressions 
of ${\mathcal{I}^L}({\rho_{\text{eqn}}},{\hat{S}_z})$, 
the quantum Fisher information ${\mathcal{F}_Q}({\rho_{\text{eqn}}},{\hat{S}_z})$, 
the standard variance $V({\rho_{\text{eqn}}},{S_z})$, 
and the Wigner-Yanase skew information ${\mathcal{I}_{1/2}}({\rho_{\text{eqn}}},{\hat{S}_z})$. 
In Fig.~\ref{fig:examples02} we plot Eq.~\eqref{eq:alphaMQCUnifSup000xxx000xxx001} 
for the system sizes $N = 3$, $N = 4$ and $N = 5$, 
and mixing parameter values $p = 0.25$ and $p = 0.5$.

\begin{figure}[!htb]
\begin{center}
\includegraphics[scale=0.9]{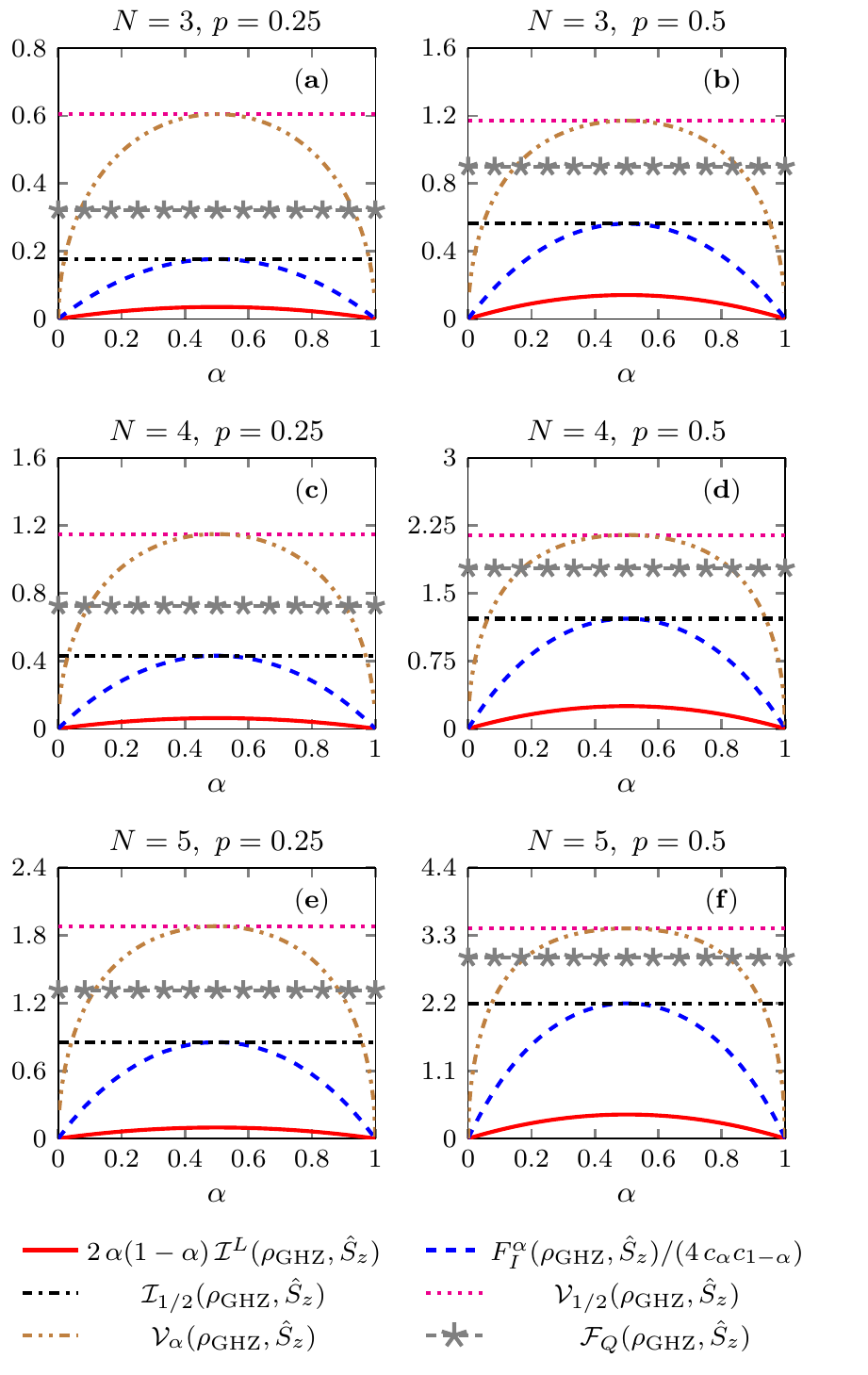}
\caption{(Color online) Plot of  $2\, \alpha(1 - \alpha)\, {\mathcal{I}^L}({\rho_{\text{GHZ}}},{\hat{S}_z})$ (red solid line), $\alpha$-MQI ${F_I^{\alpha}}({\rho_{\text{GHZ}}},{\hat{S}_z})/(4{c_{\alpha}}{c_{1 - \alpha}})$ (blue dashed line), Wigner-Yanase skew information ${\mathcal{I}_{1/2}}({\rho_{\text{GHZ}}},{\hat{S}_z})$ (black dot dashed line), $1/2$-variance ${\mathcal{V}_{1/2}}({\rho_{\text{GHZ}}},{\hat{S}_z})$ (magenta dotted line), $\alpha$-variance ${\mathcal{V}_{\alpha}}({\rho_{\text{GHZ}}},{\hat{S}_z})$ (brown dashed and double-dotted line), and quantum Fisher information ${\mathcal{F}_Q}({\rho_{\text{GHZ}}},{\hat{S}_z})$ (gray star dashed line). Here we choose the mixed state ${\rho_{\text{GHZ}}} = \left({(1 - p)}/{2^N}\right)\mathbb{I} + p\,|{\text{GHZ}_N}\rangle\langle{\text{GHZ}_N}|$, with $|{\text{GHZ}_N}\rangle = ({1}/{\sqrt{2}})\left(\,{|0\rangle^{\otimes N}} + {|1\rangle^{\otimes N}}\right)$, and the generator ${\hat{S}_z} = ({1}/{2}) \,{\sum_{l=1}^{N}} ~{\mathbb{I}^{\, \otimes{l - 1}}}\otimes{\sigma_l^z}\otimes{\mathbb{I}^{\, \otimes{N - l}}}$, for values (a) $N = 3$ and $p = 0.25$; (b) $N = 3$ and $p = 0.5$; (c) $N = 4$ and $p = 0.25$; (d) $N = 4$ and $p = 0.5$; (e) $N = 5$ and $p = 0.25$; and (f) $N = 5$ and $p = 0.5$. In each panel, the plots successfully fulfill the constraints imposed by the chain of bounds given in Eqs.~\eqref{eq:renyiMQC0055c2a4bound000000x0p},~\eqref{eq:renyiMQC0055c2a4bound000000x0q}, and~\eqref{eq:renyiMQC0055c2a4bound000000x0p000003}.}
\label{fig:examples00003}
\end{center}
\end{figure}

Let us move to a different case. Now, we choose the unit vector $\hat{n} = \{0,0,1\}$, i.e., $\hat{A} = {\hat{S}_z}$, and consider the state 
\begin{equation}
\label{eq:renyiMQCexample000003ex0000001}
{\rho_{\text{GHZ}}} = \left(\frac{1 - p}{d}\right)\mathbb{I} + p\,|{\text{GHZ}_N}\rangle\langle{\text{GHZ}_N}| ~,
\end{equation}
with $d = 2^N$, $0 < p < 1$, and $|{\text{GHZ}_N}\rangle$ is the GHZ-state of $N$ particles defined as
\begin{equation}
\label{eq:renyiMQCexample000004ex0000001}
|{\text{GHZ}_N}\rangle = \frac{1}{\sqrt{2}}\left(\,{|0\rangle^{\otimes N}} + {|1\rangle^{\otimes N}}\right) ~.
\end{equation}
Based on Eq.~\eqref{eq:renyiMQCexample000004ex0000001}, for $0 < \alpha < 1$, one may verify that
\begin{equation}
\label{eq:renyiMQCexample000003ex000000xxxxx0000xxx9999}
{\rho_{\text{GHZ}}^{(\alpha)}} = {c_{\alpha}} {\left(\frac{1 - p}{d}\right)^{\alpha}}\mathbb{I} + {\xi_{\alpha}}(p,d)|{\text{GHZ}_N}\rangle\langle{\text{GHZ}_N}|  ~,
\end{equation}
where both functions $c_{\alpha}$ and ${\xi_{\alpha}}(p,d)$ are the ones defined in Eqs.~\eqref{eq:alphaMQCUnifSup000xxx000xxx004} and~\eqref{eq:alphaMQCUnifSup000xxx000xxx000xxx1111}, respectively. 
By analogy with the previous example, the expressions for $\alpha$-MQI take simple forms. 
We emphasize that $\alpha$-MQI is identically zero for all indeces $m \neq 0$ and $m \neq \pm N$. 
For $m = 0$ one obtains
\begin{equation}
\label{eq:alphamqixxxx_000_xxxx_00001_ghz}
{I_0^{\alpha}}({\rho_{\text{GHZ}}}) = {c_{\alpha}}{c_{1 - \alpha}} - \frac{1}{2} \, {\xi_{\alpha}}(d,p)\, {\xi_{1 - \alpha}}(d,p) ~,
\end{equation}
while, for $m = \pm N$, one gets
\begin{equation}
\label{eq:alphamqixxxx_000_xxxx_00002_ghz}
{I_{\pm N}^{\alpha}}({\rho_{\text{GHZ}}}) = \frac{1}{4} \, {\xi_{\alpha}}(d,p)\, {\xi_{1 - \alpha}}(d,p) ~,
\end{equation}
Therefore, from Eqs.~\eqref{eq:alphamqixxxx_000_xxxx_00001_ghz} and~\eqref{eq:alphamqixxxx_000_xxxx_00002_ghz} the second moment of $\alpha$-MQI is given by
\begin{equation}
\label{eq:alphaMQCUnifSup000xxx000xxx000xxx1111xxx222}
{F_I^{\alpha}}({\rho_{\text{GHZ}}},{S_z}) = {N^2}\, {\xi_{\alpha}}(d,p)\, {\xi_{1 - \alpha}}(d,p) ~.
\end{equation}
In Table~\ref{tab:unifiedtablequantities000xxx000xxx001} we list the expressions obtained for ${\mathcal{I}^L}({\rho_{\text{GHZ}}},{S_z})$, QFI ${\mathcal{F}_Q}({\rho_{\text{GHZ}}},{S_z})$, the standard variance $V({\rho_{\text{GHZ}}},{S_z})$, and the Wigner-Yanase skew information ${\mathcal{I}_{1/2}}({\rho_{\text{GHZ}}},{S_z})$. 
It is worthwhile to note that, fixing the generator $\hat{A} = {\hat{S}_z}$ 
as the collective magnetization along $z$-axis, ${F_I^{\alpha}}$ grows quadratically 
with system size $N$ for the mixed GHZ-state in Eq.~\eqref{eq:renyiMQCexample000003ex0000001}, 
while it grows linearly for the state ${\rho_{\text{eqn}}}$ in Eq.~\eqref{eq:renyiMQCexample000001ex000001}. 
In Fig.~\ref{fig:examples00003} we plot Eq.~\eqref{eq:alphaMQCUnifSup000xxx000xxx000xxx1111xxx222} 
for the values of system size $N = 3$, $N = 4$ and $N = 5$, and mixing parameter $p = 0.25$ and $p = 0.5$.

\begin{figure}[!htb]
\begin{center}
\includegraphics[scale=0.9]{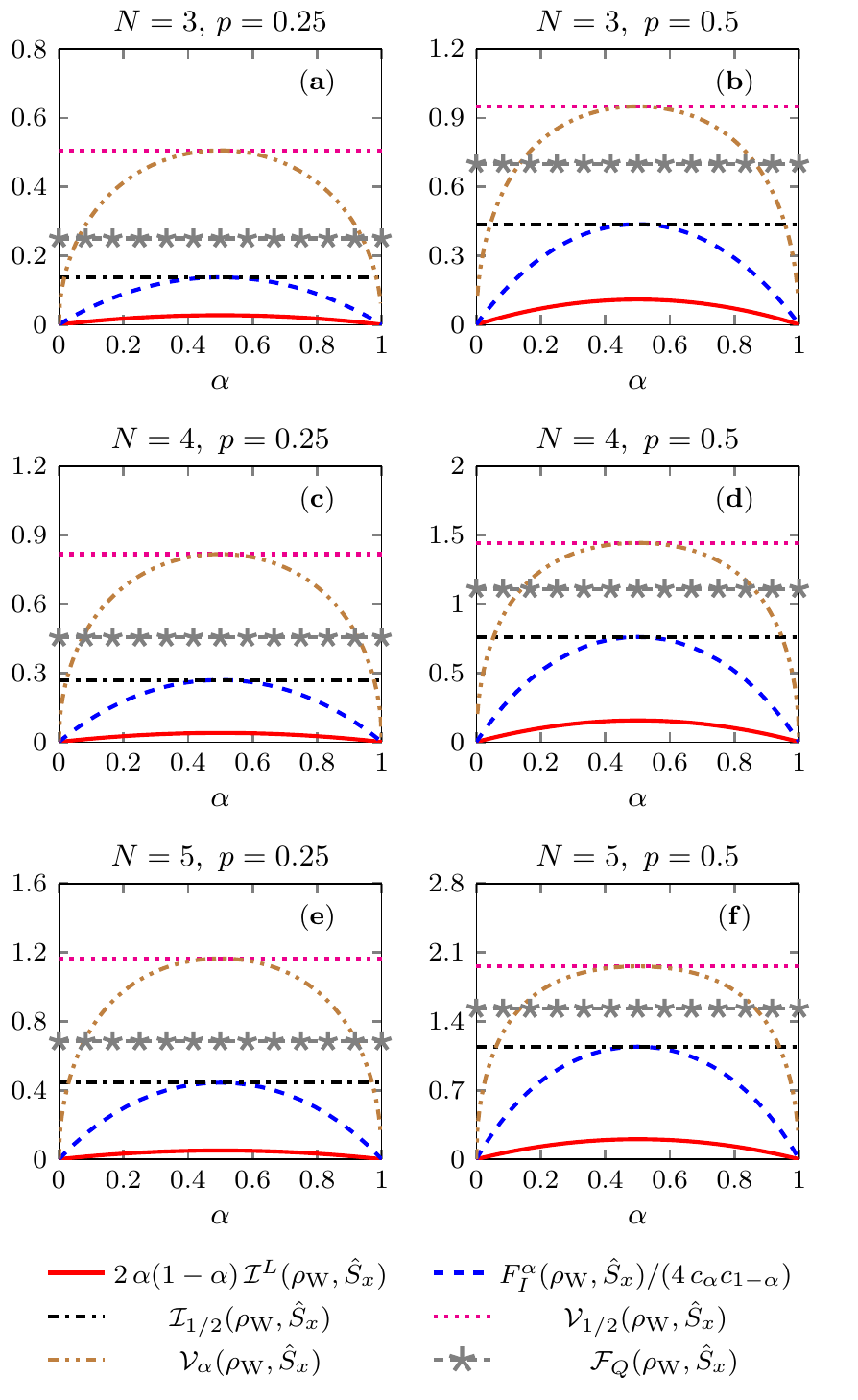}
\caption{(Color online) Plot of $2\, \alpha(1 - \alpha)\, {\mathcal{I}^L}({\rho_1},{\hat{S}_x})$ (red solid line), $\alpha$-MQI ${F_I^{\alpha}}({\rho_{\text{W}}},{\hat{S}_x})/(4{c_{\alpha}}{c_{1 - \alpha}})$ (blue dashed line), Wigner-Yanase skew information ${\mathcal{I}_{1/2}}({\rho_{\text{W}}},{\hat{S}_x})$ (black dot dashed line), $1/2$-variance ${\mathcal{V}_{1/2}}({\rho_{\text{W}}},{\hat{S}_x})$ (magenta dotted line), $\alpha$-variance ${\mathcal{V}_{\alpha}}({\rho_{\text{W}}},{\hat{S}_x})$ (brown dashed and double-dotted line), and quantum Fisher information ${\mathcal{F}_Q}({\rho_{\text{W}}},{\hat{S}_x})$ (gray star dashed line). Here we choose the mixed state ${\rho_{\text{W}}} = \left({(1 - p)}/{2^N}\right)\mathbb{I} + p\,|{W}\rangle\langle{W}|$, with $|{W}\rangle = ({1}/{\sqrt{N}})\, {\sum_{l = 1}^N} ~ {{|0\rangle}^{\, \otimes{l - 1}}}\otimes{{|1\rangle}^l}\otimes{{|0\rangle}^{\, \otimes{N - l}}} $, and the generator ${\hat{S}_x} = ({1}/{2}) \,{\sum_{l=1}^{N}} ~{\mathbb{I}^{\, \otimes{l - 1}}}\otimes{\sigma_l^x}\otimes{\mathbb{I}^{\, \otimes{N - l}}}$, for values (a) $N = 3$ and $p = 0.25$; (b) $N = 3$ and $p = 0.5$; (c) $N = 4$ and $p = 0.25$; (d) $N = 4$ and $p = 0.5$; (e) $N = 5$ and $p = 0.25$; and (f) $N = 5$ and $p = 0.5$. In each panel, the plots successfully fulfill the constraints imposed by the chain of bounds given in Eqs.~\eqref{eq:renyiMQC0055c2a4bound000000x0p},~\eqref{eq:renyiMQC0055c2a4bound000000x0q}, and~\eqref{eq:renyiMQC0055c2a4bound000000x0p000003}.}
\label{fig:examples0000320101}
\end{center}
\end{figure}

Finally, we turn to our third example. We begin by specifying the unit vector $\hat{n} = \{1,0,0\}$ related to the generator $\hat{A} = {\hat{S}_x}$, and define the probe state
\begin{equation}
\label{eq:renyiMQCexample000005x000001000xxx}
{\rho_{\text{W}}} = \left(\frac{1 - p}{d}\right)\mathbb{I} + p\,|{W}\rangle\langle{W}| ~,
\end{equation}
where $d = 2^N$, $0 < p < 1$, and $|W\rangle$ is the $W$-state of $N$ particles given by~\cite{PhysRevA.62.062314}
\begin{equation}
\label{eq:renyiMQCexample000006x00001}
|{W}\rangle = \frac{1}{\sqrt{N}}\, {\sum_{l = 1}^N} ~ {{|0\rangle}^{\, \otimes{l - 1}}}\otimes{{|1\rangle}^l}\otimes{{|0\rangle}^{\, \otimes{N - l}}} ~.
\end{equation}
For $0 < \alpha < 1$, it follows that 
\begin{equation}
\label{eq:renyiMQCexample000005x000001}
{\rho_{\text{W}}^{(\alpha)}} = {c_{\alpha}} {\left(\frac{1 - p}{d}\right)^{\alpha}}\mathbb{I} + {\xi_{\alpha}}(p,d) |{W}\rangle\langle{W}| ~,
\end{equation}
where both functions $c_{\alpha}$ and ${\xi_{\alpha}}(p,d)$ are exactly the same as defined in Eqs.~\eqref{eq:alphaMQCUnifSup000xxx000xxx004} and~\eqref{eq:alphaMQCUnifSup000xxx000xxx000xxx1111}, respectively. 
In spite of the complexity of the expressions of the coherence orders 
decomposition ${\rho_{\text{W}}^{(\alpha)}} = {\sum_m}\, {({\rho_{\text{W}}^{(\alpha)}})_m}$, 
it is possible to derive analytically the second moment of $\alpha$-MQI, which reads
\begin{equation}
\label{eq:alphaMQCUnifSup000xxx000xxx000xxx1111xxx333xxx001}
{F_I^{\alpha}}({\rho_{\text{W}}},{S_x}) = \left(\frac{3N - 2}{d - 1}\right)\left(d\, {c_{\alpha}}{c_{1 - \alpha}} - 1\right) ~.
\end{equation}
Table~\ref{tab:unifiedtablequantities000xxx000xxx001} reports the expressions obtained 
for ${\mathcal{I}^L}({\rho_{\text{W}}},{\hat{S}_x})$, quantum Fisher information 
${\mathcal{F}_Q}({\rho_{\text{W}}},{\hat{S}_x})$, standard variance $V({\rho_{\text{W}}},{\hat{S}_x})$, 
and also Wigner-Yanase skew information ${\mathcal{I}_{1/2}}({\rho_{\text{W}}},{\hat{S}_x})$. 
In Fig.~\ref{fig:examples0000320101} we plot ${F_I^{\alpha}}({\rho_{\text{W}}},{\hat{S}_x})$ 
for the values of system size $N = 3$, $N = 4$ and $N = 5$, and mixing parameter $p = 0.25$ and $p = 0.5$.

\subsection{Long-range Quantum Ising model}
\label{sec:examples000xxx0003}
\begin{figure}[t]
\begin{center}
\includegraphics[scale=0.85]{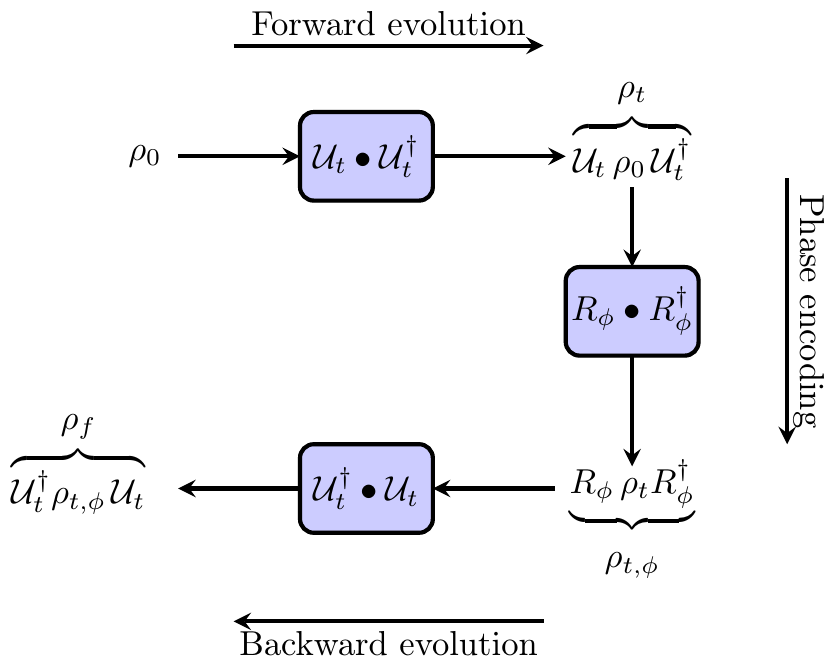}
\end{center}
\caption{(Color online) Depiction of the quantum protocol discussed in Section~\ref{sec:examples000xxx0003}. In the forward process, the initial state $\rho_0$ of the system undergoes a unitary evolution and reaches the intermediate state ${\rho_t} = {\mathcal{U}_t}\, {\rho_0}\, {\mathcal{U}_t^{\dagger}}$. Then, the ope\-ra\-tor ${R_{\phi}}$ imprints a phase shift $\phi$ into $\rho_t$, and the system is subsequently described by the state ${\rho_{t,\phi}} = {R_{\phi}}\, {\rho_t} {R_{\phi}^{\dagger}}$. In the last step of the protocol, the system evolves backward in time according to the reversed unitary dynamics and is finally described by the final state ${\rho_f} = {\mathcal{U}_t^{\dagger}} {\rho_{t,\phi}}\, {\mathcal{U}_t}$.}
\label{fig:picturepaperv001}
\end{figure}

Now we move to the dynamical scenario and consider the protocol depicted in Fig.~\ref{fig:picturepaperv001}. Such in\-ter\-fe\-ro\-me\-tric scheme is equivalent to the Loschmidt-echo protocol proposed for the creation and detection of entangled non-Gaussian states~\cite{PhysRevA.94.010102} with an Ising model with long-range interactions recently realized in a dilute gas of Rydberg-dressed cesium atoms~\cite{1910_arxiv_1910.13687}. This protocol is also analogous to time-reversal dynamics simulating Loschmidt echo in NMR many-spin systems~\cite{PhysRevB.3.684,10.1098_rsta.2015.0163}. The protocol was implemented in a trapped ion quantum simulator, and used to detect the buildup of quantum correlations in many-body systems via multiple quantum coherences~\cite{101038nphys4119v01}. The Hamiltonian of the system is a fully connected Ising model
\begin{equation}
\label{eq:fullyIsing000xxx001}
{H_{zz}} = \frac{J}{N}\,{\sum_{j < l}}\,{\sigma_j^z}{\sigma_l^z} ~,
\end{equation}
where $J$ is the coupling strength, $N$ is the number of spins, and $\sigma^z_j$ are the Pauli spin matrices. For simplicity, the system is initialized in the state
\begin{equation}
\label{eq:renyiMQCexample000001ex000001exampleVID}
{\rho_0} = \left(\frac{1 - p}{d}\right)\mathbb{I} + p\,{\left(|+\rangle\langle{+}|\right)^{\otimes N}} ~,
\end{equation}
with $d = 2^N$, $0 < p < 1$, and $|{+}\rangle = ({1}/{\sqrt{2}} \, )\left(|0\rangle + |1\rangle \right)$ being the equal superposition state.

In the forward step of the protocol of Fig.~\ref{fig:picturepaperv001}, the initial state $\rho_0$ of the system evolves unitarily according to ${\mathcal{U}_t} = {e^{-it {H_{zz}}}}$ and reaches the intermediate state ${\rho_t} = {\mathcal{U}_t}\, {\rho_0}\, {\mathcal{U}_t^{\dagger}}$. Just to clarify, here we set $\hbar = 1$. Sub\-se\-quently, the operator ${\hat{R}_{\phi}} = {e^{-i\phi{\hat{S}_{x}}}}$ rotates the system about $x$-axis, with ${\hat{S}_x} = ({1}/{2}) \,{\sum_{l=1}^{N}} ~{\mathbb{I}^{\, \otimes{l - 1}}}\otimes{\sigma_l^x}\otimes{\mathbb{I}^{\, \otimes{N - l}}}$, and thus the system is characterized by the state ${\rho_{t,\phi}} = {R_{\phi}} \, {\rho_t} {R_{\phi}^{\dagger}}$. Finally, the system evolves unitarily backward and reaches the final state ${\rho_f} = {\mathcal{U}_t^{\dagger}} {\rho_{t,\phi}}\, {\mathcal{U}_t}$. We stress that, in practice, the backward protocol is implemented inverting the sign of $H$ by changing $J \rightarrow - J$.

In the following we will apply $\alpha$-relative purity to distinguish input and output states after running the quantum protocol. Interestingly, the relative purity involving states $\rho_0$ and ${\rho_f}$ becomes 
\begin{equation}
\label{eq:renyiMQC0023example000xxxx0000xxxx000xxxx11111prev}
{f_{\alpha}}({\rho_0},{\rho_f}) = \text{Tr}({\rho_0^{\alpha}} \, {\rho_f^{1 -\alpha}}) = 
\text{Tr}({\rho_t^{\alpha}} {\rho_{t,\phi}^{1 - \alpha}}) = {f_{\alpha}}({\rho_t},{\rho_{t,\phi}}) ~, 
\end{equation}
where we have used that ${\rho_f^{1 -\alpha}} = {\mathcal{U}_t^{\dagger}} {\rho_{t,\phi}^{1 - \alpha}}\, {\mathcal{U}_t}$, since ${\mathcal{U}_t}$ is a unitary operator~\cite{PhysRevA.91.042330}. Note that, for $\phi = 0$, we thus have ${\rho_{t,0}} = {\rho_t}$ and $\alpha$-relative purity is equal to $1$. We point out that $\alpha$-relative purity will play the role of revival pro\-ba\-bi\-li\-ty exhibited by the quantum system undergoing the time-reversal evolution. Indeed, RHS of Eq.~\eqref{eq:renyiMQC0023example000xxxx0000xxxx000xxxx11111prev} means that, for a nonzero phase shift $\phi$ encoded into the time-dependent state $\rho_t$ by the rotation ${R_{\phi}} = {e^{-i\phi{\hat{S}_{x}}}}$ inserted between forward and backward time evolutions, thus the $\alpha$-relative purity $ {f_{\alpha}}({\rho_t},{\rho_{t,\phi}}) $ will deviate from the unity as a function of time $t$. Moreover, such a revival can be interpreted as a signature of the buildup of correlations of the many-body state $\rho_t$~\cite{101038nphys4119v01}.

According to Eq.~\eqref{eq:renyiMQC0029}, one may write the $\alpha$-relative purity in terms of the $\alpha$-MQI as
\begin{equation}
\label{eq:renyiMQC0023example000xxxx0000xxxx000xxxx11111}
{f_{\alpha}}({\rho_0},{\rho_f}) = {( {{c_{\alpha}}\, {c_{1 - \alpha}}} )^{-1}} \, 
{\sum_m} \,{e^{- i m \phi}} \, {I_m^{\alpha}}({\rho_t}) ~,
\end{equation}
where 
\begin{equation}
\label{eq:renyiMQC0023example0001}
{I_m^{\alpha}}({\rho_t}) = \text{Tr}\left( { [ {({\rho_t})_m^{(\alpha)}} ]^{\dagger}} 
{({\rho_t})_m^{(1 - \alpha)}} \right) ~,
\end{equation}
with 
\begin{equation}
\label{eq:renyiMQC005examplexxxx0001}
{({\rho_t})_m^{(\alpha)}} := {\sum_{{\lambda_j} - {\lambda_{\ell}} = m}}\, 
\langle{j}|{\rho_t^{(\alpha)}}|{\ell}\rangle |{j}\rangle \langle{\ell}| ~.
\end{equation}
From Sec.~\ref{sec:reviewCoherenceOrders00xx0011}, we recall that ${c_{\alpha}^{-1}} = \text{Tr}({\rho_t^{\alpha}}) = \text{Tr}({\rho_0^{\alpha}})$, where we have used that ${\rho_t^{\alpha}} = {\mathcal{U}_t}\, {\rho_0^{\alpha}}\, {\mathcal{U}_t^{\dagger}}$. Furthermore, we stress that ${\{ |{j}\rangle\}_{j = 1,\ldots,{2^N}}}$ describe the {\it reference basis} ge\-ne\-ra\-ted by the eigenstates of ${\hat{S}_x}$, with their respective set of eigenvalues $\vec{\lambda} = \{-N/2, -N/2 + 1,\ldots, N/2 - 1, N/2 \}$ which exhibits degeneracy ${\textsf{g}_{\lambda_j}} = N!/[\, (N/2 + {\lambda_j})! \, (N/2 - {\lambda_j})! \,]$, and thus $m = \{-N, -N +1, \ldots, N - 1, N\}$.
\begin{figure}[t]
\begin{center}
\includegraphics[scale=0.8]{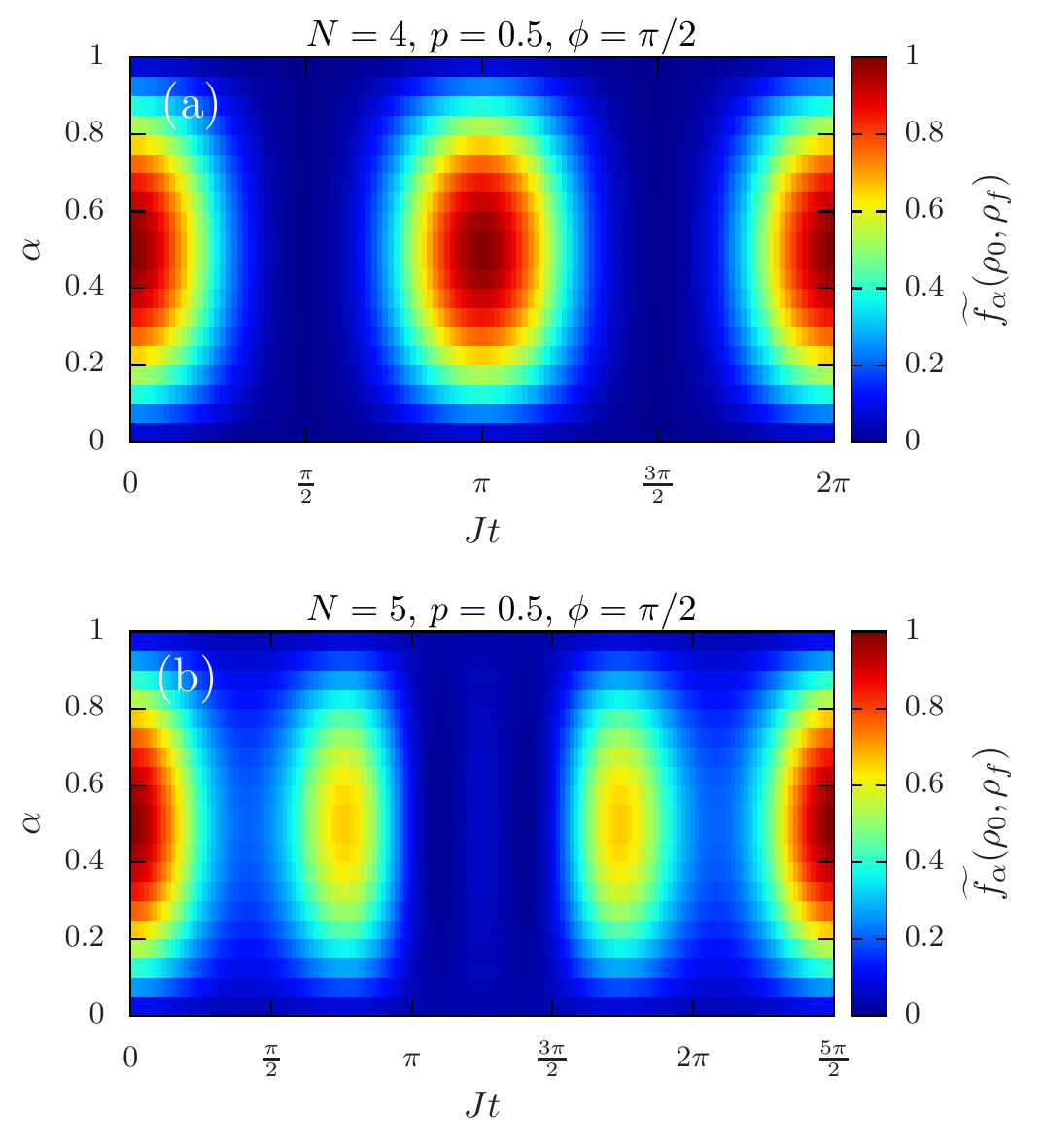}
\caption{(Color online) Density plot of normalized re\-la\-ti\-ve pu\-ri\-ty, 
$\widetilde{f_{\alpha}}({\rho_0},{\rho_f})$, for states $\rho_0$ and ${\rho_f} = 
{\mathcal{U}_t^{\dagger}} {R_{\phi}}\, {\mathcal{U}_t}\, {\rho_0}\, 
{\mathcal{U}_t^{\dagger}} {R_{\phi}^{\dagger}} {\mathcal{U}_t}$. 
Here we have ${\mathcal{U}_t} = {e^{- i t {H_{zz}}}}$, with 
${H_{zz}} = ({J}/{N})\,{\sum_{j < l}}\,{\sigma_j^z}{\sigma_l^z}$ standing as the 
fully connected Ising Hamiltonian, and also ${R_{\phi}} = {e^{- i \phi {\hat{S}_x}}}$, 
where ${\hat{S}_x} = ({1}/{2}) \,{\sum_{l=1}^{N}} ~{\mathbb{I}^{\, \otimes{l - 1}}}\otimes{\sigma_l^x}\otimes{\mathbb{I}^{\, \otimes{N - l}}}$. The input state is ${\rho_0} = \left({(1 - p)}/d\right)\mathbb{I} + p\,{\left(|+\rangle\langle{+}|\right)^{\otimes N}}$, with $d = {2^N}$ and $|{+}\rangle = ({1}/{\sqrt{2}} \, )\left(|0\rangle + |1\rangle \right)$. For simplicity, here we set $p = 0.5$ and $\phi = \pi/2$, and increase the size of the system as (a) $N = 4$, and (b) $N = 5$.}
\label{fig:pictureMQCalphaRenyi0000000xxxxx00000xxx1}
\end{center}
\end{figure}

We now apply the above discussion to numerically study the time-evolution of normalized $\alpha$-MQI spectrum, $\{ \widetilde{I_{m}^{\alpha}}({\rho_t}) \}$, and its the second moment $\widetilde{F_I^{\alpha}}({\rho_t},{\hat{S}_x})$. Wi\-thout loss of generality, here we have adopted the normalization $\widetilde{\textsf{G}} := (\textsf{G} - \min\{\textsf{G}\})/(\max\{ \textsf{G} \} - \min\{ \textsf{G} \})$. 

\begin{figure*}[t]
\begin{center}
\hspace{-0.9cm}
\includegraphics[scale=0.65]{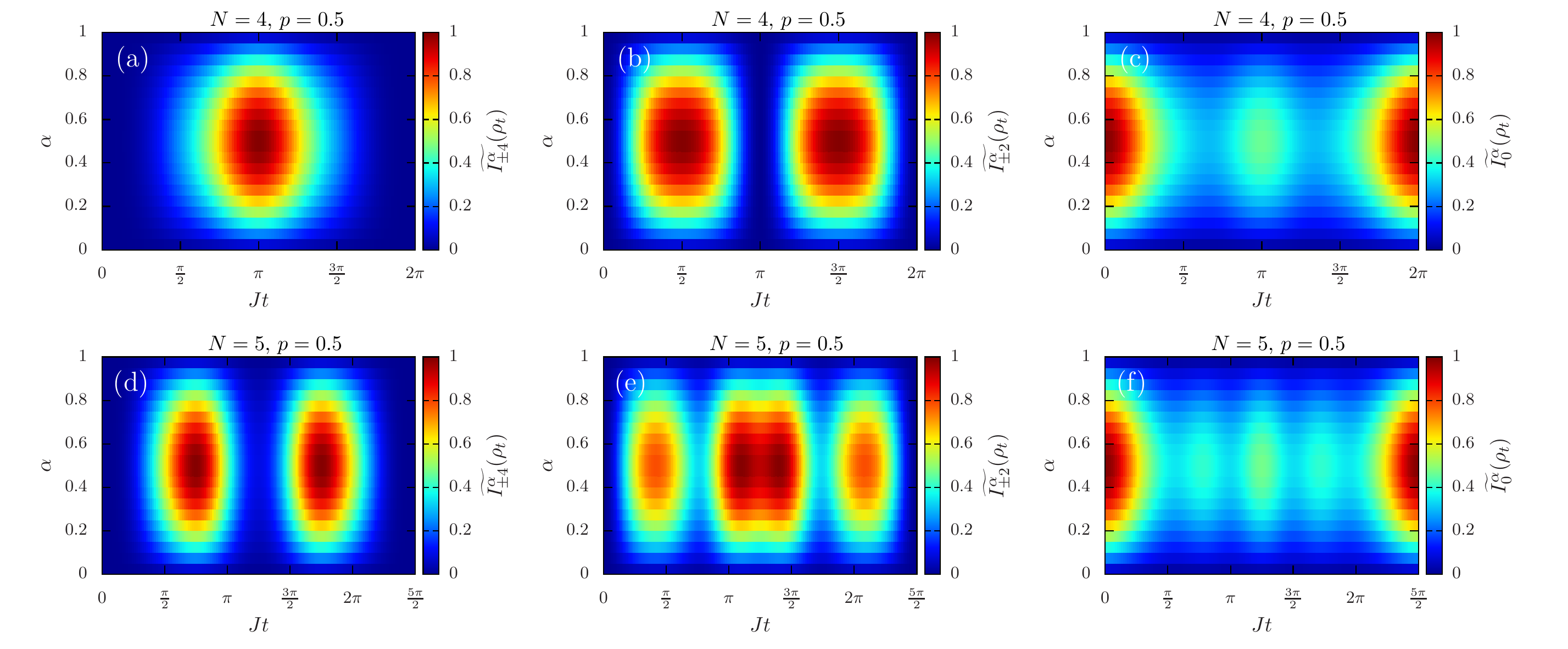}
\caption{(Color online) Density plot of normalized $\alpha$-MQI, $\widetilde{I_{m}^{\alpha}}({\rho_t})$, for the state ${\rho_t} = {e^{- i t {H_{zz}}}} \, {\rho}\, {e^{ i t {H_{zz}}}}$, where ${H_{zz}} = ({J}/{N})\,{\sum_{j < l}}\,{\sigma_j^z}{\sigma_l^z}$ is the fully connected Ising Hamiltonian, and the probe state ${\rho} = \left({(1 - p)}/d\right)\mathbb{I} + p\,{\left(|+\rangle\langle{+}|\right)^{\otimes N}}$, with $d = {2^N}$ and $|{+}\rangle = ({1}/{\sqrt{2}} \, )\left(|0\rangle + |1\rangle \right)$. For simplicity, here we fix the mixing parameter $p = 0.5$. The set of non-zero $\alpha$-MQI is given by $\widetilde{I_{\pm 4}^{\alpha}}({\rho_t})$, for (a) $N =4$ and (d) $N = 5$; $\widetilde{I_{\pm 2}^{\alpha}}({\rho_t})$, for (b) $N =4$ and (e) $N = 5$; and $\widetilde{I_{0}^{\alpha}}({\rho_t})$, for (c) $N =4$ and (f) $N = 5$.}
\label{fig:pictureMQCalphaRenyi0000000x8}
\end{center}
\end{figure*}

In Fig.~\ref{fig:pictureMQCalphaRenyi0000000xxxxx00000xxx1} we plot the normalized 
re\-la\-ti\-ve pu\-ri\-ty, $\widetilde{f_{\alpha}}({\rho_0},{\rho_f}) \equiv \widetilde{f_{\alpha}}({\rho_t},{\rho_{t,\phi}})$ 
(cf. Eq.~\eqref{eq:renyiMQC0023example000xxxx0000xxxx000xxxx11111prev}), 
as a function of $t$ and $\alpha$. Just to clarify, here 
${\rho_t} = {e^{- i t {H_{zz}}}} \, {\rho_0}\, {e^{ i t {H_{zz}}}}$, where ${H_{zz}}$ 
is given in Eq.~\eqref{eq:fullyIsing000xxx001} and $\rho_0$ is the probe state 
in Eq.~\eqref{eq:renyiMQCexample000001ex000001exampleVID},
and ${\rho_{t,\phi}} = {R_{\phi}} \, {\rho_t} {R_{\phi}^{\dagger}}$, with 
${R_{\phi}} = {e^{- i \phi {\hat{S}_x}}}$ and ${\hat{S}_x} = ({1}/{2}) 
\,{\sum_{l=1}^{N}} ~{\mathbb{I}^{\, \otimes{l - 1}}}\otimes{\sigma_l^x}\otimes{\mathbb{I}^{\, \otimes{N - l}}}$. 
We fix the mixing parameter $p = 0.5$, and the phase $\phi = \pi/2$. 

In Fig.~\ref{fig:pictureMQCalphaRenyi0000000x8} we plot the time-evolution of the normalized $\alpha$-MQI spectrum $\{ \widetilde{I_{m}^{\alpha}}({\rho_t}) \}$ (cf. Eq.~\eqref{eq:renyiMQC0023example0001}) for $N = 4$ and $N = 5$. Given the evolved state ${\rho_t} = {e^{- i t {H_{zz}}}} \, {\rho_0}\, {e^{ i t {H_{zz}}}}$, for $N = 4$ the nonzero $\alpha$-MQI are given by (a) $\widetilde{I_{\pm 4}^{\alpha}}({\rho_t})$, (b) $\widetilde{I_{\pm 2}^{\alpha}}({\rho_t})$, and (c) $\widetilde{I_{0}^{\alpha}}({\rho_t})$. Similarly, for the system size $N = 5$ the nonzero $\alpha$-MQI are given by (d) $\widetilde{I_{\pm 4}^{\alpha}}({\rho_t})$, (e) $\widetilde{I_{\pm 2}^{\alpha}}({\rho_t})$, and (f) $\widetilde{I_{0}^{\alpha}}({\rho_t})$. 
%

\begin{figure*}[thb]
\begin{center}
\includegraphics[scale=0.8]{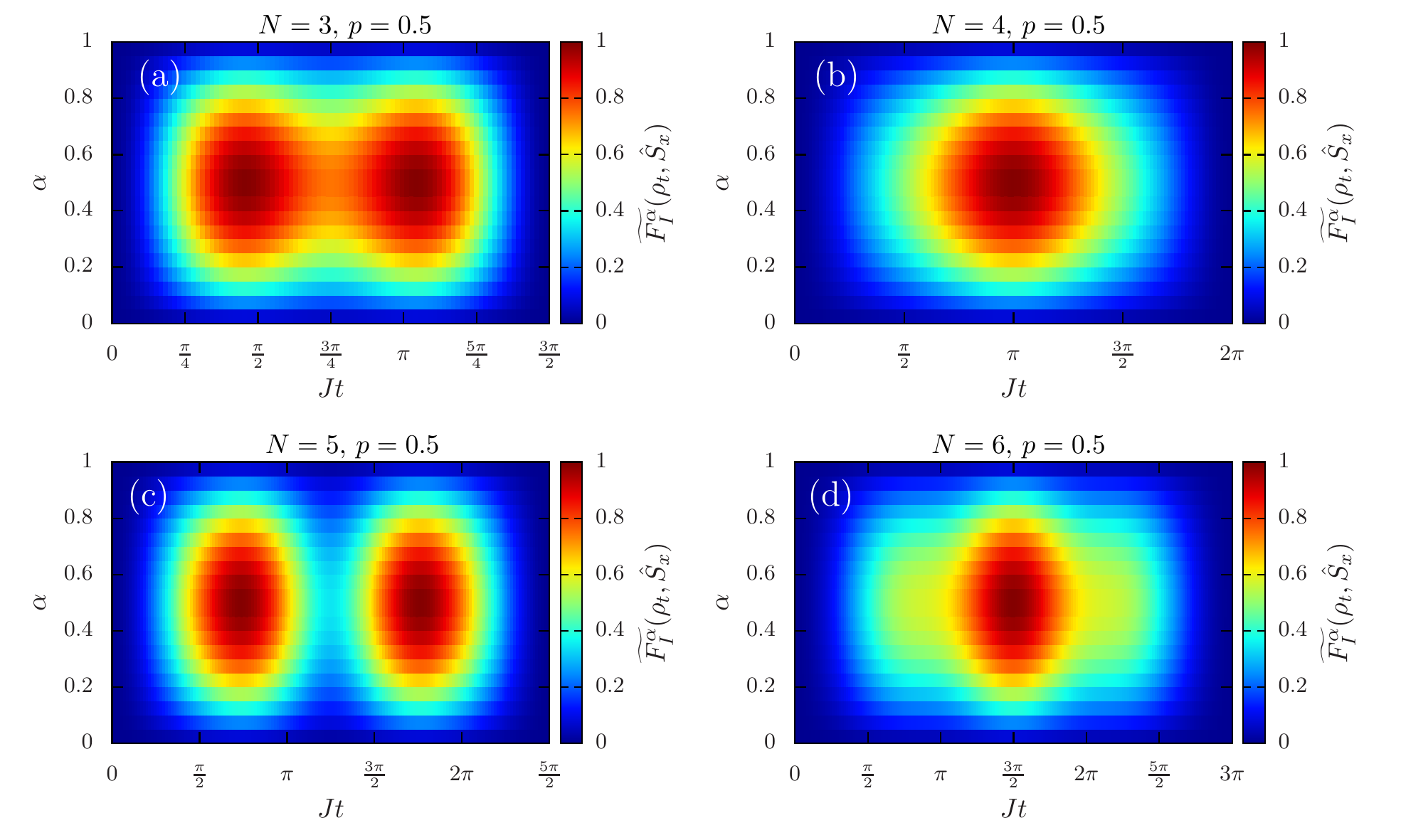}
\caption{(Color online) Density plot of normalized second moment of $\alpha$-MQI, i.e., $\widetilde{F_I^{\alpha}}({\rho_t},{\hat{S}_x})$, related to the generator ${{S}_x} = ({1}/{2}) \,{\sum_{l=1}^{N}} ~{\mathbb{I}^{\, \otimes{l - 1}}}\otimes{\sigma_l^x}\otimes{\mathbb{I}^{\, \otimes{N - l}}}$ and the evolved state ${\rho_t} = {e^{- i t {H_{zz}}}} \, {\rho}\, {e^{ i t {H_{zz}}}}$, where ${H_{zz}} = ({J}/{N})\,{\sum_{j < l}}\,{\sigma_j^z}{\sigma_l^z}$ is the fully connected Ising Hamiltonian. Here we choose the initial state of the system as ${\rho} = \left({(1 - p)}/d\right)\mathbb{I} + p\,{\left(|+\rangle\langle{+}|\right)^{\otimes N}}$, with $d = {2^N}$ and $|{+}\rangle = ({1}/{\sqrt{2}} \, )\left(|0\rangle + |1\rangle \right)$. For simplicity, here we fix $p = 0.5$ and increase the size of the system as (a) $N = 3$, (b) $N = 4$, (c) $N = 5$, and (d) $N = 6$.}
\label{fig:pictureMQCalphaRenyi0000000x222222}
\end{center}
\end{figure*}

Finally, in Fig.~\ref{fig:pictureMQCalphaRenyi0000000x222222} we plot the normalized second moment of $\alpha$-MQI spectrum, $\widetilde{F_I^{\alpha}}({\rho_t},{\hat{S}_x})$, as a function of $t$ and $\alpha$, by varying the size of the system as (a) $N = 3$, (b) $N = 4$, (c) $N = 5$, and (d) $N = 6$. 
As can be seen, time evolution of $\widetilde{F_I^{\alpha}}({\rho_t},{\hat{S}_x})$ oscillates with period $\pi N/2$.
%


\section{Conclusions}
\label{sec:conclusions000xxx001}

In conclusion, we have shown that, by considering a quantum system undergoing a unitary phase encoding process, the R\'{e}nyi relative entropy ($\alpha$-RRE) is linked to the well known Wigner-Yanase-Dyson skew information (WYDSI). We further provided a novel framework addressing the coherence orders of a quantum state with respect to the eigenbasis of an observable $\hat{A}$. We introduced the $\alpha$-Multiple Quantum Intensity ($\alpha$-MQI), ${I^{\alpha}_m}(\rho)$, which is intimately linked to $\alpha$-RRE, and thus proved that WYDSI can be also written as the second moment of multiple quantum coherence spectrum ($\alpha$-MQC), ${F_I^{\alpha}}(\rho,\hat{A})$. 

The second main result concerns the derivation of a family of lower and upper bounds to the second moment of $\alpha$-MQI. Interestingly, we have shown that ${F_I^{\alpha}}(\rho,\hat{A})$ provides a lower bound on the quantum Fisher information. Noteworthy, bridging $\alpha$-MQC and QFI has a number of implications. On one hand, this link unveil the role of the second moment of $\alpha$-MQI
in quantum phase estimation and metrology.
On the other hand, it demonstrates that the second moment of $\alpha$-MQI can also witnesses multiparticle entanglement. 

Finally, we illustrate our main results by inves\-ti\-ga\-ting the single qubit state, Bell-diagonal states, and some paradigmatic multiparticle states. We numerically stu\-died the time evolution of $\alpha$-MQC spectrum and the overall signal of relative purity, by simulating the time reversal dynamics of a many-body all-to-all Ising Hamiltonian. Interestingly, dynamical behaviour of $\alpha$-MQC unveils information about buildup of many-body correlations, and also signals the recently claimed property of quantum information scrambling~\cite{PhysRevLett.120.040402,101038nphys4119v01}. Our results might also find applications in the field of quantum thermodynamics, regarding the family of second laws of thermodynamics parametrized by $\alpha$-RRE which was a\-ddressed in Refs.~\cite{Brandao3275,PhysRevLett.115.210403}.


\begin{acknowledgments}
We thank T. Roscilde for fruitful discussions. D. P. P. and T. M. acknowledges the financial support from the Brazilian ministries MEC and MCTIC, funding agencies CAPES and CNPq. T. M. acknowledges CNPq for support through Bolsa de produtividade em Pesquisa n.311079/2015-6. This work was supported by the Serrapilheira Institute (grant number Serra-1812-27802), CAPES-NUFFIC project number 88887.156521/2017-00. This study was financed by the Coordena\c{c}\~{a}o de Aperfei\c{c}oamento de Pessoal de N\'{i}vel Superior -- Brasil (CAPES) -- Finance Code 001. This work was also supported by the QuantERA ERA-NET Cofund in Quantum Technologies projects CEBBEC.
\end{acknowledgments}

\setcounter{equation}{0}
\setcounter{figure}{0}
\setcounter{table}{0}
\setcounter{section}{0}
\numberwithin{equation}{section}
\makeatletter
\renewcommand{\thesection}{\Alph{section}} 
\renewcommand{\thesubsection}{\thesection.\arabic{subsection}}
\renewcommand{\theequation}{\Alph{section}\arabic{equation}}
\renewcommand{\thefigure}{\arabic{figure}}
\renewcommand{\bibnumfmt}[1]{[#1]}
\renewcommand{\citenumfont}[1]{#1}

\section*{Appendix}


\section{Properties of $\alpha$-MQC}
\label{sec:appendix0001a}

In this Appendix we prove Eqs.~\eqref{eq:renyiMQC0020aaaa1aa},~\eqref{eq:renyiMQC0020} and~\eqref{eq:renyiMQC0027} of the main text. 
First, starting from Eq.~\eqref{eq:renyiMQC005}, it is possible to conclude that
\begin{align}
{({\rho_m^{(\alpha)}})^{\dagger}} &= {\sum_{{\lambda_j} - {\lambda_{\ell}} = m}}\, {\langle{j}|{\rho^{(\alpha)}}|{\ell}\rangle^*} |{\ell}\rangle\langle{j}| \nonumber\\
&= {\sum_{{\lambda_j} - {\lambda_{\ell}} = m}}\, {\langle{\ell}|\, {({\rho^{(\alpha)}})^{\dagger}}|{j}\rangle} |{\ell}\rangle\langle{j}| \nonumber\\
&=  {\sum_{{\lambda_{\ell}} - {\lambda_{j}} = - m}}\, {\langle{\ell}|{\rho^{(\alpha)}}|{j}\rangle} |{\ell}\rangle\langle{j}| \nonumber\\
&= {\rho_{-m}^{(\alpha)}}  ~.
\end{align}
From second to the third line we have used that ${\rho^{(\alpha)}}$ is Hermitian, and from the third to the fourth line we have changed the summation labels. 

Now, we show that ${\rho_m^{(\alpha)}}$ and ${\rho_n^{(\beta)}}$ satisfies an orthogonality constraint with respect to the Hilbert-Schmidt inner product. In order to verify explicitly Eq.~\eqref{eq:renyiMQC0020}, one may proceed as
\begin{align}
\label{eq:renyiMQC0021app}
&{\langle {\rho_m^{(\alpha)}} {\rho_n^{(\beta)}} \rangle_{\text{HS}}} = \nonumber\\
&= {\sum_{{\lambda_j} - {\lambda_{\ell}} = m}}\,\,{\sum_{{\lambda_p} - {\lambda_q} = n}}\, {\langle{j}|{\rho^{(\alpha)}}|{\ell}\rangle^*} {\langle{p}|{\rho^{(\beta)}}|{q}\rangle} \langle{q}|{\ell}\rangle\langle{j}|p\rangle \nonumber\\
&= {\sum_{{\lambda_j} - {\lambda_{\ell}} = m}}\,\,{\sum_{{\lambda_p} - {\lambda_q} = n}}\, {\delta_{q,\ell}}{\delta_{j,p}} \, {\langle{j}|{\rho^{(\alpha)}}|{\ell}\rangle^*} {\langle{p}|{\rho^{(\beta)}}|{q}\rangle}  \nonumber\\
&= {\sum_{{\lambda_j} - {\lambda_{\ell}} = m}}\,\,{\sum_{{\lambda_j} - {\lambda_l} = n}}\, {\langle{j}|{\rho^{(\alpha)}}|{\ell}\rangle^*} {\langle{j}|{\rho^{(\beta)}}|{\ell}\rangle} ~.
\end{align}
where ${\langle{A,B}\rangle_{\text{HS}}} : = \text{Tr}({A^{\dagger}}B)$, for $A,B\in \mathcal{B}(\mathcal{H})$, denotes the Hilbert-Schmidt inner product. Going into details, from the first to the second line we have applied the cyclic permutation under the trace, and from the second to the third line we used $\langle{r}|s\rangle = {\delta_{r,s}}$. From Eq.~\eqref{eq:renyiMQC0021app}, one may conclude that the double summation is nonzero, only for $m = n$. Indeed, given two fixed integers $m$ and $n$, such selection rule comes from the fact that both constraints ${\lambda_j} - {\lambda_l} = m$ and ${\lambda_j} - {\lambda_l} = n$ are simultaneously fulfilled if, and only if, $m = n$. 
Therefore, we readily obtain
\begin{align}
\label{eq:renyiMQC0022app}
{\langle {\rho_m^{(\alpha)}} {\rho_n^{(\beta)}} \rangle_{\text{HS}}} &= {\delta_{m,n}}\, {\sum_{{\lambda_j} - {\lambda_{\ell}} = m}}\, {\langle{j}|{\rho^{(\alpha)}}|{\ell}\rangle^*} {\langle{j}|{\rho^{(\beta)}}|{\ell}\rangle} \nonumber\\
&= {\delta_{m,n}}\,  {\langle {\rho_m^{(\alpha)}} {\rho_m^{(\beta)}} \rangle_{\text{HS}}} ~.
\end{align}

Finally, we will conclude by proving Eq.~\eqref{eq:renyiMQC0027}. 
Suppose now that the density matrix ${\rho^{(\alpha)}}$ undergoes the translationally-covariant evolution 
${\mathcal{U}_{\phi}}(\bullet) := {e^{-i\phi\hat{A}}} \bullet {e^{i\phi\hat{A}}}$ generated by the observable 
$\hat{A}$. 
Hence, starting from Eq.~\eqref{eq:renyiMQC004} in the main text, one gets
\begin{equation}
\label{eq:renyiMQC0025app}
{\mathcal{U}_{\phi}} ({\rho^{(\alpha)}})  = {\sum_m}\, {\mathcal{U}_{\phi}} ({\rho_m^{(\alpha)}})  ~.
\end{equation}
By using Eq.~\eqref{eq:renyiMQC005} it is possible to write 
\begin{align}
\label{eq:renyiMQC0026app}
{\mathcal{U}_{\phi}} ({\rho_m^{(\alpha)}})  &= {\sum_{{\lambda_j} - {\lambda_{\ell}} = m}}\, \langle{j}|{\rho^{(\alpha)}}|{\ell}\rangle \, {e^{-i\phi\hat{A}}}|{j}\rangle\langle{\ell}|{e^{i\phi\hat{A}}} \nonumber\\ 
&= {e^{- i m \phi}}\, {\sum_{{\lambda_j} - {\lambda_{\ell}} = m}}\, \langle{j}|{\rho^{(\alpha)}}|{\ell}\rangle |{j}\rangle\langle{\ell}| \nonumber\\
&= {e^{- i m \phi}} \, {\rho_m^{(\alpha)}} ~,
\end{align}
where $m = {\lambda_j} - {\lambda_{\ell}}$. We stress that from the second to the third line, we used that $\hat{A}|\ell\rangle = {\lambda_{\ell}}|\ell\rangle$ since $|\ell\rangle$ is an eigenstate of the operator $\hat{A}$. Therefore, by substituting Eq.~\eqref{eq:renyiMQC0026app} into~\eqref{eq:renyiMQC0025app} we finally obtain the result
\begin{equation}
\label{eq:renyiMQC0027app}
{\mathcal{U}_{\phi}} ({\rho^{(\alpha)}}) = {\sum_m}\,{e^{- i m \phi}} \, {\rho_m^{(\alpha)}}  ~.
\end{equation}


\section{Limiting case of relative R\'{e}nyi entropy for $\alpha \rightarrow 1$}
\label{sec:criticalcase000xxx000111}

In this Appendix we investigate the behaviour of Eq.~\eqref{eq:renyiMQC0059a} when taking the limit $\alpha \rightarrow 1$. Given the states $\rho$ and ${\rho_{\phi}} = {e^{i\phi\hat{A}}}\,\rho {e^{-i\phi\hat{A}}}$, the Taylor expansion of $\alpha$-relative R\'{e}nyi entropy up to second order in $\phi$, around $\phi = 0$, becomes
\begin{equation}
\label{eq:renyiMQC0059apppendix}
{\text{D}_{\alpha}}(\rho\|{\rho_{\phi}}) \approx - \frac{\phi^2}{\alpha - 1}\, {\mathcal{I}_{\alpha}}(\rho,\hat{A}) + {O}({\phi^3}) ~,
\end{equation}
where ${\mathcal{I}_{\alpha}}(\rho,\hat{A})$ stands for the Wigner-Yanase-Dyson skew information (WYDSI) and, according Eq.~\eqref{eq:renyiMQC0055}, is also written as
\begin{equation}
\label{eq:renyiMQC0055c}
{\mathcal{I}_{\alpha}}(\rho,\hat{A}) = \text{Tr}\left(\rho{\hat{A}^2}\right) - \text{Tr}\left({\rho^{\alpha}}\hat{A}{\rho^{1 - \alpha}}{\hat{A}}\right) ~.
\end{equation}
In particular, note that WYDSI vanishes for $\alpha = 1$. In this case, for $\alpha \rightarrow 1$ the right-hand side of Eq.~\eqref{eq:renyiMQC0059apppendix} will exhibit an indeterminacy form as $\frac{0}{0}$. Notably, one may formally circumvent this issue by applying l'H\^{o}pital rule, which implies the prior differentiation of both numerator and denominator with respect to $\alpha$, and finally take the limit $\alpha \rightarrow 1$. Therefore, one gets
\begin{equation}
\label{eq:renyiMQC0059appppendix000xxx000yyy001ccccc}
{ \lim_{\alpha \rightarrow 1}} \, {\text{D}_{\alpha}}(\rho\|{\rho_{\phi}}) \approx - {\phi^2} \, { \lim_{\alpha \rightarrow 1}} \,  \frac{\frac{d}{d\alpha}{\mathcal{I}_{\alpha}}(\rho,\hat{A}) }{\frac{d}{d\alpha}(\alpha - 1)} + {O}({\phi^3}) ~.
\end{equation}
The denominator in the right-hand side of Eq.~\eqref{eq:renyiMQC0059appppendix000xxx000yyy001ccccc} is well behaved and approaches $1$ as $\alpha \rightarrow 1$. Moving to the numerator, to determine explicitly the derivative of WYDSI with respect to $\alpha$, we shall begin by simplifying the quantity ${\mathcal{I}_{\alpha}}(\rho,\hat{A})$. Let $\rho = {\sum_{\ell}}\, {p_{\ell}}|{\psi_{\ell}}\rangle\langle{\psi_{\ell}}|$ be the spectral decomposition of the density matrix into the basis ${\{|{\psi_{\ell}}\rangle \}_{\ell = 1,\ldots ,d}}$, with $0 \leq {p_{\ell}} \leq 1$, $\text{Tr}(\rho) = {\sum_{\ell}}\,{p_{\ell}} = 1$, and $\langle{\psi_j}|{\psi_{\ell}}\rangle = {\delta_{j,\ell}}$ for all $j,\ell$. In this case, it is straightforward to verify that 
\begin{equation}
\label{eq:renyiMQC0055c1}
\text{Tr}\left(\rho{\hat{A}^2}\right) 
= {\sum_{j,\ell}}\, {p_j} {|\langle{\psi_j}|\hat{A}|{\psi_{\ell}}\rangle|^2} ~,
\end{equation}
and 
\begin{equation}
\label{eq:renyiMQC0055c2}
\text{Tr}\left({\rho^{\alpha}}\hat{A}{\rho^{1 - \alpha}}{\hat{A}}\right) = {\sum_{j,\ell}} \, {p_j^{\alpha}} {p_{\ell}^{1 - \alpha}} {|\langle{\psi_j}|\hat{A}|{\psi_{\ell}}\rangle|^2} ~.
\end{equation}
By substituting Eqs.~\eqref{eq:renyiMQC0055c1} and~\eqref{eq:renyiMQC0055c2} into Eq.~\eqref{eq:renyiMQC0055c}, and also using that ${p_j} - {p_j^{\alpha}} {p_{\ell}^{1 - \alpha}} = {p_j^{\alpha}}({p_{j}^{1 - \alpha}} - {p_{\ell}^{1 - \alpha}})$, one obtains
\begin{equation}
\label{eq:renyiMQC0055c6xxxx00001}
{\mathcal{I}_{\alpha}}(\rho,\hat{A}) = {\sum_{j,\ell}}  \, {p_j^{\alpha}} \left( {p_{j}^{1 - \alpha}} - {p_{\ell}^{1 - \alpha}} \right) {|\langle{\psi_j}|\hat{A}|{\psi_{\ell}}\rangle|^2} ~.
\end{equation}
To di\-ffe\-ren\-tiate WYDSI with respect to $\alpha$, 
we will take advantage from the algebraic identity 
$d\, {p_j^{\alpha}} / d\alpha = {p_j^{\alpha}} \ln{p_j}$. 
Hence, by combining this result with the derivative of 
Eq.~\eqref{eq:renyiMQC0055c6xxxx00001}, it is straightforward to conclude that
\begin{align}
\label{eq:renyiMQC0055c6xxxx00001xxxx00001}
{ \lim_{\alpha \rightarrow 1}} \, \frac{d}{d\alpha}{\mathcal{I}_{\alpha}}(\rho,\hat{A}) &= {\sum_{j,\ell}}  \, {p_j} ( \ln{p_{\ell}} - \ln{p_j})\,  {|\langle{\psi_j}|\hat{A}|{\psi_{\ell}}\rangle|^2} \nonumber\\ 
&= \text{Tr}(\hat{A}\rho\hat{A}\ln\rho) - \text{Tr}({\hat{A}^2}\rho\ln\rho) ~.
\end{align}
Finally, from Eq.~\eqref{eq:renyiMQC0055c6xxxx00001xxxx00001} 
one may readily simplify Eq.~\eqref{eq:renyiMQC0059appppendix000xxx000yyy001ccccc} 
and obtain the limiting case $\alpha \rightarrow 1$ of Taylor expansion of 
relative R\'{e}nyi entropy as follows
\begin{equation}
\label{eq:renyiMQC0059appppendix00002}
{ \lim_{\alpha \rightarrow 1}} \, {\text{D}_{\alpha}}(\rho\|{\rho_{\phi}}) 
\approx {\phi^2}\left( \text{Tr}({\hat{A}^2}\rho\ln\rho) - \text{Tr}(\hat{A}\rho\hat{A}\ln\rho)  \right) 
+ {O}({\phi^3}) ~.
\end{equation}


\section{Lower bound for WYDSI}
\label{sec:boundsWYDSIxxx000xxx111}

In this Appendix we will investigate some bounds on Wigner-Yanase-Dyson 
skew information (WYDSI). To begin, we notice that, from 
Eq.~\eqref{eq:renyiMQC0055c6xxxx00001}, WYDSI also read as 
\begin{equation}
\label{eq:renyiMQC0055c6}
{\mathcal{I}_{\alpha}}(\rho,\hat{A}) = {\sum_{j < \ell}}  \left( {p_j^{\alpha}} - {p_{\ell}^{\alpha}}\right) \left( {p_{j}^{1 - \alpha}} - {p_{\ell}^{1 - \alpha}} \right) {|\langle{\psi_j}|\hat{A}|{\psi_{\ell}}\rangle|^2} ~,
\end{equation}
which comes from the fact that, since $\hat{A}$ is a Hermitian operator, thus the amplitude ${|\langle{j}|\hat{A}|{\ell}\rangle|^2}$ remains invariant under changing labels $j \longrightarrow \ell$. In particular, for $\alpha = 1/2$ Eq.~\eqref{eq:renyiMQC0055c6} becomes 
\begin{equation}
\label{eq:renyiMQC0055c6a0001}
{\mathcal{I}_{1/2}}(\rho,\hat{A}) = {\sum_{j < \ell}}  {\left( \sqrt{p_j} - \sqrt{p_{\ell}} \, \right)^2} {|\langle{\psi_j}|\hat{A}|{\psi_{\ell}}\rangle|^2} ~.
\end{equation}
Now, we address the quantifier ${\mathcal{I}^L}(\rho,\hat{A})$, which can be written as 
\begin{equation}
\label{eq:renyiMQC00550d02d32appppp}
{\mathcal{I}^L}(\rho,\hat{A}) = \frac{1}{2}\left(\text{Tr}({\rho^2}{\hat{A}^2}) - \text{Tr}(\rho\hat{A}\rho\hat{A})\right) ~.
\end{equation}
In turn, notice that
\begin{equation}
\label{eq:renyiMQC0055c1a1}
\text{Tr}\left({\rho^2}{\hat{A}^2}\right) 
= {\sum_{j,\ell}}\, {p_j^2} \, {|\langle{\psi_j}|\hat{A}|{\psi_{\ell}}\rangle|^2} ~,
\end{equation}
and 
\begin{equation}
\label{eq:renyiMQC0055c2a2}
\text{Tr}\left({\rho}\hat{A}{\rho}{\hat{A}}\right) = {\sum_{j,\ell}} \, {p_j} {p_{\ell}} \, {|\langle{\psi_j}|\hat{A}|{\psi_{\ell}}\rangle|^2} ~.
\end{equation}
Thus, by substituting Eqs.~\eqref{eq:renyiMQC0055c1a1} and~\eqref{eq:renyiMQC0055c2a2} into Eq.~\eqref{eq:renyiMQC00550d02d32appppp}, we obtain
\begin{equation}
\label{eq:renyiMQC0055c2a3}
{\mathcal{I}^L}(\rho,\hat{A}) = \frac{1}{2} \, {\sum_{j,\ell}} \, {p_j} ({p_j} - {p_{\ell}} ) {|\langle{\psi_j}|\hat{A}|{\psi_{\ell}}\rangle|^2} ~.
\end{equation}
Once more, as the amplitude ${|\langle{j}|\hat{A}|{\ell}\rangle|^2}$ 
is invariant under changing labels $j \rightarrow \ell$, one gets
\begin{equation}
\label{eq:renyiMQC0055c2a4}
{\mathcal{I}^L}(\rho,\hat{A}) = \frac{1}{2} \, {\sum_{j < \ell}} \, {({p_j} - {p_{\ell}})^2} {|\langle{\psi_j}|\hat{A}|{\psi_{\ell}}\rangle|^2} ~.
\end{equation}

Some remarks are now in order. Yanagi~\cite{Yanagi_2010} (see Lemma $3.3$) 
has proved that for any $x  > 0$ and $0 \leq \alpha \leq 1$, the following inequality holds
\begin{equation}
\label{eq:renyiMQC0055c2a4bound000000x0a}
{(1 - 2\alpha)^2}{(x - 1)^2} - {({x^{\alpha}} - {x^{1 - \alpha}})^2} \geq 0 ~.
\end{equation}
Interestingly, we stress that Eq.~\eqref{eq:renyiMQC0055c2a4bound000000x0a} 
can be also written as 
\begin{equation}
\label{eq:renyiMQC0055c2a4bound000000x0c}
 4\, \alpha(1 - \alpha){(1 - x)^2} \leq (1 - {x^{\alpha}})(1 - {x^{1 - \alpha}})\,{\kappa_{\alpha}}(x) ~,
\end{equation}
where we define
\begin{equation}
\label{eq:renyiMQC0055c2a4bound000000x0cxxxxxx0000001111}
{\kappa_{\alpha}}(x) := 1 + x + {x^{\alpha}} + {x^{1 - \alpha}}
\end{equation}
From now on, we will focus mainly on 
Eq.~\eqref{eq:renyiMQC0055c2a4bound000000x0cxxxxxx0000001111} 
in the search of a new class of bounds to WYDSI. According to Heinz 
inequality~\cite{BHATIA2006355,Audenaert2007}, for $a > 0$, $b > 0$ 
and $0 < \alpha < 1$, the following inequality holds
\begin{equation}
\label{eq:renyiMQC0055c2a4bound000000x0e}
{a^{\alpha}}{b^{1- \alpha}} + {a^{1 - \alpha}}{b^{ \alpha}} \leq a + b ~.
\end{equation}
In special, by choosing $x = a/b$, with $x > 0$, 
Eq.~\eqref{eq:renyiMQC0055c2a4bound000000x0e} becomes
\begin{equation}
\label{eq:renyiMQC0055c2a4bound000000x0f}
{x^{\alpha}} + {x^{1 - \alpha}} \leq 1 + x ~.
\end{equation}
Hence, Eq.~\eqref{eq:renyiMQC0055c2a4bound000000x0f} allows us to conclude the bound
\begin{equation}
\label{eq:renyiMQC0055c2a4bound000000x0g}
{\kappa_{\alpha}}(x) \leq 2\, (1 + x) ~.
\end{equation}
By substituting Eq.~\eqref{eq:renyiMQC0055c2a4bound000000x0g} into Eq.~\eqref{eq:renyiMQC0055c2a4bound000000x0c}, it yields the new bound
\begin{equation}
\label{eq:renyiMQC0055c2a4bound000000x0i}
 2\, \alpha(1 - \alpha){(1 - x)^2} \leq (1 + x)(1 - {x^{\alpha}})(1 - {x^{1 - \alpha}}) ~.
\end{equation}
We would like to stress that bound in Eq.~\eqref{eq:renyiMQC0055c2a4bound000000x0i} 
applies to any $x > 0$ and $0 \leq \alpha \leq 1$.

Starting from Eq.~\eqref{eq:renyiMQC0055c2a4bound000000x0i}, 
let us choose $x = {p_j}/{p_{\ell}}$, with $x > 0$, and $0 < {p_j} \leq 1$ 
and $0 < {p_{\ell}} \leq 1$. In this case, it is straightforward to write down the inequality 
\begin{equation}
\label{eq:renyiMQC0055c2a4bound000000x0j}
2\,\alpha(1 - \alpha){({p_j} - {p_{\ell}})^2} \leq ({p_j} + {p_{\ell}})({p_j^{\alpha}} - 
{p_{\ell}^{\alpha}})({p_j^{1 - \alpha}} - {p_{\ell}^{1 - \alpha}}) ~.
\end{equation}
Hence, by substituting Eq.~\eqref{eq:renyiMQC0055c2a4bound000000x0j} 
into Eq.~\eqref{eq:renyiMQC0055c2a4}, one may conclude that 
\begin{align}
\label{eq:renyiMQC0055c2a4bound000000x0k}
&2\, \alpha(1 - \alpha)\, {\mathcal{I}^L}(\rho,\hat{A}) =  {\sum_{j < \ell}} 
\,\alpha(1 - \alpha){({p_j} - {p_{\ell}})^2} {|\langle{\psi_j}|\hat{A}|{\psi_{\ell}}\rangle|^2} \nonumber\\
&\leq \frac{1}{2} \, {\sum_{j < \ell}} \, ({p_j} + {p_{\ell}})({p_j^{\alpha}} - 
{p_{\ell}^{\alpha}})({p_j^{1 - \alpha}} - {p_{\ell}^{1 - \alpha}}) {|\langle{\psi_j}|\hat{A}|{\psi_{\ell}}\rangle|^2} ~.
\end{align}
Now we approach a crucial point in our derivation. Going into details, RHS in Eq.~\eqref{eq:renyiMQC0055c2a4bound000000x0k} will exactly recover 
Wigner-Yanase-Dyson skew information in Eq.~\eqref{eq:renyiMQC0055c6} 
if we turn to the fact that ${p_j} + {p_{\ell}} \leq 2$ for all $0 < {p_j} \leq 1$ 
and $0 < {p_{\ell}} \leq 1$. Therefore, applying such a result into 
Eq.~\eqref{eq:renyiMQC0055c2a4bound000000x0k}, we obtain
\begin{align}
\label{eq:renyiMQC0055c2a4bound000000x0l}
&2\, \alpha(1 - \alpha)\, {\mathcal{I}^L}(\rho,\hat{A}) \leq \nonumber\\ 
&{\sum_{j < \ell}} \,({p_j^{\alpha}} - {p_{\ell}^{\alpha}})({p_j^{1 - \alpha}} - {p_{\ell}^{1 - \alpha}}) {|\langle{\psi_j}|\hat{A}|{\psi_{\ell}}\rangle|^2} ~.
\end{align}
Finally, it is straightforward to obtain the lower bound
\begin{equation}
\label{eq:renyiMQC0055c2a4bound000000x0mappp}
{\mathcal{I}_{\alpha}}(\rho,\hat{A}) \geq 2\alpha(1 - \alpha) \, 
{\mathcal{I}^L}(\rho,\hat{A}) ~.
\end{equation}

%

\begin{thebibliography}{98}%
\makeatletter
\providecommand \@ifxundefined [1]{%
 \@ifx{#1\undefined}
}%
\providecommand \@ifnum [1]{%
 \ifnum #1\expandafter \@firstoftwo
 \else \expandafter \@secondoftwo
 \fi
}%
\providecommand \@ifx [1]{%
 \ifx #1\expandafter \@firstoftwo
 \else \expandafter \@secondoftwo
 \fi
}%
\providecommand \natexlab [1]{#1}%
\providecommand \enquote  [1]{``#1''}%
\providecommand \bibnamefont  [1]{#1}%
\providecommand \bibfnamefont [1]{#1}%
\providecommand \citenamefont [1]{#1}%
\providecommand \href@noop [0]{\@secondoftwo}%
\providecommand \href [0]{\begingroup \@sanitize@url \@href}%
\providecommand \@href[1]{\@@startlink{#1}\@@href}%
\providecommand \@@href[1]{\endgroup#1\@@endlink}%
\providecommand \@sanitize@url [0]{\catcode `\\12\catcode `\$12\catcode
  `\&12\catcode `\#12\catcode `\^12\catcode `\_12\catcode `\%12\relax}%
\providecommand \@@startlink[1]{}%
\providecommand \@@endlink[0]{}%
\providecommand \url  [0]{\begingroup\@sanitize@url \@url }%
\providecommand \@url [1]{\endgroup\@href {#1}{\urlprefix }}%
\providecommand \urlprefix  [0]{URL }%
\providecommand \Eprint [0]{\href }%
\providecommand \doibase [0]{http://dx.doi.org/}%
\providecommand \selectlanguage [0]{\@gobble}%
\providecommand \bibinfo  [0]{\@secondoftwo}%
\providecommand \bibfield  [0]{\@secondoftwo}%
\providecommand \translation [1]{[#1]}%
\providecommand \BibitemOpen [0]{}%
\providecommand \bibitemStop [0]{}%
\providecommand \bibitemNoStop [0]{.\EOS\space}%
\providecommand \EOS [0]{\spacefactor3000\relax}%
\providecommand \BibitemShut  [1]{\csname bibitem#1\endcsname}%
\let\auto@bib@innerbib\@empty
\bibitem [{\citenamefont {Jaeger}(2018)}]{LARS_2018}%
  \BibitemOpen
  \bibfield  {author} {\bibinfo {author} {\bibfnamefont {L.}~\bibnamefont
  {Jaeger}},\ }\href {\doibase 10.1007/978-3-319-98824-5} {\emph {\bibinfo
  {title} {The {S}econd {Q}uantum {R}evolution}}}\ (\bibinfo  {publisher}
  {Springer},\ \bibinfo {address} {New York},\ \bibinfo {year}
  {2018})\BibitemShut {NoStop}%
\bibitem [{\citenamefont {\AA{}berg}(2014)}]{PhysRevLett.113.150402}%
  \BibitemOpen
  \bibfield  {author} {\bibinfo {author} {\bibfnamefont {J.}~\bibnamefont
  {\AA{}berg}},\ }\bibfield  {title} {\enquote {\bibinfo {title} {Catalytic
  coherence},}\ }\href {\doibase 10.1103/PhysRevLett.113.150402} {\bibfield
  {journal} {\bibinfo  {journal} {Phys. Rev. Lett.}\ }\textbf {\bibinfo
  {volume} {113}},\ \bibinfo {pages} {150402} (\bibinfo {year}
  {2014})}\BibitemShut {NoStop}%
\bibitem [{\citenamefont {\ifmmode \acute{C}\else
  \'{C}\fi{}wikli\ifmmode~\acute{n}\else \'{n}\fi{}ski}\ \emph
  {et~al.}(2015)\citenamefont {\ifmmode \acute{C}\else
  \'{C}\fi{}wikli\ifmmode~\acute{n}\else \'{n}\fi{}ski}, \citenamefont
  {Studzi\ifmmode~\acute{n}\else \'{n}\fi{}ski}, \citenamefont {Horodecki},\
  and\ \citenamefont {Oppenheim}}]{PhysRevLett.115.210403}%
  \BibitemOpen
  \bibfield  {author} {\bibinfo {author} {\bibfnamefont {P.}~\bibnamefont
  {\ifmmode \acute{C}\else \'{C}\fi{}wikli\ifmmode~\acute{n}\else
  \'{n}\fi{}ski}}, \bibinfo {author} {\bibfnamefont {M.}~\bibnamefont
  {Studzi\ifmmode~\acute{n}\else \'{n}\fi{}ski}}, \bibinfo {author}
  {\bibfnamefont {M.}~\bibnamefont {Horodecki}}, \ and\ \bibinfo {author}
  {\bibfnamefont {J.}~\bibnamefont {Oppenheim}},\ }\bibfield  {title} {\enquote
  {\bibinfo {title} {{L}imitations on the {E}volution of {Q}uantum
  {C}oherences: {T}owards {F}ully {Q}uantum {S}econd {L}aws of
  {T}hermodynamics},}\ }\href {\doibase 10.1103/PhysRevLett.115.210403}
  {\bibfield  {journal} {\bibinfo  {journal} {Phys. Rev. Lett.}\ }\textbf
  {\bibinfo {volume} {115}},\ \bibinfo {pages} {210403} (\bibinfo {year}
  {2015})}\BibitemShut {NoStop}%
\bibitem [{\citenamefont {Karpat}\ \emph {et~al.}(2014)\citenamefont {Karpat},
  \citenamefont {\ifmmode~\mbox{\c{C}}\else \c{C}\fi{}akmak},\ and\
  \citenamefont {Fanchini}}]{PhysRevB.90.104431}%
  \BibitemOpen
  \bibfield  {author} {\bibinfo {author} {\bibfnamefont {G.}~\bibnamefont
  {Karpat}}, \bibinfo {author} {\bibfnamefont {B.}~\bibnamefont
  {\ifmmode~\mbox{\c{C}}\else \c{C}\fi{}akmak}}, \ and\ \bibinfo {author}
  {\bibfnamefont {F.~F.}\ \bibnamefont {Fanchini}},\ }\bibfield  {title}
  {\enquote {\bibinfo {title} {Quantum coherence and uncertainty in the
  anisotropic {X}{Y} chain},}\ }\href {\doibase 10.1103/PhysRevB.90.104431}
  {\bibfield  {journal} {\bibinfo  {journal} {Phys. Rev. B}\ }\textbf {\bibinfo
  {volume} {90}},\ \bibinfo {pages} {104431} (\bibinfo {year}
  {2014})}\BibitemShut {NoStop}%
\bibitem [{\citenamefont {Malvezzi}\ \emph {et~al.}(2016)\citenamefont
  {Malvezzi}, \citenamefont {Karpat}, \citenamefont {\ifmmode~\mbox{\c{C}}\else
  \c{C}\fi{}akmak}, \citenamefont {Fanchini}, \citenamefont {Debarba},\ and\
  \citenamefont {Vianna}}]{PhysRevB.93.184428}%
  \BibitemOpen
  \bibfield  {author} {\bibinfo {author} {\bibfnamefont {A.~L.}\ \bibnamefont
  {Malvezzi}}, \bibinfo {author} {\bibfnamefont {G.}~\bibnamefont {Karpat}},
  \bibinfo {author} {\bibfnamefont {B.}~\bibnamefont
  {\ifmmode~\mbox{\c{C}}\else \c{C}\fi{}akmak}}, \bibinfo {author}
  {\bibfnamefont {F.~F.}\ \bibnamefont {Fanchini}}, \bibinfo {author}
  {\bibfnamefont {T.}~\bibnamefont {Debarba}}, \ and\ \bibinfo {author}
  {\bibfnamefont {R.~O.}\ \bibnamefont {Vianna}},\ }\bibfield  {title}
  {\enquote {\bibinfo {title} {Quantum correlations and coherence in spin-$1$
  {H}eisenberg chains},}\ }\href {\doibase 10.1103/PhysRevB.93.184428}
  {\bibfield  {journal} {\bibinfo  {journal} {Phys. Rev. B}\ }\textbf {\bibinfo
  {volume} {93}},\ \bibinfo {pages} {184428} (\bibinfo {year}
  {2016})}\BibitemShut {NoStop}%
\bibitem [{\citenamefont {Huelga}\ and\ \citenamefont
  {Plenio}(2014)}]{2014_Nature_10_621}%
  \BibitemOpen
  \bibfield  {author} {\bibinfo {author} {\bibfnamefont {S.~F.}\ \bibnamefont
  {Huelga}}\ and\ \bibinfo {author} {\bibfnamefont {M.~B.}\ \bibnamefont
  {Plenio}},\ }\bibfield  {title} {\enquote {\bibinfo {title} {Quantum biology:
  {A} vibrant environment},}\ }\href {\doibase 10.1038/nphys3047} {\bibfield
  {journal} {\bibinfo  {journal} {Nat. Phys.}\ }\textbf {\bibinfo {volume}
  {10}},\ \bibinfo {pages} {621} (\bibinfo {year} {2014})}\BibitemShut
  {NoStop}%
\bibitem [{\citenamefont {Roden}\ \emph {et~al.}(2016)\citenamefont {Roden},
  \citenamefont {Bennett},\ and\ \citenamefont {Whaley}}]{10.1063_1.4953243}%
  \BibitemOpen
  \bibfield  {author} {\bibinfo {author} {\bibfnamefont {J.~J.~J.}\
  \bibnamefont {Roden}}, \bibinfo {author} {\bibfnamefont {D.~I.~G.}\
  \bibnamefont {Bennett}}, \ and\ \bibinfo {author} {\bibfnamefont {K.~B.}\
  \bibnamefont {Whaley}},\ }\bibfield  {title} {\enquote {\bibinfo {title}
  {Long-range energy transport in photosystem {I}{I}},}\ }\href {\doibase
  10.1063/1.4953243} {\bibfield  {journal} {\bibinfo  {journal} {J. Chem.
  Phys.}\ }\textbf {\bibinfo {volume} {144}},\ \bibinfo {pages} {245101}
  (\bibinfo {year} {2016})}\BibitemShut {NoStop}%
\bibitem [{\citenamefont {Streltsov}\ \emph {et~al.}(2017)\citenamefont
  {Streltsov}, \citenamefont {Adesso},\ and\ \citenamefont
  {Plenio}}]{RevModPhys.89.041003}%
  \BibitemOpen
  \bibfield  {author} {\bibinfo {author} {\bibfnamefont {A.}~\bibnamefont
  {Streltsov}}, \bibinfo {author} {\bibfnamefont {G.}~\bibnamefont {Adesso}}, \
  and\ \bibinfo {author} {\bibfnamefont {M.~B.}\ \bibnamefont {Plenio}},\
  }\bibfield  {title} {\enquote {\bibinfo {title} {Colloquium: {Q}uantum
  coherence as a resource},}\ }\href {\doibase 10.1103/RevModPhys.89.041003}
  {\bibfield  {journal} {\bibinfo  {journal} {Rev. Mod. Phys.}\ }\textbf
  {\bibinfo {volume} {89}},\ \bibinfo {pages} {041003} (\bibinfo {year}
  {2017})}\BibitemShut {NoStop}%
\bibitem [{\citenamefont {Munowitz}(1988)}]{Munowitz}%
  \BibitemOpen
  \bibfield  {author} {\bibinfo {author} {\bibfnamefont {M.}~\bibnamefont
  {Munowitz}},\ }\href@noop {} {\emph {\bibinfo {title} {Coherence and
  {N}{M}{R}}}}\ (\bibinfo  {publisher} {John Wiley \& Sons},\ \bibinfo
  {address} {Cambridge},\ \bibinfo {year} {1988})\BibitemShut {NoStop}%
\bibitem [{\citenamefont {Keeler}(2010)}]{Keeler}%
  \BibitemOpen
  \bibfield  {author} {\bibinfo {author} {\bibfnamefont {J.}~\bibnamefont
  {Keeler}},\ }\href@noop {} {\emph {\bibinfo {title} {Understanding {NMR}
  {S}pectroscopy}}},\ \bibinfo {edition} {2nd}\ ed.\ (\bibinfo  {publisher}
  {John Wiley \& Sons},\ \bibinfo {address} {Cambridge},\ \bibinfo {year}
  {2010})\BibitemShut {NoStop}%
\bibitem [{\citenamefont {Baum}\ \emph {et~al.}(1985)\citenamefont {Baum},
  \citenamefont {Munowitz}, \citenamefont {Garroway},\ and\ \citenamefont
  {Pines}}]{10.1063_1.449344}%
  \BibitemOpen
  \bibfield  {author} {\bibinfo {author} {\bibfnamefont {J.}~\bibnamefont
  {Baum}}, \bibinfo {author} {\bibfnamefont {M.}~\bibnamefont {Munowitz}},
  \bibinfo {author} {\bibfnamefont {A.~N.}\ \bibnamefont {Garroway}}, \ and\
  \bibinfo {author} {\bibfnamefont {A.}~\bibnamefont {Pines}},\ }\bibfield
  {title} {\enquote {\bibinfo {title} {Multiple{-}quantum dynamics in solid
  state {N}{M}{R}},}\ }\href {\doibase 10.1063/1.449344} {\bibfield  {journal}
  {\bibinfo  {journal} {J. Chem. Phys.}\ }\textbf {\bibinfo {volume} {83}},\
  \bibinfo {pages} {2015} (\bibinfo {year} {1985})}\BibitemShut {NoStop}%
\bibitem [{\citenamefont {Munowitz}\ and\ \citenamefont
  {Pines}(1986)}]{MUNOWITZ525}%
  \BibitemOpen
  \bibfield  {author} {\bibinfo {author} {\bibfnamefont {M.}~\bibnamefont
  {Munowitz}}\ and\ \bibinfo {author} {\bibfnamefont {A.}~\bibnamefont
  {Pines}},\ }\bibfield  {title} {\enquote {\bibinfo {title}
  {Multiple{-}{Q}uantum {N}uclear {M}agnetic {R}esonance {S}pectroscopy},}\
  }\href {\doibase 10.1126/science.233.4763.525} {\bibfield  {journal}
  {\bibinfo  {journal} {Science}\ }\textbf {\bibinfo {volume} {233}},\ \bibinfo
  {pages} {525} (\bibinfo {year} {1986})}\BibitemShut {NoStop}%
\bibitem [{\citenamefont {Baum}\ and\ \citenamefont
  {Pines}(1986)}]{10.1021_ja00284a001}%
  \BibitemOpen
  \bibfield  {author} {\bibinfo {author} {\bibfnamefont {J.}~\bibnamefont
  {Baum}}\ and\ \bibinfo {author} {\bibfnamefont {A.}~\bibnamefont {Pines}},\
  }\bibfield  {title} {\enquote {\bibinfo {title} {{N}{M}{R} studies of
  clustering in solids},}\ }\href {\doibase 10.1021/ja00284a001} {\bibfield
  {journal} {\bibinfo  {journal} {J. Am. Chem. Soc.}\ }\textbf {\bibinfo
  {volume} {108}},\ \bibinfo {pages} {7447} (\bibinfo {year}
  {1986})}\BibitemShut {NoStop}%
\bibitem [{\citenamefont {Khitrin}(1997)}]{KHITRIN1997217}%
  \BibitemOpen
  \bibfield  {author} {\bibinfo {author} {\bibfnamefont {A.~K.}\ \bibnamefont
  {Khitrin}},\ }\bibfield  {title} {\enquote {\bibinfo {title} {Growth of
  {N}{M}{R} multiple-quantum coherences in quasi-one-dimensional systems},}\
  }\href {\doibase 10.1016/S0009-2614(97)00661-1} {\bibfield  {journal}
  {\bibinfo  {journal} {Chem. Phys. Lett.}\ }\textbf {\bibinfo {volume}
  {274}},\ \bibinfo {pages} {217} (\bibinfo {year} {1997})}\BibitemShut
  {NoStop}%
\bibitem [{\citenamefont {\'Alvarez}\ and\ \citenamefont
  {Suter}(2011)}]{PhysRevA.84.012320}%
  \BibitemOpen
  \bibfield  {author} {\bibinfo {author} {\bibfnamefont {G.~A.}\ \bibnamefont
  {\'Alvarez}}\ and\ \bibinfo {author} {\bibfnamefont {D.}~\bibnamefont
  {Suter}},\ }\bibfield  {title} {\enquote {\bibinfo {title} {Localization
  effects induced by decoherence in superpositions of many-spin quantum
  states},}\ }\href {\doibase 10.1103/PhysRevA.84.012320} {\bibfield  {journal}
  {\bibinfo  {journal} {Phys. Rev. A}\ }\textbf {\bibinfo {volume} {84}},\
  \bibinfo {pages} {012320} (\bibinfo {year} {2011})}\BibitemShut {NoStop}%
\bibitem [{\citenamefont {Furman}\ \emph {et~al.}(2008)\citenamefont {Furman},
  \citenamefont {Meerovich},\ and\ \citenamefont
  {Sokolovsky}}]{PhysRevA.78.042301}%
  \BibitemOpen
  \bibfield  {author} {\bibinfo {author} {\bibfnamefont {G.~B.}\ \bibnamefont
  {Furman}}, \bibinfo {author} {\bibfnamefont {V.~M.}\ \bibnamefont
  {Meerovich}}, \ and\ \bibinfo {author} {\bibfnamefont {V.~L.}\ \bibnamefont
  {Sokolovsky}},\ }\bibfield  {title} {\enquote {\bibinfo {title} {Multiple
  quantum {N}{M}{R} and entanglement dynamics in dipolar coupling spin
  systems},}\ }\href {\doibase 10.1103/PhysRevA.78.042301} {\bibfield
  {journal} {\bibinfo  {journal} {Phys. Rev. A}\ }\textbf {\bibinfo {volume}
  {78}},\ \bibinfo {pages} {042301} (\bibinfo {year} {2008})}\BibitemShut
  {NoStop}%
\bibitem [{\citenamefont {Pires}\ \emph {et~al.}(2018)\citenamefont {Pires},
  \citenamefont {Silva}, \citenamefont {deAzevedo}, \citenamefont
  {Soares-Pinto},\ and\ \citenamefont {Filgueiras}}]{PhysRevA.98.032101}%
  \BibitemOpen
  \bibfield  {author} {\bibinfo {author} {\bibfnamefont {D.~P.}\ \bibnamefont
  {Pires}}, \bibinfo {author} {\bibfnamefont {I.~A.}\ \bibnamefont {Silva}},
  \bibinfo {author} {\bibfnamefont {E.~R.}\ \bibnamefont {deAzevedo}}, \bibinfo
  {author} {\bibfnamefont {D.~O.}\ \bibnamefont {Soares-Pinto}}, \ and\
  \bibinfo {author} {\bibfnamefont {J.~G.}\ \bibnamefont {Filgueiras}},\
  }\bibfield  {title} {\enquote {\bibinfo {title} {Coherence orders,
  decoherence, and quantum metrology},}\ }\href {\doibase
  10.1103/PhysRevA.98.032101} {\bibfield  {journal} {\bibinfo  {journal} {Phys.
  Rev. A}\ }\textbf {\bibinfo {volume} {98}},\ \bibinfo {pages} {032101}
  (\bibinfo {year} {2018})}\BibitemShut {NoStop}%
\bibitem [{\citenamefont {G\"{a}rttner}\ \emph {et~al.}(2018)\citenamefont
  {G\"{a}rttner}, \citenamefont {Hauke},\ and\ \citenamefont
  {Rey}}]{PhysRevLett.120.040402}%
  \BibitemOpen
  \bibfield  {author} {\bibinfo {author} {\bibfnamefont {M.}~\bibnamefont
  {G\"{a}rttner}}, \bibinfo {author} {\bibfnamefont {P.}~\bibnamefont {Hauke}},
  \ and\ \bibinfo {author} {\bibfnamefont {A.~M.}\ \bibnamefont {Rey}},\
  }\bibfield  {title} {\enquote {\bibinfo {title} {Relating
  {O}ut-of-{T}ime-{O}rder {C}orrelations to {E}ntanglement via
  {M}ultiple-{Q}uantum {C}oherences},}\ }\href {\doibase
  10.1103/PhysRevLett.120.040402} {\bibfield  {journal} {\bibinfo  {journal}
  {Phys. Rev. Lett.}\ }\textbf {\bibinfo {volume} {120}},\ \bibinfo {pages}
  {040402} (\bibinfo {year} {2018})}\BibitemShut {NoStop}%
\bibitem [{\citenamefont {G\"{a}rttner}\ \emph {et~al.}(2017)\citenamefont
  {G\"{a}rttner}, \citenamefont {Bohnet}, \citenamefont {Safavi-Naini},
  \citenamefont {Wall}, \citenamefont {Bollinger},\ and\ \citenamefont
  {Rey}}]{101038nphys4119v01}%
  \BibitemOpen
  \bibfield  {author} {\bibinfo {author} {\bibfnamefont {M.}~\bibnamefont
  {G\"{a}rttner}}, \bibinfo {author} {\bibfnamefont {J.}~\bibnamefont
  {Bohnet}}, \bibinfo {author} {\bibfnamefont {A.}~\bibnamefont
  {Safavi-Naini}}, \bibinfo {author} {\bibfnamefont {M.~L.}\ \bibnamefont
  {Wall}}, \bibinfo {author} {\bibfnamefont {J.~J.}\ \bibnamefont {Bollinger}},
  \ and\ \bibinfo {author} {\bibfnamefont {A.~M.}\ \bibnamefont {Rey}},\
  }\bibfield  {title} {\enquote {\bibinfo {title} {Measuring out-of-time-order
  correlations and multiple quantum spectra in a trapped-ion quantum magnet},}\
  }\href {\doibase 10.1038/nphys4119} {\bibfield  {journal} {\bibinfo
  {journal} {Nature Phys.}\ }\textbf {\bibinfo {volume} {13}},\ \bibinfo
  {pages} {781} (\bibinfo {year} {2017})}\BibitemShut {NoStop}%
\bibitem [{\citenamefont {Wei}\ \emph {et~al.}(2018)\citenamefont {Wei},
  \citenamefont {Ramanathan},\ and\ \citenamefont
  {Cappellaro}}]{PhysRevLett.120.070501}%
  \BibitemOpen
  \bibfield  {author} {\bibinfo {author} {\bibfnamefont {K.~X.}\ \bibnamefont
  {Wei}}, \bibinfo {author} {\bibfnamefont {C.}~\bibnamefont {Ramanathan}}, \
  and\ \bibinfo {author} {\bibfnamefont {P.}~\bibnamefont {Cappellaro}},\
  }\bibfield  {title} {\enquote {\bibinfo {title} {Exploring {L}ocalization in
  {N}uclear {S}pin {C}hains},}\ }\href {\doibase
  10.1103/PhysRevLett.120.070501} {\bibfield  {journal} {\bibinfo  {journal}
  {Phys. Rev. Lett.}\ }\textbf {\bibinfo {volume} {120}},\ \bibinfo {pages}
  {070501} (\bibinfo {year} {2018})}\BibitemShut {NoStop}%
\bibitem [{\citenamefont {S\'anchez}\ \emph {et~al.}(2020)\citenamefont
  {S\'anchez}, \citenamefont {Chattah}, \citenamefont {Wei}, \citenamefont
  {Buljubasich}, \citenamefont {Cappellaro},\ and\ \citenamefont
  {Pastawski}}]{PhysRevLett.124.030601}%
  \BibitemOpen
  \bibfield  {author} {\bibinfo {author} {\bibfnamefont {C.~M.}\ \bibnamefont
  {S\'anchez}}, \bibinfo {author} {\bibfnamefont {A.~K.}\ \bibnamefont
  {Chattah}}, \bibinfo {author} {\bibfnamefont {K.~X.}\ \bibnamefont {Wei}},
  \bibinfo {author} {\bibfnamefont {L.}~\bibnamefont {Buljubasich}}, \bibinfo
  {author} {\bibfnamefont {P.}~\bibnamefont {Cappellaro}}, \ and\ \bibinfo
  {author} {\bibfnamefont {H.~M.}\ \bibnamefont {Pastawski}},\ }\bibfield
  {title} {\enquote {\bibinfo {title} {Perturbation {I}ndependent {D}ecay of
  the {L}oschmidt {E}cho in a {M}any-{B}ody {S}ystem},}\ }\href {\doibase
  10.1103/PhysRevLett.124.030601} {\bibfield  {journal} {\bibinfo  {journal}
  {Phys. Rev. Lett.}\ }\textbf {\bibinfo {volume} {124}},\ \bibinfo {pages}
  {030601} (\bibinfo {year} {2020})}\BibitemShut {NoStop}%
\bibitem [{\citenamefont {Joshi}\ \emph {et~al.}(2020)\citenamefont {Joshi},
  \citenamefont {Elben}, \citenamefont {Vermersch}, \citenamefont {Brydges},
  \citenamefont {Maier}, \citenamefont {Zoller}, \citenamefont {Blatt},\ and\
  \citenamefont {Roos}}]{2001_arxiv_2001.02176}%
  \BibitemOpen
  \bibfield  {author} {\bibinfo {author} {\bibfnamefont {M.~K.}\ \bibnamefont
  {Joshi}}, \bibinfo {author} {\bibfnamefont {A.}~\bibnamefont {Elben}},
  \bibinfo {author} {\bibfnamefont {B.}~\bibnamefont {Vermersch}}, \bibinfo
  {author} {\bibfnamefont {T.}~\bibnamefont {Brydges}}, \bibinfo {author}
  {\bibfnamefont {C.}~\bibnamefont {Maier}}, \bibinfo {author} {\bibfnamefont
  {P.}~\bibnamefont {Zoller}}, \bibinfo {author} {\bibfnamefont
  {R.}~\bibnamefont {Blatt}}, \ and\ \bibinfo {author} {\bibfnamefont {C.~F.}\
  \bibnamefont {Roos}},\ }\bibfield  {title} {\enquote {\bibinfo {title}
  {Quantum {I}nformation {S}crambling in a {T}rapped-ion {Q}uantum {S}imulator
  with {T}unable {R}ange {I}nteractions},}\ }\href {\doibase
  10.1103/PhysRevLett.124.240505} {\bibfield  {journal} {\bibinfo  {journal}
  {Phys. Rev. Lett.}\ }\textbf {\bibinfo {volume} {124}},\ \bibinfo {pages}
  {240505} (\bibinfo {year} {2020})}\BibitemShut {NoStop}%
\bibitem [{\citenamefont {Wei}\ \emph {et~al.}(2019)\citenamefont {Wei},
  \citenamefont {Peng}, \citenamefont {Shtanko}, \citenamefont {Marvian},
  \citenamefont {Lloyd}, \citenamefont {Ramanathan},\ and\ \citenamefont
  {Cappellaro}}]{PhysRevLett.123.090605}%
  \BibitemOpen
  \bibfield  {author} {\bibinfo {author} {\bibfnamefont {K.~X.}\ \bibnamefont
  {Wei}}, \bibinfo {author} {\bibfnamefont {P.}~\bibnamefont {Peng}}, \bibinfo
  {author} {\bibfnamefont {O.}~\bibnamefont {Shtanko}}, \bibinfo {author}
  {\bibfnamefont {I.}~\bibnamefont {Marvian}}, \bibinfo {author} {\bibfnamefont
  {S.}~\bibnamefont {Lloyd}}, \bibinfo {author} {\bibfnamefont
  {C.}~\bibnamefont {Ramanathan}}, \ and\ \bibinfo {author} {\bibfnamefont
  {P.}~\bibnamefont {Cappellaro}},\ }\bibfield  {title} {\enquote {\bibinfo
  {title} {Emergent {P}rethermalization {S}ignatures in {O}ut-of-{T}ime
  {O}rdered {C}orrelations},}\ }\href {\doibase 10.1103/PhysRevLett.123.090605}
  {\bibfield  {journal} {\bibinfo  {journal} {Phys. Rev. Lett.}\ }\textbf
  {\bibinfo {volume} {123}},\ \bibinfo {pages} {090605} (\bibinfo {year}
  {2019})}\BibitemShut {NoStop}%
\bibitem [{\citenamefont {Li}\ \emph {et~al.}(2017)\citenamefont {Li},
  \citenamefont {Fan}, \citenamefont {Wang}, \citenamefont {Ye}, \citenamefont
  {Zeng}, \citenamefont {Zhai}, \citenamefont {Peng},\ and\ \citenamefont
  {Du}}]{PhysRevX.7.031011}%
  \BibitemOpen
  \bibfield  {author} {\bibinfo {author} {\bibfnamefont {J.}~\bibnamefont
  {Li}}, \bibinfo {author} {\bibfnamefont {R.}~\bibnamefont {Fan}}, \bibinfo
  {author} {\bibfnamefont {H.}~\bibnamefont {Wang}}, \bibinfo {author}
  {\bibfnamefont {B.}~\bibnamefont {Ye}}, \bibinfo {author} {\bibfnamefont
  {B.}~\bibnamefont {Zeng}}, \bibinfo {author} {\bibfnamefont {H.}~\bibnamefont
  {Zhai}}, \bibinfo {author} {\bibfnamefont {X.}~\bibnamefont {Peng}}, \ and\
  \bibinfo {author} {\bibfnamefont {J.}~\bibnamefont {Du}},\ }\bibfield
  {title} {\enquote {\bibinfo {title} {Measuring {O}ut-of-{T}ime-{O}rder
  {C}orrelators on a {N}uclear {M}agnetic {R}esonance {Q}uantum {S}imulator},}\
  }\href {\doibase 10.1103/PhysRevX.7.031011} {\bibfield  {journal} {\bibinfo
  {journal} {Phys. Rev. X}\ }\textbf {\bibinfo {volume} {7}},\ \bibinfo {pages}
  {031011} (\bibinfo {year} {2017})}\BibitemShut {NoStop}%
\bibitem [{\citenamefont {Landsman}\ \emph {et~al.}(2019)\citenamefont
  {Landsman}, \citenamefont {Figgatt}, \citenamefont {Schuster}, \citenamefont
  {Linke}, \citenamefont {Yoshida}, \citenamefont {Yao},\ and\ \citenamefont
  {Monroe}}]{10.1038nphys4119}%
  \BibitemOpen
  \bibfield  {author} {\bibinfo {author} {\bibfnamefont {K.~A.}\ \bibnamefont
  {Landsman}}, \bibinfo {author} {\bibfnamefont {C.}~\bibnamefont {Figgatt}},
  \bibinfo {author} {\bibfnamefont {T.}~\bibnamefont {Schuster}}, \bibinfo
  {author} {\bibfnamefont {N.~M.}\ \bibnamefont {Linke}}, \bibinfo {author}
  {\bibfnamefont {B.}~\bibnamefont {Yoshida}}, \bibinfo {author} {\bibfnamefont
  {N.~Y.}\ \bibnamefont {Yao}}, \ and\ \bibinfo {author} {\bibfnamefont
  {C.}~\bibnamefont {Monroe}},\ }\bibfield  {title} {\enquote {\bibinfo {title}
  {Verified quantum information scrambling},}\ }\href {\doibase
  10.1038/s41586-019-0952-6} {\bibfield  {journal} {\bibinfo  {journal}
  {Nature}\ }\textbf {\bibinfo {volume} {567}},\ \bibinfo {pages} {61}
  (\bibinfo {year} {2019})}\BibitemShut {NoStop}%
\bibitem [{\citenamefont {Brand\~{a}o}\ \emph {et~al.}(2015)\citenamefont
  {Brand\~{a}o}, \citenamefont {Horodecki}, \citenamefont {Ng}, \citenamefont
  {Oppenheim},\ and\ \citenamefont {Wehner}}]{Brandao3275}%
  \BibitemOpen
  \bibfield  {author} {\bibinfo {author} {\bibfnamefont {F.}~\bibnamefont
  {Brand\~{a}o}}, \bibinfo {author} {\bibfnamefont {M.}~\bibnamefont
  {Horodecki}}, \bibinfo {author} {\bibfnamefont {N.}~\bibnamefont {Ng}},
  \bibinfo {author} {\bibfnamefont {J.}~\bibnamefont {Oppenheim}}, \ and\
  \bibinfo {author} {\bibfnamefont {S.}~\bibnamefont {Wehner}},\ }\bibfield
  {title} {\enquote {\bibinfo {title} {The second laws of quantum
  thermodynamics},}\ }\href {\doibase 10.1073/pnas.1411728112} {\bibfield
  {journal} {\bibinfo  {journal} {Proc. Natl. Acad. Sci.}\ }\textbf {\bibinfo
  {volume} {112}},\ \bibinfo {pages} {3275} (\bibinfo {year}
  {2015})}\BibitemShut {NoStop}%
\bibitem [{\citenamefont {Wei}\ and\ \citenamefont {Plenio}(2017)}]{Wei_2017}%
  \BibitemOpen
  \bibfield  {author} {\bibinfo {author} {\bibfnamefont {B.-B.}\ \bibnamefont
  {Wei}}\ and\ \bibinfo {author} {\bibfnamefont {M.~B.}\ \bibnamefont
  {Plenio}},\ }\bibfield  {title} {\enquote {\bibinfo {title} {Relations
  between dissipated work in non-equilibrium process and the family of
  {R}{\'{e}}nyi divergences},}\ }\href {\doibase 10.1088/1367-2630/aa579e}
  {\bibfield  {journal} {\bibinfo  {journal} {New J. Phys.}\ }\textbf {\bibinfo
  {volume} {19}},\ \bibinfo {pages} {023002} (\bibinfo {year}
  {2017})}\BibitemShut {NoStop}%
\bibitem [{\citenamefont {Guarnieri}\ \emph {et~al.}(2019)\citenamefont
  {Guarnieri}, \citenamefont {Ng}, \citenamefont {Modi}, \citenamefont
  {Eisert}, \citenamefont {Paternostro},\ and\ \citenamefont
  {Goold}}]{PhysRevE.99.050101}%
  \BibitemOpen
  \bibfield  {author} {\bibinfo {author} {\bibfnamefont {G.}~\bibnamefont
  {Guarnieri}}, \bibinfo {author} {\bibfnamefont {N.~H.~Y.}\ \bibnamefont
  {Ng}}, \bibinfo {author} {\bibfnamefont {K.}~\bibnamefont {Modi}}, \bibinfo
  {author} {\bibfnamefont {J.}~\bibnamefont {Eisert}}, \bibinfo {author}
  {\bibfnamefont {M.}~\bibnamefont {Paternostro}}, \ and\ \bibinfo {author}
  {\bibfnamefont {J.}~\bibnamefont {Goold}},\ }\bibfield  {title} {\enquote
  {\bibinfo {title} {Quantum work statistics and resource theories: {B}ridging
  the gap through {R}\'enyi divergences},}\ }\href {\doibase
  10.1103/PhysRevE.99.050101} {\bibfield  {journal} {\bibinfo  {journal} {Phys.
  Rev. E}\ }\textbf {\bibinfo {volume} {99}},\ \bibinfo {pages} {050101}
  (\bibinfo {year} {2019})}\BibitemShut {NoStop}%
\bibitem [{\citenamefont {Leditzky}\ \emph {et~al.}(2016)\citenamefont
  {Leditzky}, \citenamefont {Wilde},\ and\ \citenamefont
  {Datta}}]{10.1063_1.4960099}%
  \BibitemOpen
  \bibfield  {author} {\bibinfo {author} {\bibfnamefont {F.}~\bibnamefont
  {Leditzky}}, \bibinfo {author} {\bibfnamefont {M.~M.}\ \bibnamefont {Wilde}},
  \ and\ \bibinfo {author} {\bibfnamefont {N.}~\bibnamefont {Datta}},\
  }\bibfield  {title} {\enquote {\bibinfo {title} {Strong converse theorems
  using {R}\'{e}nyi entropies},}\ }\href {\doibase 10.1063/1.4960099}
  {\bibfield  {journal} {\bibinfo  {journal} {J. Math. Phys.}\ }\textbf
  {\bibinfo {volume} {57}},\ \bibinfo {pages} {082202} (\bibinfo {year}
  {2016})}\BibitemShut {NoStop}%
\bibitem [{\citenamefont {Chitambar}\ and\ \citenamefont
  {Gour}(2016{\natexlab{a}})}]{PhysRevLett.117.030401}%
  \BibitemOpen
  \bibfield  {author} {\bibinfo {author} {\bibfnamefont {E.}~\bibnamefont
  {Chitambar}}\ and\ \bibinfo {author} {\bibfnamefont {G.}~\bibnamefont
  {Gour}},\ }\bibfield  {title} {\enquote {\bibinfo {title} {Critical
  {E}xamination of {I}ncoherent {O}perations and a {P}hysically {C}onsistent
  {R}esource {T}heory of {Q}uantum {C}oherence},}\ }\href {\doibase
  10.1103/PhysRevLett.117.030401} {\bibfield  {journal} {\bibinfo  {journal}
  {Phys. Rev. Lett.}\ }\textbf {\bibinfo {volume} {117}},\ \bibinfo {pages}
  {030401} (\bibinfo {year} {2016}{\natexlab{a}})}\BibitemShut {NoStop}%
\bibitem [{\citenamefont {Chitambar}\ and\ \citenamefont
  {Gour}(2016{\natexlab{b}})}]{PhysRevA.94.052336}%
  \BibitemOpen
  \bibfield  {author} {\bibinfo {author} {\bibfnamefont {E.}~\bibnamefont
  {Chitambar}}\ and\ \bibinfo {author} {\bibfnamefont {G.}~\bibnamefont
  {Gour}},\ }\bibfield  {title} {\enquote {\bibinfo {title} {Comparison of
  incoherent operations and measures of coherence},}\ }\href {\doibase
  10.1103/PhysRevA.94.052336} {\bibfield  {journal} {\bibinfo  {journal} {Phys.
  Rev. A}\ }\textbf {\bibinfo {volume} {94}},\ \bibinfo {pages} {052336}
  (\bibinfo {year} {2016}{\natexlab{b}})}\BibitemShut {NoStop}%
\bibitem [{\citenamefont {Rastegin}(2016)}]{PhysRevA.93.032136}%
  \BibitemOpen
  \bibfield  {author} {\bibinfo {author} {\bibfnamefont {A.~E.}\ \bibnamefont
  {Rastegin}},\ }\bibfield  {title} {\enquote {\bibinfo {title}
  {Quantum-coherence quantifiers based on the {T}sallis relative
  $\ensuremath{\alpha}$ entropies},}\ }\href {\doibase
  10.1103/PhysRevA.93.032136} {\bibfield  {journal} {\bibinfo  {journal} {Phys.
  Rev. A}\ }\textbf {\bibinfo {volume} {93}},\ \bibinfo {pages} {032136}
  (\bibinfo {year} {2016})}\BibitemShut {NoStop}%
\bibitem [{\citenamefont {Streltsov}\ \emph {et~al.}(2018)\citenamefont
  {Streltsov}, \citenamefont {Kampermann}, \citenamefont {W\"{o}lk},
  \citenamefont {Gessner},\ and\ \citenamefont {Bru{\ss}}}]{Streltsov_2018}%
  \BibitemOpen
  \bibfield  {author} {\bibinfo {author} {\bibfnamefont {A.}~\bibnamefont
  {Streltsov}}, \bibinfo {author} {\bibfnamefont {H.}~\bibnamefont
  {Kampermann}}, \bibinfo {author} {\bibfnamefont {S.}~\bibnamefont
  {W\"{o}lk}}, \bibinfo {author} {\bibfnamefont {M.}~\bibnamefont {Gessner}}, \
  and\ \bibinfo {author} {\bibfnamefont {D.}~\bibnamefont {Bru{\ss}}},\
  }\bibfield  {title} {\enquote {\bibinfo {title} {Maximal coherence and the
  resource theory of purity},}\ }\href {\doibase 10.1088/1367-2630/aac484}
  {\bibfield  {journal} {\bibinfo  {journal} {New J. Phys.}\ }\textbf {\bibinfo
  {volume} {20}},\ \bibinfo {pages} {053058} (\bibinfo {year}
  {2018})}\BibitemShut {NoStop}%
\bibitem [{\citenamefont {Seshadreesan}\ \emph {et~al.}(2018)\citenamefont
  {Seshadreesan}, \citenamefont {Lami},\ and\ \citenamefont
  {Wilde}}]{10.1063_1.5007167}%
  \BibitemOpen
  \bibfield  {author} {\bibinfo {author} {\bibfnamefont {K.~P.}\ \bibnamefont
  {Seshadreesan}}, \bibinfo {author} {\bibfnamefont {L.}~\bibnamefont {Lami}},
  \ and\ \bibinfo {author} {\bibfnamefont {M.~M.}\ \bibnamefont {Wilde}},\
  }\bibfield  {title} {\enquote {\bibinfo {title} {R\'{e}nyi relative entropies
  of quantum {G}aussian states},}\ }\href {\doibase 10.1063/1.5007167}
  {\bibfield  {journal} {\bibinfo  {journal} {J. Math. Phys.}\ }\textbf
  {\bibinfo {volume} {59}},\ \bibinfo {pages} {072204} (\bibinfo {year}
  {2018})}\BibitemShut {NoStop}%
\bibitem [{\citenamefont {Horodecki}\ and\ \citenamefont
  {Ekert}(2002)}]{PhysRevLett.89.127902}%
  \BibitemOpen
  \bibfield  {author} {\bibinfo {author} {\bibfnamefont {P.}~\bibnamefont
  {Horodecki}}\ and\ \bibinfo {author} {\bibfnamefont {A.}~\bibnamefont
  {Ekert}},\ }\bibfield  {title} {\enquote {\bibinfo {title} {Method for
  {D}irect {D}etection of {Q}uantum {E}ntanglement},}\ }\href {\doibase
  10.1103/PhysRevLett.89.127902} {\bibfield  {journal} {\bibinfo  {journal}
  {Phys. Rev. Lett.}\ }\textbf {\bibinfo {volume} {89}},\ \bibinfo {pages}
  {127902} (\bibinfo {year} {2002})}\BibitemShut {NoStop}%
\bibitem [{\citenamefont {Cardy}(2011)}]{PhysRevLett.106.150404}%
  \BibitemOpen
  \bibfield  {author} {\bibinfo {author} {\bibfnamefont {J.}~\bibnamefont
  {Cardy}},\ }\bibfield  {title} {\enquote {\bibinfo {title} {Measuring
  {E}ntanglement {U}sing {Q}uantum {Q}uenches},}\ }\href {\doibase
  10.1103/PhysRevLett.106.150404} {\bibfield  {journal} {\bibinfo  {journal}
  {Phys. Rev. Lett.}\ }\textbf {\bibinfo {volume} {106}},\ \bibinfo {pages}
  {150404} (\bibinfo {year} {2011})}\BibitemShut {NoStop}%
\bibitem [{\citenamefont {Abanin}\ and\ \citenamefont
  {Demler}(2012)}]{PhysRevLett.109.020504}%
  \BibitemOpen
  \bibfield  {author} {\bibinfo {author} {\bibfnamefont {D.~A.}\ \bibnamefont
  {Abanin}}\ and\ \bibinfo {author} {\bibfnamefont {E.}~\bibnamefont
  {Demler}},\ }\bibfield  {title} {\enquote {\bibinfo {title} {Measuring
  {E}ntanglement {E}ntropy of a {G}eneric {M}any-{B}ody {S}ystem with a
  {Q}uantum {S}witch},}\ }\href {\doibase 10.1103/PhysRevLett.109.020504}
  {\bibfield  {journal} {\bibinfo  {journal} {Phys. Rev. Lett.}\ }\textbf
  {\bibinfo {volume} {109}},\ \bibinfo {pages} {020504} (\bibinfo {year}
  {2012})}\BibitemShut {NoStop}%
\bibitem [{\citenamefont {Elben}\ \emph {et~al.}()\citenamefont {Elben},
  \citenamefont {Kueng}, \citenamefont {Huang}, \citenamefont {van Bijnen},
  \citenamefont {Kokail}, \citenamefont {Dalmonte}, \citenamefont {Calabrese},
  \citenamefont {Kraus}, \citenamefont {Preskill}, \citenamefont {Zoller},\
  and\ \citenamefont {Vermersch}}]{2020_arxiv_2007.06305}%
  \BibitemOpen
  \bibfield  {author} {\bibinfo {author} {\bibfnamefont {A.}~\bibnamefont
  {Elben}}, \bibinfo {author} {\bibfnamefont {R.}~\bibnamefont {Kueng}},
  \bibinfo {author} {\bibfnamefont {H.-Y.}\ \bibnamefont {Huang}}, \bibinfo
  {author} {\bibfnamefont {R.}~\bibnamefont {van Bijnen}}, \bibinfo {author}
  {\bibfnamefont {C.}~\bibnamefont {Kokail}}, \bibinfo {author} {\bibfnamefont
  {M.}~\bibnamefont {Dalmonte}}, \bibinfo {author} {\bibfnamefont
  {P.}~\bibnamefont {Calabrese}}, \bibinfo {author} {\bibfnamefont
  {B.}~\bibnamefont {Kraus}}, \bibinfo {author} {\bibfnamefont
  {J.}~\bibnamefont {Preskill}}, \bibinfo {author} {\bibfnamefont
  {P.}~\bibnamefont {Zoller}}, \ and\ \bibinfo {author} {\bibfnamefont
  {B.}~\bibnamefont {Vermersch}},\ }\bibfield  {title} {\enquote {\bibinfo
  {title} {Mixed-state entanglement from local randomized measurements},}\
  }\href@noop {} {\ }\Eprint {http://arxiv.org/abs/2007.06305}
  {arXiv:2007.06305} \BibitemShut {NoStop}%
\bibitem [{\citenamefont {Elben}\ \emph {et~al.}(2018)\citenamefont {Elben},
  \citenamefont {Vermersch}, \citenamefont {Dalmonte}, \citenamefont {Cirac},\
  and\ \citenamefont {Zoller}}]{PhysRevLett.120.050406}%
  \BibitemOpen
  \bibfield  {author} {\bibinfo {author} {\bibfnamefont {A.}~\bibnamefont
  {Elben}}, \bibinfo {author} {\bibfnamefont {B.}~\bibnamefont {Vermersch}},
  \bibinfo {author} {\bibfnamefont {M.}~\bibnamefont {Dalmonte}}, \bibinfo
  {author} {\bibfnamefont {J.~I.}\ \bibnamefont {Cirac}}, \ and\ \bibinfo
  {author} {\bibfnamefont {P.}~\bibnamefont {Zoller}},\ }\bibfield  {title}
  {\enquote {\bibinfo {title} {R\'enyi {E}ntropies from {R}andom {Q}uenches in
  {A}tomic {H}ubbard and {S}pin {M}odels},}\ }\href {\doibase
  10.1103/PhysRevLett.120.050406} {\bibfield  {journal} {\bibinfo  {journal}
  {Phys. Rev. Lett.}\ }\textbf {\bibinfo {volume} {120}},\ \bibinfo {pages}
  {050406} (\bibinfo {year} {2018})}\BibitemShut {NoStop}%
\bibitem [{\citenamefont {Islam}\ \emph {et~al.}(2015)\citenamefont {Islam},
  \citenamefont {Ma}, \citenamefont {Preiss}, \citenamefont {Tai},
  \citenamefont {Lukin}, \citenamefont {Rispoli},\ and\ \citenamefont
  {Greiner}}]{20159_arxiv_1509.01160}%
  \BibitemOpen
  \bibfield  {author} {\bibinfo {author} {\bibfnamefont {R.}~\bibnamefont
  {Islam}}, \bibinfo {author} {\bibfnamefont {R.}~\bibnamefont {Ma}}, \bibinfo
  {author} {\bibfnamefont {P.~M.}\ \bibnamefont {Preiss}}, \bibinfo {author}
  {\bibfnamefont {M.~E.}\ \bibnamefont {Tai}}, \bibinfo {author} {\bibfnamefont
  {A.}~\bibnamefont {Lukin}}, \bibinfo {author} {\bibfnamefont
  {M.}~\bibnamefont {Rispoli}}, \ and\ \bibinfo {author} {\bibfnamefont
  {M.}~\bibnamefont {Greiner}},\ }\bibfield  {title} {\enquote {\bibinfo
  {title} {Measuring entanglement entropy in a quantum many-body system},}\
  }\href {\doibase 10.1038/nature15750} {\bibfield  {journal} {\bibinfo
  {journal} {Nature (London)}\ }\textbf {\bibinfo {volume} {528}},\ \bibinfo
  {pages} {77} (\bibinfo {year} {2015})}\BibitemShut {NoStop}%
\bibitem [{\citenamefont {Linke}\ \emph {et~al.}(2018)\citenamefont {Linke},
  \citenamefont {Johri}, \citenamefont {Figgatt}, \citenamefont {Landsman},
  \citenamefont {Matsuura},\ and\ \citenamefont {Monroe}}]{PhysRevA.98.052334}%
  \BibitemOpen
  \bibfield  {author} {\bibinfo {author} {\bibfnamefont {N.~M.}\ \bibnamefont
  {Linke}}, \bibinfo {author} {\bibfnamefont {S.}~\bibnamefont {Johri}},
  \bibinfo {author} {\bibfnamefont {C.}~\bibnamefont {Figgatt}}, \bibinfo
  {author} {\bibfnamefont {K.~A.}\ \bibnamefont {Landsman}}, \bibinfo {author}
  {\bibfnamefont {A.~Y.}\ \bibnamefont {Matsuura}}, \ and\ \bibinfo {author}
  {\bibfnamefont {C.}~\bibnamefont {Monroe}},\ }\bibfield  {title} {\enquote
  {\bibinfo {title} {Measuring the {R}\'enyi entropy of a two-site
  {F}ermi-{H}ubbard model on a trapped ion quantum computer},}\ }\href
  {\doibase 10.1103/PhysRevA.98.052334} {\bibfield  {journal} {\bibinfo
  {journal} {Phys. Rev. A}\ }\textbf {\bibinfo {volume} {98}},\ \bibinfo
  {pages} {052334} (\bibinfo {year} {2018})}\BibitemShut {NoStop}%
\bibitem [{\citenamefont {Brydges}\ \emph {et~al.}(2019)\citenamefont
  {Brydges}, \citenamefont {Elben}, \citenamefont {Jurcevic}, \citenamefont
  {Vermersch}, \citenamefont {Maier}, \citenamefont {Lanyon}, \citenamefont
  {Zoller}, \citenamefont {Blatt},\ and\ \citenamefont {Roos}}]{Brydges260}%
  \BibitemOpen
  \bibfield  {author} {\bibinfo {author} {\bibfnamefont {T.}~\bibnamefont
  {Brydges}}, \bibinfo {author} {\bibfnamefont {A.}~\bibnamefont {Elben}},
  \bibinfo {author} {\bibfnamefont {P.}~\bibnamefont {Jurcevic}}, \bibinfo
  {author} {\bibfnamefont {B.}~\bibnamefont {Vermersch}}, \bibinfo {author}
  {\bibfnamefont {C.}~\bibnamefont {Maier}}, \bibinfo {author} {\bibfnamefont
  {B.~P.}\ \bibnamefont {Lanyon}}, \bibinfo {author} {\bibfnamefont
  {P.}~\bibnamefont {Zoller}}, \bibinfo {author} {\bibfnamefont
  {R.}~\bibnamefont {Blatt}}, \ and\ \bibinfo {author} {\bibfnamefont {C.~F.}\
  \bibnamefont {Roos}},\ }\bibfield  {title} {\enquote {\bibinfo {title}
  {Probing {R}{\'e}nyi entanglement entropy via randomized measurements},}\
  }\href {\doibase 10.1126/science.aau4963} {\bibfield  {journal} {\bibinfo
  {journal} {Science}\ }\textbf {\bibinfo {volume} {364}},\ \bibinfo {pages}
  {260} (\bibinfo {year} {2019})}\BibitemShut {NoStop}%
\bibitem [{\citenamefont {Daley}\ \emph {et~al.}(2012)\citenamefont {Daley},
  \citenamefont {Pichler}, \citenamefont {Schachenmayer},\ and\ \citenamefont
  {Zoller}}]{PhysRevLett.109.020505}%
  \BibitemOpen
  \bibfield  {author} {\bibinfo {author} {\bibfnamefont {A.~J.}\ \bibnamefont
  {Daley}}, \bibinfo {author} {\bibfnamefont {H.}~\bibnamefont {Pichler}},
  \bibinfo {author} {\bibfnamefont {J.}~\bibnamefont {Schachenmayer}}, \ and\
  \bibinfo {author} {\bibfnamefont {P.}~\bibnamefont {Zoller}},\ }\bibfield
  {title} {\enquote {\bibinfo {title} {Measuring {E}ntanglement {G}rowth in
  {Q}uench {D}ynamics of {B}osons in an {O}ptical {L}attice},}\ }\href
  {\doibase 10.1103/PhysRevLett.109.020505} {\bibfield  {journal} {\bibinfo
  {journal} {Phys. Rev. Lett.}\ }\textbf {\bibinfo {volume} {109}},\ \bibinfo
  {pages} {020505} (\bibinfo {year} {2012})}\BibitemShut {NoStop}%
\bibitem [{\citenamefont {Fr\'erot}\ and\ \citenamefont
  {Roscilde}(2016)}]{PhysRevB.94.075121}%
  \BibitemOpen
  \bibfield  {author} {\bibinfo {author} {\bibfnamefont {I.}~\bibnamefont
  {Fr\'erot}}\ and\ \bibinfo {author} {\bibfnamefont {T.}~\bibnamefont
  {Roscilde}},\ }\bibfield  {title} {\enquote {\bibinfo {title} {Quantum
  variance: {A} measure of quantum coherence and quantum correlations for
  many-body systems},}\ }\href {\doibase 10.1103/PhysRevB.94.075121} {\bibfield
   {journal} {\bibinfo  {journal} {Phys. Rev. B}\ }\textbf {\bibinfo {volume}
  {94}},\ \bibinfo {pages} {075121} (\bibinfo {year} {2016})}\BibitemShut
  {NoStop}%
\bibitem [{\citenamefont {Sidhu}\ and\ \citenamefont
  {Kok}(2020)}]{10.1116_1.5119961}%
  \BibitemOpen
  \bibfield  {author} {\bibinfo {author} {\bibfnamefont {J.~S.}\ \bibnamefont
  {Sidhu}}\ and\ \bibinfo {author} {\bibfnamefont {P.}~\bibnamefont {Kok}},\
  }\bibfield  {title} {\enquote {\bibinfo {title} {Geometric perspective on
  quantum parameter estimation},}\ }\href {\doibase 10.1116/1.5119961}
  {\bibfield  {journal} {\bibinfo  {journal} {{A}{V}{S} {Q}uantum {S}ci.}\
  }\textbf {\bibinfo {volume} {2}},\ \bibinfo {pages} {014701} (\bibinfo {year}
  {2020})}\BibitemShut {NoStop}%
\bibitem [{\citenamefont {Pezz\'e}\ and\ \citenamefont
  {Smerzi}(2009)}]{PhysRevLett.102.100401}%
  \BibitemOpen
  \bibfield  {author} {\bibinfo {author} {\bibfnamefont {L.}~\bibnamefont
  {Pezz\'e}}\ and\ \bibinfo {author} {\bibfnamefont {A.}~\bibnamefont
  {Smerzi}},\ }\bibfield  {title} {\enquote {\bibinfo {title} {Entanglement,
  {N}onlinear {D}ynamics, and the {H}eisenberg {L}imit},}\ }\href {\doibase
  10.1103/PhysRevLett.102.100401} {\bibfield  {journal} {\bibinfo  {journal}
  {Phys. Rev. Lett.}\ }\textbf {\bibinfo {volume} {102}},\ \bibinfo {pages}
  {100401} (\bibinfo {year} {2009})}\BibitemShut {NoStop}%
\bibitem [{\citenamefont {Pezz\'{e}}\ and\ \citenamefont
  {Smerzi}(2014)}]{pezze_smerzi_188_691_2014}%
  \BibitemOpen
  \bibfield  {author} {\bibinfo {author} {\bibfnamefont {L.}~\bibnamefont
  {Pezz\'{e}}}\ and\ \bibinfo {author} {\bibfnamefont {A.}~\bibnamefont
  {Smerzi}},\ }\bibfield  {title} {\enquote {\bibinfo {title} {Quantum theory
  of phase estimation},}\ }in\ \href {\doibase 10.3254/978-1-61499-448-0-691}
  {\emph {\bibinfo {booktitle} {Atom Interferometry, Proceedings of the
  International School of Physics ``Enrico Fermi''}}},\ Vol.\ \bibinfo {volume}
  {188},\ \bibinfo {editor} {edited by\ \bibinfo {editor} {\bibfnamefont
  {G.~M.}\ \bibnamefont {Tino}}\ and\ \bibinfo {editor} {\bibfnamefont {M.~A.}\
  \bibnamefont {Kasevich}}}\ (\bibinfo  {publisher} {IOS Press},\ \bibinfo
  {address} {Amsterdam},\ \bibinfo {year} {2014})\ pp.\ \bibinfo {pages}
  {691--741}\BibitemShut {NoStop}%
\bibitem [{\citenamefont {Petz}(1985)}]{PETZ1985_21_PublResInstKyoto}%
  \BibitemOpen
  \bibfield  {author} {\bibinfo {author} {\bibfnamefont {D.}~\bibnamefont
  {Petz}},\ }\bibfield  {title} {\enquote {\bibinfo {title} {Quasi-entropies
  for {S}tates of a von {N}eumann {A}lgebra},}\ }\href {\doibase
  10.2977/prims/1195178929} {\bibfield  {journal} {\bibinfo  {journal} {Publ.
  Res. Inst. Math. Sci.}\ }\textbf {\bibinfo {volume} {21}},\ \bibinfo {pages}
  {787} (\bibinfo {year} {1985})}\BibitemShut {NoStop}%
\bibitem [{\citenamefont {Petz}(1986)}]{PETZ198657}%
  \BibitemOpen
  \bibfield  {author} {\bibinfo {author} {\bibfnamefont {D.}~\bibnamefont
  {Petz}},\ }\bibfield  {title} {\enquote {\bibinfo {title} {Quasi-entropies
  for finite quantum systems},}\ }\href {\doibase 10.1016/0034-4877(86)90067-4}
  {\bibfield  {journal} {\bibinfo  {journal} {Rep. Math. Phys.}\ }\textbf
  {\bibinfo {volume} {23}},\ \bibinfo {pages} {57} (\bibinfo {year}
  {1986})}\BibitemShut {NoStop}%
\bibitem [{\citenamefont {Mosonyi}\ and\ \citenamefont
  {Hiai}(2011)}]{IEEE.57.2474.2011}%
  \BibitemOpen
  \bibfield  {author} {\bibinfo {author} {\bibfnamefont {M.}~\bibnamefont
  {Mosonyi}}\ and\ \bibinfo {author} {\bibfnamefont {F.}~\bibnamefont {Hiai}},\
  }\bibfield  {title} {\enquote {\bibinfo {title} {On the {Q}uantum {R}\'{e}nyi
  {R}elative {E}ntropies and {R}elated {C}apacity {F}ormulas},}\ }\href
  {\doibase 10.1109/TIT.2011.2110050} {\bibfield  {journal} {\bibinfo
  {journal} {IEEE Trans. Inf. Theory}\ }\textbf {\bibinfo {volume} {57}},\
  \bibinfo {pages} {2474} (\bibinfo {year} {2011})}\BibitemShut {NoStop}%
\bibitem [{\citenamefont {M\"{u}ller-Lennert}\ \emph
  {et~al.}(2013)\citenamefont {M\"{u}ller-Lennert}, \citenamefont {Dupuis},
  \citenamefont {Szehr}, \citenamefont {Fehr},\ and\ \citenamefont
  {Tomamichel}}]{10.10631.4838856}%
  \BibitemOpen
  \bibfield  {author} {\bibinfo {author} {\bibfnamefont {M.}~\bibnamefont
  {M\"{u}ller-Lennert}}, \bibinfo {author} {\bibfnamefont {F.}~\bibnamefont
  {Dupuis}}, \bibinfo {author} {\bibfnamefont {O.}~\bibnamefont {Szehr}},
  \bibinfo {author} {\bibfnamefont {S.}~\bibnamefont {Fehr}}, \ and\ \bibinfo
  {author} {\bibfnamefont {M.}~\bibnamefont {Tomamichel}},\ }\bibfield  {title}
  {\enquote {\bibinfo {title} {On quantum {R}\'{e}nyi entropies: {A} new
  generalization and some properties},}\ }\href {\doibase 10.1063/1.4838856}
  {\bibfield  {journal} {\bibinfo  {journal} {J. Math. Phys.}\ }\textbf
  {\bibinfo {volume} {54}},\ \bibinfo {pages} {122203} (\bibinfo {year}
  {2013})}\BibitemShut {NoStop}%
\bibitem [{\citenamefont {Leditzky}\ \emph {et~al.}(2017)\citenamefont
  {Leditzky}, \citenamefont {Rouz{\'e}},\ and\ \citenamefont
  {Datta}}]{Leditzky2017}%
  \BibitemOpen
  \bibfield  {author} {\bibinfo {author} {\bibfnamefont {F.}~\bibnamefont
  {Leditzky}}, \bibinfo {author} {\bibfnamefont {C.}~\bibnamefont {Rouz{\'e}}},
  \ and\ \bibinfo {author} {\bibfnamefont {N.}~\bibnamefont {Datta}},\
  }\bibfield  {title} {\enquote {\bibinfo {title} {Data processing for the
  sandwiched {R}\'{e}nyi divergence: a condition for equality},}\ }\href
  {\doibase 10.1007/s11005-016-0896-9} {\bibfield  {journal} {\bibinfo
  {journal} {Lett. Math. Phys.}\ }\textbf {\bibinfo {volume} {107}},\ \bibinfo
  {pages} {61} (\bibinfo {year} {2017})}\BibitemShut {NoStop}%
\bibitem [{\citenamefont {Mosonyi}\ and\ \citenamefont
  {Ogawa}(2015)}]{Mosonyi2015}%
  \BibitemOpen
  \bibfield  {author} {\bibinfo {author} {\bibfnamefont {M.}~\bibnamefont
  {Mosonyi}}\ and\ \bibinfo {author} {\bibfnamefont {T.}~\bibnamefont
  {Ogawa}},\ }\bibfield  {title} {\enquote {\bibinfo {title} {Quantum
  {H}ypothesis {T}esting and the {O}perational {I}nterpretation of the
  {Q}uantum {R}\'{e}nyi {R}elative {E}ntropies},}\ }\href {\doibase
  10.1007/s00220-014-2248-x} {\bibfield  {journal} {\bibinfo  {journal}
  {Commun. Math. Phys.}\ }\textbf {\bibinfo {volume} {334}},\ \bibinfo {pages}
  {1617} (\bibinfo {year} {2015})}\BibitemShut {NoStop}%
\bibitem [{\citenamefont {Csisz\'{a}r}(1967)}]{csiszar1967}%
  \BibitemOpen
  \bibfield  {author} {\bibinfo {author} {\bibfnamefont {I.}~\bibnamefont
  {Csisz\'{a}r}},\ }\bibfield  {title} {\enquote {\bibinfo {title}
  {Information-type measures of difference of probability distributions and
  indirect observations},}\ }\href@noop {} {\bibfield  {journal} {\bibinfo
  {journal} {Studia Sci. Math. Hungar}\ }\textbf {\bibinfo {volume} {2}},\
  \bibinfo {pages} {299} (\bibinfo {year} {1967})}\BibitemShut {NoStop}%
\bibitem [{\citenamefont {Hiai}\ \emph {et~al.}(1981)\citenamefont {Hiai},
  \citenamefont {Ohya},\ and\ \citenamefont {Tsukada}}]{hiai1981}%
  \BibitemOpen
  \bibfield  {author} {\bibinfo {author} {\bibfnamefont {F.}~\bibnamefont
  {Hiai}}, \bibinfo {author} {\bibfnamefont {M.}~\bibnamefont {Ohya}}, \ and\
  \bibinfo {author} {\bibfnamefont {M.}~\bibnamefont {Tsukada}},\ }\bibfield
  {title} {\enquote {\bibinfo {title} {Sufficiency, {K}{M}{S} condition and
  relative entropy in von {N}eumann algebras},}\ }\href@noop {} {\bibfield
  {journal} {\bibinfo  {journal} {Pacific J. Math.}\ }\textbf {\bibinfo
  {volume} {96}},\ \bibinfo {pages} {99} (\bibinfo {year} {1981})}\BibitemShut
  {NoStop}%
\bibitem [{\citenamefont {Gilardoni}(2010)}]{5605338}%
  \BibitemOpen
  \bibfield  {author} {\bibinfo {author} {\bibfnamefont {G.~L.}\ \bibnamefont
  {Gilardoni}},\ }\bibfield  {title} {\enquote {\bibinfo {title} {On
  {P}insker's and {V}ajda's {T}ype {I}nequalities for {C}sisz\'{a}r's
  $f$-{D}ivergences},}\ }\href {\doibase 10.1109/TIT.2010.2068710} {\bibfield
  {journal} {\bibinfo  {journal} {IEEE Trans. Inf. Theory}\ }\textbf {\bibinfo
  {volume} {56}},\ \bibinfo {pages} {5377} (\bibinfo {year}
  {2010})}\BibitemShut {NoStop}%
\bibitem [{\citenamefont {Rastegin}(2013)}]{ARastegin_MathPhysAnalGeom_16_213}%
  \BibitemOpen
  \bibfield  {author} {\bibinfo {author} {\bibfnamefont {A.~E.}\ \bibnamefont
  {Rastegin}},\ }\bibfield  {title} {\enquote {\bibinfo {title} {Bounds of the
  {P}insker and {F}annes {T}ypes on the {T}sallis {R}elative {E}ntropy},}\
  }\href {\doibase 10.1007/s11040-013-9128-z} {\bibfield  {journal} {\bibinfo
  {journal} {Math. Phys. Anal. Geom.}\ }\textbf {\bibinfo {volume} {16}},\
  \bibinfo {pages} {213} (\bibinfo {year} {2013})}\BibitemShut {NoStop}%
\bibitem [{\citenamefont {Wilde}\ \emph {et~al.}(2014)\citenamefont {Wilde},
  \citenamefont {Winter},\ and\ \citenamefont {Yang}}]{Winter_2014_331}%
  \BibitemOpen
  \bibfield  {author} {\bibinfo {author} {\bibfnamefont {M.~M.}\ \bibnamefont
  {Wilde}}, \bibinfo {author} {\bibfnamefont {A.}~\bibnamefont {Winter}}, \
  and\ \bibinfo {author} {\bibfnamefont {D.}~\bibnamefont {Yang}},\ }\bibfield
  {title} {\enquote {\bibinfo {title} {Strong {C}onverse for the {C}lassical
  {C}apacity of {E}ntanglement-{B}reaking and {H}adamard {C}hannels via a
  {S}andwiched {R}\'{e}nyi {R}elative {E}ntropy},}\ }\href {\doibase
  10.1007/s00220-014-2122-x} {\bibfield  {journal} {\bibinfo  {journal}
  {Commun. Math. Phys.}\ }\textbf {\bibinfo {volume} {331}},\ \bibinfo {pages}
  {593} (\bibinfo {year} {2014})}\BibitemShut {NoStop}%
\bibitem [{\citenamefont {Audenaert}(2014)}]{DBLPjournals_qic_Audenaert01}%
  \BibitemOpen
  \bibfield  {author} {\bibinfo {author} {\bibfnamefont {K.~M.~R.}\
  \bibnamefont {Audenaert}},\ }\bibfield  {title} {\enquote {\bibinfo {title}
  {Comparisons between quantum state distinguishability measures},}\ }\href
  {\doibase 10.26421/QIC14.1-2} {\bibfield  {journal} {\bibinfo  {journal}
  {Quantum {I}nf. {C}omput.}\ }\textbf {\bibinfo {volume} {14}},\ \bibinfo
  {pages} {31} (\bibinfo {year} {2014})}\BibitemShut {NoStop}%
\bibitem [{\citenamefont {Audenaert}\ \emph {et~al.}(2007)\citenamefont
  {Audenaert}, \citenamefont {Calsamiglia}, \citenamefont {Mu\~noz Tapia},
  \citenamefont {Bagan}, \citenamefont {Masanes}, \citenamefont {Acin},\ and\
  \citenamefont {Verstraete}}]{PhysRevLett.98.160501}%
  \BibitemOpen
  \bibfield  {author} {\bibinfo {author} {\bibfnamefont {K.~M.~R.}\
  \bibnamefont {Audenaert}}, \bibinfo {author} {\bibfnamefont {J.}~\bibnamefont
  {Calsamiglia}}, \bibinfo {author} {\bibfnamefont {R.}~\bibnamefont {Mu\~noz
  Tapia}}, \bibinfo {author} {\bibfnamefont {E.}~\bibnamefont {Bagan}},
  \bibinfo {author} {\bibfnamefont {Ll.}\ \bibnamefont {Masanes}}, \bibinfo
  {author} {\bibfnamefont {A.}~\bibnamefont {Acin}}, \ and\ \bibinfo {author}
  {\bibfnamefont {F.}~\bibnamefont {Verstraete}},\ }\bibfield  {title}
  {\enquote {\bibinfo {title} {Discriminating {S}tates: {T}he {Q}uantum
  {C}hernoff {B}ound},}\ }\href {\doibase 10.1103/PhysRevLett.98.160501}
  {\bibfield  {journal} {\bibinfo  {journal} {Phys. Rev. Lett.}\ }\textbf
  {\bibinfo {volume} {98}},\ \bibinfo {pages} {160501} (\bibinfo {year}
  {2007})}\BibitemShut {NoStop}%
\bibitem [{\citenamefont {Powers}\ and\ \citenamefont
  {St{\o}rmer}(1970)}]{Powers1970}%
  \BibitemOpen
  \bibfield  {author} {\bibinfo {author} {\bibfnamefont {R.~T.}\ \bibnamefont
  {Powers}}\ and\ \bibinfo {author} {\bibfnamefont {E.}~\bibnamefont
  {St{\o}rmer}},\ }\bibfield  {title} {\enquote {\bibinfo {title} {Free states
  of the canonical anticommutation relations},}\ }\href {\doibase
  10.1007/BF01645492} {\bibfield  {journal} {\bibinfo  {journal} {Commun. Math.
  Phys.}\ }\textbf {\bibinfo {volume} {16}},\ \bibinfo {pages} {1} (\bibinfo
  {year} {1970})}\BibitemShut {NoStop}%
\bibitem [{\citenamefont {Jak\v{s}i\'{c}}\ \emph {et~al.}(2012)\citenamefont
  {Jak\v{s}i\'{c}}, \citenamefont {Ogata}, \citenamefont {Pillet},\ and\
  \citenamefont {Seiringer}}]{doi:10.1142_S0129055X12300026}%
  \BibitemOpen
  \bibfield  {author} {\bibinfo {author} {\bibfnamefont {V.}~\bibnamefont
  {Jak\v{s}i\'{c}}}, \bibinfo {author} {\bibfnamefont {Y.}~\bibnamefont
  {Ogata}}, \bibinfo {author} {\bibfnamefont {C.-A.}\ \bibnamefont {Pillet}}, \
  and\ \bibinfo {author} {\bibfnamefont {R.}~\bibnamefont {Seiringer}},\
  }\bibfield  {title} {\enquote {\bibinfo {title} {Quantum {H}ypothesis
  {T}esting and {N}on-{E}quilibrium {S}tatistical {M}echanics},}\ }\href
  {\doibase 10.1142/S0129055X12300026} {\bibfield  {journal} {\bibinfo
  {journal} {Rev. Math. Phys.}\ }\textbf {\bibinfo {volume} {24}},\ \bibinfo
  {pages} {1230002} (\bibinfo {year} {2012})}\BibitemShut {NoStop}%
\bibitem [{\citenamefont {Umegaki}(1962)}]{umegaki1962}%
  \BibitemOpen
  \bibfield  {author} {\bibinfo {author} {\bibfnamefont {H.}~\bibnamefont
  {Umegaki}},\ }\bibfield  {title} {\enquote {\bibinfo {title} {Conditional
  expectation in an operator algebra. {I}{V}. {E}ntropy and information},}\
  }\href {\doibase 10.2996/kmj/1138844604} {\bibfield  {journal} {\bibinfo
  {journal} {Kodai Math. Sem. Rep.}\ }\textbf {\bibinfo {volume} {14}},\
  \bibinfo {pages} {59} (\bibinfo {year} {1962})}\BibitemShut {NoStop}%
\bibitem [{\citenamefont {van Erven}\ and\ \citenamefont
  {Harremos}(2014)}]{6832827ErvenHarremos}%
  \BibitemOpen
  \bibfield  {author} {\bibinfo {author} {\bibfnamefont {T}~\bibnamefont {van
  Erven}}\ and\ \bibinfo {author} {\bibfnamefont {P.}~\bibnamefont
  {Harremos}},\ }\bibfield  {title} {\enquote {\bibinfo {title} {R\'{e}nyi
  {D}ivergence and {K}ullback-{L}eibler {D}ivergence},}\ }\href {\doibase
  10.1109/TIT.2014.2320500} {\bibfield  {journal} {\bibinfo  {journal} {IEEE
  Trans. Inf. Theory}\ }\textbf {\bibinfo {volume} {60}},\ \bibinfo {pages}
  {3797--3820} (\bibinfo {year} {2014})}\BibitemShut {NoStop}%
\bibitem [{\citenamefont {{Datta}}(2009)}]{4957651_Datta}%
  \BibitemOpen
  \bibfield  {author} {\bibinfo {author} {\bibfnamefont {N.}~\bibnamefont
  {{Datta}}},\ }\bibfield  {title} {\enquote {\bibinfo {title} {Min- and
  {M}ax-{R}elative {E}ntropies and a {N}ew {E}ntanglement {M}onotone},}\ }\href
  {\doibase 10.1109/TIT.2009.2018325} {\bibfield  {journal} {\bibinfo
  {journal} {IEEE Trans. Inf. Theory}\ }\textbf {\bibinfo {volume} {55}},\
  \bibinfo {pages} {2816} (\bibinfo {year} {2009})}\BibitemShut {NoStop}%
\bibitem [{\citenamefont {Marvian}\ and\ \citenamefont
  {Spekkens}(2016)}]{PhysRevA.94.052324}%
  \BibitemOpen
  \bibfield  {author} {\bibinfo {author} {\bibfnamefont {I.}~\bibnamefont
  {Marvian}}\ and\ \bibinfo {author} {\bibfnamefont {R.~W.}\ \bibnamefont
  {Spekkens}},\ }\bibfield  {title} {\enquote {\bibinfo {title} {How to
  quantify coherence: {D}istinguishing speakable and unspeakable notions},}\
  }\href {\doibase 10.1103/PhysRevA.94.052324} {\bibfield  {journal} {\bibinfo
  {journal} {Phys. Rev. A}\ }\textbf {\bibinfo {volume} {94}},\ \bibinfo
  {pages} {052324} (\bibinfo {year} {2016})}\BibitemShut {NoStop}%
\bibitem [{\citenamefont {Bathia}(1997)}]{Bathia_Rajendra}%
  \BibitemOpen
  \bibfield  {author} {\bibinfo {author} {\bibfnamefont {Rajendra}\
  \bibnamefont {Bathia}},\ }\href@noop {} {\emph {\bibinfo {title} {Matrix
  Analysis}}}\ (\bibinfo  {publisher} {Springer-Verlag},\ \bibinfo {address}
  {New York},\ \bibinfo {year} {1997})\BibitemShut {NoStop}%
\bibitem [{\citenamefont {Pires}\ \emph {et~al.}(2015)\citenamefont {Pires},
  \citenamefont {C\'eleri},\ and\ \citenamefont
  {Soares-Pinto}}]{PhysRevA.91.042330}%
  \BibitemOpen
  \bibfield  {author} {\bibinfo {author} {\bibfnamefont {D.~P.}\ \bibnamefont
  {Pires}}, \bibinfo {author} {\bibfnamefont {L.~C.}\ \bibnamefont {C\'eleri}},
  \ and\ \bibinfo {author} {\bibfnamefont {D.~O.}\ \bibnamefont
  {Soares-Pinto}},\ }\bibfield  {title} {\enquote {\bibinfo {title} {Geometric
  lower bound for a quantum coherence measure},}\ }\href {\doibase
  10.1103/PhysRevA.91.042330} {\bibfield  {journal} {\bibinfo  {journal} {Phys.
  Rev. A}\ }\textbf {\bibinfo {volume} {91}},\ \bibinfo {pages} {042330}
  (\bibinfo {year} {2015})}\BibitemShut {NoStop}%
\bibitem [{\citenamefont {Giovannetti}\ \emph {et~al.}(2006)\citenamefont
  {Giovannetti}, \citenamefont {Lloyd},\ and\ \citenamefont
  {Maccone}}]{PhysRevLett.96_010401}%
  \BibitemOpen
  \bibfield  {author} {\bibinfo {author} {\bibfnamefont {V.}~\bibnamefont
  {Giovannetti}}, \bibinfo {author} {\bibfnamefont {S.}~\bibnamefont {Lloyd}},
  \ and\ \bibinfo {author} {\bibfnamefont {L.}~\bibnamefont {Maccone}},\
  }\bibfield  {title} {\enquote {\bibinfo {title} {Quantum {M}etrology},}\
  }\href {\doibase 10.1103/PhysRevLett.96.010401} {\bibfield  {journal}
  {\bibinfo  {journal} {Phys. Rev. Lett.}\ }\textbf {\bibinfo {volume} {96}},\
  \bibinfo {pages} {010401} (\bibinfo {year} {2006})}\BibitemShut {NoStop}%
\bibitem [{\citenamefont {Pezz\`{e}}\ \emph {et~al.}(2018)\citenamefont
  {Pezz\`{e}}, \citenamefont {Smerzi}, \citenamefont {Oberthaler},
  \citenamefont {Schmied},\ and\ \citenamefont
  {Treutlein}}]{RevModPhys.90.035005}%
  \BibitemOpen
  \bibfield  {author} {\bibinfo {author} {\bibfnamefont {L.}~\bibnamefont
  {Pezz\`{e}}}, \bibinfo {author} {\bibfnamefont {A.}~\bibnamefont {Smerzi}},
  \bibinfo {author} {\bibfnamefont {M.~K.}\ \bibnamefont {Oberthaler}},
  \bibinfo {author} {\bibfnamefont {R.}~\bibnamefont {Schmied}}, \ and\
  \bibinfo {author} {\bibfnamefont {P.}~\bibnamefont {Treutlein}},\ }\bibfield
  {title} {\enquote {\bibinfo {title} {Quantum metrology with nonclassical
  states of atomic ensembles},}\ }\href {\doibase 10.1103/RevModPhys.90.035005}
  {\bibfield  {journal} {\bibinfo  {journal} {Rev. Mod. Phys.}\ }\textbf
  {\bibinfo {volume} {90}},\ \bibinfo {pages} {035005} (\bibinfo {year}
  {2018})}\BibitemShut {NoStop}%
\bibitem [{\citenamefont {T{\'{o}}th}\ and\ \citenamefont
  {Apellaniz}(2014)}]{T_th_2014}%
  \BibitemOpen
  \bibfield  {author} {\bibinfo {author} {\bibfnamefont {G.}~\bibnamefont
  {T{\'{o}}th}}\ and\ \bibinfo {author} {\bibfnamefont {I.}~\bibnamefont
  {Apellaniz}},\ }\bibfield  {title} {\enquote {\bibinfo {title} {Quantum
  metrology from a quantum information science perspective},}\ }\href {\doibase
  10.1088/1751-8113/47/42/424006} {\bibfield  {journal} {\bibinfo  {journal}
  {J. Phys. A: Math. Theor.}\ }\textbf {\bibinfo {volume} {47}},\ \bibinfo
  {pages} {424006} (\bibinfo {year} {2014})}\BibitemShut {NoStop}%
\bibitem [{\citenamefont {Jozsa}(1994)}]{1994_JModOpt_41_2315}%
  \BibitemOpen
  \bibfield  {author} {\bibinfo {author} {\bibfnamefont {R.}~\bibnamefont
  {Jozsa}},\ }\bibfield  {title} {\enquote {\bibinfo {title} {Fidelity for
  {M}ixed {Q}uantum {S}tates},}\ }\href {\doibase 10.1080/09500349414552171}
  {\bibfield  {journal} {\bibinfo  {journal} {J. Mod. Opt.}\ }\textbf {\bibinfo
  {volume} {41}},\ \bibinfo {pages} {2315} (\bibinfo {year}
  {1994})}\BibitemShut {NoStop}%
\bibitem [{\citenamefont {Wigner}\ and\ \citenamefont
  {Yanase}(1963)}]{Wigner910}%
  \BibitemOpen
  \bibfield  {author} {\bibinfo {author} {\bibfnamefont {E.~P.}\ \bibnamefont
  {Wigner}}\ and\ \bibinfo {author} {\bibfnamefont {Mutsuo~M.}\ \bibnamefont
  {Yanase}},\ }\bibfield  {title} {\enquote {\bibinfo {title} {Information
  contents of distributions},}\ }\href {\doibase 10.1073/pnas.49.6.910}
  {\bibfield  {journal} {\bibinfo  {journal} {Proc. Natl. Acad. Sci.}\ }\textbf
  {\bibinfo {volume} {49}},\ \bibinfo {pages} {910} (\bibinfo {year}
  {1963})}\BibitemShut {NoStop}%
\bibitem [{\citenamefont {Lieb}(1973)}]{LIEB1973267}%
  \BibitemOpen
  \bibfield  {author} {\bibinfo {author} {\bibfnamefont {E.~H.}\ \bibnamefont
  {Lieb}},\ }\bibfield  {title} {\enquote {\bibinfo {title} {Convex trace
  functions and the {W}igner-{Y}anase-{D}yson conjecture},}\ }\href {\doibase
  10.1016/0001-8708(73)90011-X} {\bibfield  {journal} {\bibinfo  {journal}
  {Adv. Math.}\ }\textbf {\bibinfo {volume} {11}},\ \bibinfo {pages} {267}
  (\bibinfo {year} {1973})}\BibitemShut {NoStop}%
\bibitem [{\citenamefont {Lieb}\ and\ \citenamefont
  {Ruskai}(1973)}]{PhysRevLett.30.434}%
  \BibitemOpen
  \bibfield  {author} {\bibinfo {author} {\bibfnamefont {E.~H.}\ \bibnamefont
  {Lieb}}\ and\ \bibinfo {author} {\bibfnamefont {M.~B.}\ \bibnamefont
  {Ruskai}},\ }\bibfield  {title} {\enquote {\bibinfo {title} {A {F}undamental
  {P}roperty of {Q}uantum-{M}echanical {E}ntropy},}\ }\href {\doibase
  10.1103/PhysRevLett.30.434} {\bibfield  {journal} {\bibinfo  {journal} {Phys.
  Rev. Lett.}\ }\textbf {\bibinfo {volume} {30}},\ \bibinfo {pages} {434}
  (\bibinfo {year} {1973})}\BibitemShut {NoStop}%
\bibitem [{\citenamefont {Takagi}(2019)}]{Takagi_SciRep14562}%
  \BibitemOpen
  \bibfield  {author} {\bibinfo {author} {\bibfnamefont {R.}~\bibnamefont
  {Takagi}},\ }\bibfield  {title} {\enquote {\bibinfo {title} {Skew
  informations from an operational view via resource theory of asymmetry},}\
  }\href {\doibase 10.1038/s41598-019-50279-w} {\bibfield  {journal} {\bibinfo
  {journal} {Sci. Rep.}\ }\textbf {\bibinfo {volume} {9}},\ \bibinfo {pages}
  {14562} (\bibinfo {year} {2019})}\BibitemShut {NoStop}%
\bibitem [{\citenamefont {Marvian}\ and\ \citenamefont
  {Spekkens}(2014)}]{Marvian_nature}%
  \BibitemOpen
  \bibfield  {author} {\bibinfo {author} {\bibfnamefont {I.}~\bibnamefont
  {Marvian}}\ and\ \bibinfo {author} {\bibfnamefont {R.~W.}\ \bibnamefont
  {Spekkens}},\ }\bibfield  {title} {\enquote {\bibinfo {title} {Extending
  {N}oether's theorem by quantifying the asymmetry of quantum states},}\ }\href
  {\doibase 10.1038/ncomms4821} {\bibfield  {journal} {\bibinfo  {journal}
  {Nature Commun.}\ }\textbf {\bibinfo {volume} {5}},\ \bibinfo {pages} {3821}
  (\bibinfo {year} {2014})}\BibitemShut {NoStop}%
\bibitem [{\citenamefont {Miller}\ \emph {et~al.}(2019)\citenamefont {Miller},
  \citenamefont {Scandi}, \citenamefont {Anders},\ and\ \citenamefont
  {Perarnau-Llobet}}]{PhysRevLett.123.230603}%
  \BibitemOpen
  \bibfield  {author} {\bibinfo {author} {\bibfnamefont {H.~J.~D.}\
  \bibnamefont {Miller}}, \bibinfo {author} {\bibfnamefont {M.}~\bibnamefont
  {Scandi}}, \bibinfo {author} {\bibfnamefont {J.}~\bibnamefont {Anders}}, \
  and\ \bibinfo {author} {\bibfnamefont {M.}~\bibnamefont {Perarnau-Llobet}},\
  }\bibfield  {title} {\enquote {\bibinfo {title} {Work {F}luctuations in
  {S}low {P}rocesses: {Q}uantum {S}ignatures and {O}ptimal {C}ontrol},}\ }\href
  {\doibase 10.1103/PhysRevLett.123.230603} {\bibfield  {journal} {\bibinfo
  {journal} {Phys. Rev. Lett.}\ }\textbf {\bibinfo {volume} {123}},\ \bibinfo
  {pages} {230603} (\bibinfo {year} {2019})}\BibitemShut {NoStop}%
\bibitem [{\citenamefont {Scandi}\ \emph {et~al.}(2020)\citenamefont {Scandi},
  \citenamefont {Miller}, \citenamefont {Anders},\ and\ \citenamefont
  {Perarnau-Llobet}}]{PhysRevResearch.2.023377}%
  \BibitemOpen
  \bibfield  {author} {\bibinfo {author} {\bibfnamefont {M.}~\bibnamefont
  {Scandi}}, \bibinfo {author} {\bibfnamefont {H.~J.~D.}\ \bibnamefont
  {Miller}}, \bibinfo {author} {\bibfnamefont {J.}~\bibnamefont {Anders}}, \
  and\ \bibinfo {author} {\bibfnamefont {M.}~\bibnamefont {Perarnau-Llobet}},\
  }\bibfield  {title} {\enquote {\bibinfo {title} {Quantum work statistics
  close to equilibrium},}\ }\href {\doibase 10.1103/PhysRevResearch.2.023377}
  {\bibfield  {journal} {\bibinfo  {journal} {Phys. Rev. Research}\ }\textbf
  {\bibinfo {volume} {2}},\ \bibinfo {pages} {023377} (\bibinfo {year}
  {2020})}\BibitemShut {NoStop}%
\bibitem [{\citenamefont {Mashahd}()}]{MarvianThesis}%
  \BibitemOpen
  \bibfield  {author} {\bibinfo {author} {\bibfnamefont {I.~M.}\ \bibnamefont
  {Mashahd}},\ }\emph {\bibinfo {title} {Symmetry, asymmetry and quantum
  information}},\ \href {http://hdl.handle.net/10012/7088} {Ph.D. thesis},\
  \bibinfo  {school} {University of Waterloo, 2012}\BibitemShut {NoStop}%
\bibitem [{\citenamefont {Macieszczak}\ \emph {et~al.}(2019)\citenamefont
  {Macieszczak}, \citenamefont {Levi}, \citenamefont {Macr\`{\i}},
  \citenamefont {Lesanovsky},\ and\ \citenamefont
  {Garrahan}}]{PhysRevA.99.052354}%
  \BibitemOpen
  \bibfield  {author} {\bibinfo {author} {\bibfnamefont {K.}~\bibnamefont
  {Macieszczak}}, \bibinfo {author} {\bibfnamefont {E.}~\bibnamefont {Levi}},
  \bibinfo {author} {\bibfnamefont {T.}~\bibnamefont {Macr\`{\i}}}, \bibinfo
  {author} {\bibfnamefont {I.}~\bibnamefont {Lesanovsky}}, \ and\ \bibinfo
  {author} {\bibfnamefont {J.~P.}\ \bibnamefont {Garrahan}},\ }\bibfield
  {title} {\enquote {\bibinfo {title} {Coherence, entanglement, and quantumness
  in closed and open systems with conserved charge, with an application to
  many-body localization},}\ }\href {\doibase 10.1103/PhysRevA.99.052354}
  {\bibfield  {journal} {\bibinfo  {journal} {Phys. Rev. A}\ }\textbf {\bibinfo
  {volume} {99}},\ \bibinfo {pages} {052354} (\bibinfo {year}
  {2019})}\BibitemShut {NoStop}%
\bibitem [{\citenamefont {Girolami}(2014)}]{PhysRevLett.113.170401}%
  \BibitemOpen
  \bibfield  {author} {\bibinfo {author} {\bibfnamefont {D.}~\bibnamefont
  {Girolami}},\ }\bibfield  {title} {\enquote {\bibinfo {title} {Observable
  {M}easure of {Q}uantum {C}oherence in {F}inite {D}imensional {S}ystems},}\
  }\href {\doibase 10.1103/PhysRevLett.113.170401} {\bibfield  {journal}
  {\bibinfo  {journal} {Phys. Rev. Lett.}\ }\textbf {\bibinfo {volume} {113}},\
  \bibinfo {pages} {170401} (\bibinfo {year} {2014})}\BibitemShut {NoStop}%
\bibitem [{\citenamefont {Zhang}\ \emph {et~al.}(2017)\citenamefont {Zhang},
  \citenamefont {Yadin}, \citenamefont {Hou}, \citenamefont {Cao},
  \citenamefont {Liu}, \citenamefont {Huang}, \citenamefont {Maity},
  \citenamefont {Vedral}, \citenamefont {Li}, \citenamefont {Guo},\ and\
  \citenamefont {Girolami}}]{PhysRevA.96.042327}%
  \BibitemOpen
  \bibfield  {author} {\bibinfo {author} {\bibfnamefont {C.}~\bibnamefont
  {Zhang}}, \bibinfo {author} {\bibfnamefont {B.}~\bibnamefont {Yadin}},
  \bibinfo {author} {\bibfnamefont {Z.-B.}\ \bibnamefont {Hou}}, \bibinfo
  {author} {\bibfnamefont {H.}~\bibnamefont {Cao}}, \bibinfo {author}
  {\bibfnamefont {B.-H.}\ \bibnamefont {Liu}}, \bibinfo {author} {\bibfnamefont
  {Y.-F.}\ \bibnamefont {Huang}}, \bibinfo {author} {\bibfnamefont
  {R.}~\bibnamefont {Maity}}, \bibinfo {author} {\bibfnamefont
  {V.}~\bibnamefont {Vedral}}, \bibinfo {author} {\bibfnamefont {C.-F.}\
  \bibnamefont {Li}}, \bibinfo {author} {\bibfnamefont {G.-C.}\ \bibnamefont
  {Guo}}, \ and\ \bibinfo {author} {\bibfnamefont {D.}~\bibnamefont
  {Girolami}},\ }\bibfield  {title} {\enquote {\bibinfo {title} {Detecting
  metrologically useful asymmetry and entanglement by a few local
  measurements},}\ }\href {\doibase 10.1103/PhysRevA.96.042327} {\bibfield
  {journal} {\bibinfo  {journal} {Phys. Rev. A}\ }\textbf {\bibinfo {volume}
  {96}},\ \bibinfo {pages} {042327} (\bibinfo {year} {2017})}\BibitemShut
  {NoStop}%
\bibitem [{\citenamefont {Girolami}\ and\ \citenamefont
  {Yadin}(2017)}]{e19030124}%
  \BibitemOpen
  \bibfield  {author} {\bibinfo {author} {\bibfnamefont {D.}~\bibnamefont
  {Girolami}}\ and\ \bibinfo {author} {\bibfnamefont {B.}~\bibnamefont
  {Yadin}},\ }\bibfield  {title} {\enquote {\bibinfo {title} {Witnessing
  {M}ultipartite {E}ntanglement by {D}etecting {A}symmetry},}\ }\href {\doibase
  10.3390/e19030124} {\bibfield  {journal} {\bibinfo  {journal} {Entropy}\
  }\textbf {\bibinfo {volume} {19(3)}},\ \bibinfo {pages} {124} (\bibinfo
  {year} {2017})}\BibitemShut {NoStop}%
\bibitem [{\citenamefont {Yanagi}(2010)}]{Yanagi_2010}%
  \BibitemOpen
  \bibfield  {author} {\bibinfo {author} {\bibfnamefont {K.}~\bibnamefont
  {Yanagi}},\ }\bibfield  {title} {\enquote {\bibinfo {title}
  {Wigner-{Y}anase-{D}yson skew information and uncertainty relation},}\ }\href
  {\doibase 10.1088/1742-6596/201/1/012015} {\bibfield  {journal} {\bibinfo
  {journal} {J. Phys.: Conf. Ser.}\ }\textbf {\bibinfo {volume} {201}},\
  \bibinfo {pages} {012015} (\bibinfo {year} {2010})}\BibitemShut {NoStop}%
\bibitem [{\citenamefont {Braunstein}\ and\ \citenamefont
  {Caves}(1994)}]{PhysRevLett.72.3439}%
  \BibitemOpen
  \bibfield  {author} {\bibinfo {author} {\bibfnamefont {S.~L.}\ \bibnamefont
  {Braunstein}}\ and\ \bibinfo {author} {\bibfnamefont {C.~M.}\ \bibnamefont
  {Caves}},\ }\bibfield  {title} {\enquote {\bibinfo {title} {Statistical
  distance and the geometry of quantum states},}\ }\href {\doibase
  10.1103/PhysRevLett.72.3439} {\bibfield  {journal} {\bibinfo  {journal}
  {Phys. Rev. Lett.}\ }\textbf {\bibinfo {volume} {72}},\ \bibinfo {pages}
  {3439} (\bibinfo {year} {1994})}\BibitemShut {NoStop}%
\bibitem [{\citenamefont {Bengtsson}\ and\ \citenamefont
  {\.{Z}yczkowski}(2006)}]{Ingemar_Bengtsson_Zyczkowski}%
  \BibitemOpen
  \bibfield  {author} {\bibinfo {author} {\bibfnamefont {I.}~\bibnamefont
  {Bengtsson}}\ and\ \bibinfo {author} {\bibfnamefont {K.}~\bibnamefont
  {\.{Z}yczkowski}},\ }\href@noop {} {\emph {\bibinfo {title} {Geometry of
  {Q}uantum {S}tates: {A}n {I}ntroduction to {Q}uantum {E}ntanglement}}}\
  (\bibinfo  {publisher} {Cambridge University Press},\ \bibinfo {address}
  {England},\ \bibinfo {year} {2006})\BibitemShut {NoStop}%
\bibitem [{\citenamefont {Luo}(2004)}]{Luoproc1322004}%
  \BibitemOpen
  \bibfield  {author} {\bibinfo {author} {\bibfnamefont {S.}~\bibnamefont
  {Luo}},\ }\bibfield  {title} {\enquote {\bibinfo {title} {Wigner-{Y}anase
  skew information vs. quantum {F}isher information},}\ }\href {\doibase
  10.1090/S0002-9939-03-07175-2} {\bibfield  {journal} {\bibinfo  {journal}
  {Proc. Amer. Math. Soc.}\ }\textbf {\bibinfo {volume} {132}},\ \bibinfo
  {pages} {885} (\bibinfo {year} {2004})}\BibitemShut {NoStop}%
\bibitem [{\citenamefont {Gibilisco}\ \emph {et~al.}(2009)\citenamefont
  {Gibilisco}, \citenamefont {Imparato},\ and\ \citenamefont
  {Isola}}]{Gibilisco2008137}%
  \BibitemOpen
  \bibfield  {author} {\bibinfo {author} {\bibfnamefont {P.}~\bibnamefont
  {Gibilisco}}, \bibinfo {author} {\bibfnamefont {D.}~\bibnamefont {Imparato}},
  \ and\ \bibinfo {author} {\bibfnamefont {T.}~\bibnamefont {Isola}},\
  }\bibfield  {title} {\enquote {\bibinfo {title} {Inequalities for quantum
  {F}isher information},}\ }\href {\doibase 10.1090/S0002-9939-08-09447-1}
  {\bibfield  {journal} {\bibinfo  {journal} {Proc. Amer. Math. Soc.}\ }\textbf
  {\bibinfo {volume} {137}},\ \bibinfo {pages} {317} (\bibinfo {year}
  {2009})}\BibitemShut {NoStop}%
\bibitem [{\citenamefont {Horodecki}\ and\ \citenamefont
  {Horodecki}(1996)}]{PhysRevA.54.1838}%
  \BibitemOpen
  \bibfield  {author} {\bibinfo {author} {\bibfnamefont {R.}~\bibnamefont
  {Horodecki}}\ and\ \bibinfo {author} {\bibfnamefont {M.}~\bibnamefont
  {Horodecki}},\ }\bibfield  {title} {\enquote {\bibinfo {title}
  {Information-theoretic aspects of inseparability of mixed states},}\ }\href
  {\doibase 10.1103/PhysRevA.54.1838} {\bibfield  {journal} {\bibinfo
  {journal} {Phys. Rev. A}\ }\textbf {\bibinfo {volume} {54}},\ \bibinfo
  {pages} {1838} (\bibinfo {year} {1996})}\BibitemShut {NoStop}%
\bibitem [{\citenamefont {Horodecki}\ \emph {et~al.}(2009)\citenamefont
  {Horodecki}, \citenamefont {Horodecki}, \citenamefont {Horodecki},\ and\
  \citenamefont {Horodecki}}]{RevModPhys.81.865}%
  \BibitemOpen
  \bibfield  {author} {\bibinfo {author} {\bibfnamefont {R.}~\bibnamefont
  {Horodecki}}, \bibinfo {author} {\bibfnamefont {P.}~\bibnamefont
  {Horodecki}}, \bibinfo {author} {\bibfnamefont {M.}~\bibnamefont
  {Horodecki}}, \ and\ \bibinfo {author} {\bibfnamefont {K.}~\bibnamefont
  {Horodecki}},\ }\bibfield  {title} {\enquote {\bibinfo {title} {Quantum
  entanglement},}\ }\href {\doibase 10.1103/RevModPhys.81.865} {\bibfield
  {journal} {\bibinfo  {journal} {Rev. Mod. Phys.}\ }\textbf {\bibinfo {volume}
  {81}},\ \bibinfo {pages} {865} (\bibinfo {year} {2009})}\BibitemShut
  {NoStop}%
\bibitem [{\citenamefont {D\"ur}\ \emph {et~al.}(2000)\citenamefont {D\"ur},
  \citenamefont {Vidal},\ and\ \citenamefont {Cirac}}]{PhysRevA.62.062314}%
  \BibitemOpen
  \bibfield  {author} {\bibinfo {author} {\bibfnamefont {W.}~\bibnamefont
  {D\"ur}}, \bibinfo {author} {\bibfnamefont {G.}~\bibnamefont {Vidal}}, \ and\
  \bibinfo {author} {\bibfnamefont {J.~I.}\ \bibnamefont {Cirac}},\ }\bibfield
  {title} {\enquote {\bibinfo {title} {Three qubits can be entangled in two
  inequivalent ways},}\ }\href {\doibase 10.1103/PhysRevA.62.062314} {\bibfield
   {journal} {\bibinfo  {journal} {Phys. Rev. A}\ }\textbf {\bibinfo {volume}
  {62}},\ \bibinfo {pages} {062314} (\bibinfo {year} {2000})}\BibitemShut
  {NoStop}%
\bibitem [{\citenamefont {Macr\`{\i}}\ \emph {et~al.}(2016)\citenamefont
  {Macr\`{\i}}, \citenamefont {Smerzi},\ and\ \citenamefont
  {Pezz\`e}}]{PhysRevA.94.010102}%
  \BibitemOpen
  \bibfield  {author} {\bibinfo {author} {\bibfnamefont {T.}~\bibnamefont
  {Macr\`{\i}}}, \bibinfo {author} {\bibfnamefont {A.}~\bibnamefont {Smerzi}},
  \ and\ \bibinfo {author} {\bibfnamefont {L.}~\bibnamefont {Pezz\`e}},\
  }\bibfield  {title} {\enquote {\bibinfo {title} {Loschmidt echo for quantum
  metrology},}\ }\href {\doibase 10.1103/PhysRevA.94.010102} {\bibfield
  {journal} {\bibinfo  {journal} {Phys. Rev. A}\ }\textbf {\bibinfo {volume}
  {94}},\ \bibinfo {pages} {010102} (\bibinfo {year} {2016})}\BibitemShut
  {NoStop}%
\bibitem [{\citenamefont {Borish}\ \emph {et~al.}(2020)\citenamefont {Borish},
  \citenamefont {Markovi\ifmmode~\acute{c}\else \'{c}\fi{}}, \citenamefont
  {Hines}, \citenamefont {Rajagopal},\ and\ \citenamefont
  {Schleier-Smith}}]{1910_arxiv_1910.13687}%
  \BibitemOpen
  \bibfield  {author} {\bibinfo {author} {\bibfnamefont {V.}~\bibnamefont
  {Borish}}, \bibinfo {author} {\bibfnamefont {O.}~\bibnamefont
  {Markovi\ifmmode~\acute{c}\else \'{c}\fi{}}}, \bibinfo {author}
  {\bibfnamefont {J.~A.}\ \bibnamefont {Hines}}, \bibinfo {author}
  {\bibfnamefont {S.~V.}\ \bibnamefont {Rajagopal}}, \ and\ \bibinfo {author}
  {\bibfnamefont {M.}~\bibnamefont {Schleier-Smith}},\ }\bibfield  {title}
  {\enquote {\bibinfo {title} {Transverse-{F}ield {I}sing {D}ynamics in a
  {R}ydberg-{D}ressed {A}tomic {G}as},}\ }\href {\doibase
  10.1103/PhysRevLett.124.063601} {\bibfield  {journal} {\bibinfo  {journal}
  {Phys. Rev. Lett.}\ }\textbf {\bibinfo {volume} {124}},\ \bibinfo {pages}
  {063601} (\bibinfo {year} {2020})}\BibitemShut {NoStop}%
\bibitem [{\citenamefont {Rhim}\ \emph {et~al.}(1971)\citenamefont {Rhim},
  \citenamefont {Pines},\ and\ \citenamefont {Waugh}}]{PhysRevB.3.684}%
  \BibitemOpen
  \bibfield  {author} {\bibinfo {author} {\bibfnamefont {W-K.}\ \bibnamefont
  {Rhim}}, \bibinfo {author} {\bibfnamefont {A.}~\bibnamefont {Pines}}, \ and\
  \bibinfo {author} {\bibfnamefont {J.~S.}\ \bibnamefont {Waugh}},\ }\bibfield
  {title} {\enquote {\bibinfo {title} {Time-{R}eversal {E}xperiments in
  {D}ipolar-{C}oupled {S}pin {S}ystems},}\ }\href {\doibase
  10.1103/PhysRevB.3.684} {\bibfield  {journal} {\bibinfo  {journal} {Phys.
  Rev. B}\ }\textbf {\bibinfo {volume} {3}},\ \bibinfo {pages} {684} (\bibinfo
  {year} {1971})}\BibitemShut {NoStop}%
\bibitem [{\citenamefont {Zangara}\ \emph {et~al.}(2016)\citenamefont
  {Zangara}, \citenamefont {Bendersky}, \citenamefont {Levstein},\ and\
  \citenamefont {Pastawski}}]{10.1098_rsta.2015.0163}%
  \BibitemOpen
  \bibfield  {author} {\bibinfo {author} {\bibfnamefont {P.~R.}\ \bibnamefont
  {Zangara}}, \bibinfo {author} {\bibfnamefont {D.}~\bibnamefont {Bendersky}},
  \bibinfo {author} {\bibfnamefont {P.~R.}\ \bibnamefont {Levstein}}, \ and\
  \bibinfo {author} {\bibfnamefont {H.~M.}\ \bibnamefont {Pastawski}},\
  }\bibfield  {title} {\enquote {\bibinfo {title} {Loschmidt echo in many-spin
  systems: contrasting time scales of local and global measurements},}\ }\href
  {\doibase 10.1098/rsta.2015.0163} {\bibfield  {journal} {\bibinfo  {journal}
  {Phil. Trans. R. Soc. A}\ }\textbf {\bibinfo {volume} {374}},\ \bibinfo
  {pages} {20150163} (\bibinfo {year} {2016})}\BibitemShut {NoStop}%
\bibitem [{\citenamefont {Bhatia}(2006)}]{BHATIA2006355}%
  \BibitemOpen
  \bibfield  {author} {\bibinfo {author} {\bibfnamefont {R.}~\bibnamefont
  {Bhatia}},\ }\bibfield  {title} {\enquote {\bibinfo {title} {Interpolating
  the arithmetic-geometric mean inequality and its operator version},}\ }\href
  {\doibase 10.1016/j.laa.2005.03.005} {\bibfield  {journal} {\bibinfo
  {journal} {Lin. Alg. Appl.}\ }\textbf {\bibinfo {volume} {413}},\ \bibinfo
  {pages} {355} (\bibinfo {year} {2006})}\BibitemShut {NoStop}%
\bibitem [{\citenamefont {Audenaert}(2007)}]{Audenaert2007}%
  \BibitemOpen
  \bibfield  {author} {\bibinfo {author} {\bibfnamefont {K.}~\bibnamefont
  {Audenaert}},\ }\bibfield  {title} {\enquote {\bibinfo {title} {A singular
  value inequality for {H}einz means},}\ }\href {\doibase
  10.1016/j.laa.2006.10.006} {\bibfield  {journal} {\bibinfo  {journal} {Lin.
  Alg. Appl.}\ }\textbf {\bibinfo {volume} {422}},\ \bibinfo {pages} {279}
  (\bibinfo {year} {2007})}\BibitemShut {NoStop}%
\end{thebibliography}

%

%
\end{document}